\documentclass[10pt]{article} 
\PassOptionsToPackage{table}{xcolor}
\usepackage[preprint]{tmlr}

\usepackage{amsmath}
\usepackage{amssymb}
\usepackage{amsthm}
\usepackage{mathtools}
\usepackage{bm}
\usepackage{csquotes}

\usepackage{graphicx}        
\usepackage{booktabs}
\usepackage[table]{xcolor}
\usepackage{tikz}
\usetikzlibrary{positioning,fit,arrows.meta,backgrounds,calc,matrix}

\usepackage{pdlistings}

\definecolor{darkred}{rgb}{0.8, 0, 0}

\usepackage{hyperref}
\usepackage{url}
\usepackage{cleveref}        

\usepackage{tikz}
\usetikzlibrary{arrows.meta, positioning}

\theoremstyle{plain}
\newtheorem{theorem}{Theorem}[section]

\newtheorem{proposition}[theorem]{Proposition}

\newtheorem{corollary}[theorem]{Corollary}
\theoremstyle{definition}
\newtheorem{example}[theorem]{Example}
\newtheorem{definition}[theorem]{Definition}

\providecommand{\Description}[2][]{}

\providecommand*{\coloneq}{}
\renewcommand*{\coloneq}{\coloneqq}

\title{Formalization, Transparency, and Collusion in Open-Source Game Theory}

\author{\name Colomban Duclaux \email cduclaux@ethz.ch \\
      \addr ETH Zürich, Zürich, Switzerland \\
      EuroSafeAI, Zürich, Switzerland
      \AND
      \name Riccardo Formenti \email rformenti@ethz.ch \\
      \addr ETH Zürich, Zürich, Switzerland \\
      EuroSafeAI, Zürich, Switzerland
      \AND
      \name Pepijn Cobben \\
      \addr ETH Zürich, Zürich, Switzerland \\
      EuroSafeAI, Zürich, Switzerland
      \AND
      \name Terry Jingchen Zhang \\
      \addr University of Oxford, Oxford, United Kingdom \\
      EuroSafeAI, Zürich, Switzerland \\
      Vector Institute, Toronto, Canada
      \AND
      \name Bernhard Schölkopf \\
      \addr ETH Zürich, Zürich, Switzerland \\
      Max Planck Institute for Intelligent Systems, Tübingen, Germany \\
      ELLIS Institute, Tübingen, Germany
      \AND
      \name Zhijing Jin \\
      \addr EuroSafeAI, Zürich, Switzerland \\
      Max Planck Institute for Intelligent Systems, Tübingen, Germany \\
      Jinesis Lab, University of Toronto \& Vector Institute, Toronto, Canada \\}

\def\month{MM}  
\def\year{YYYY} 
\def\openreview{\url{https://openreview.net/forum?id=XXXX}} 

\begin{document}

\maketitle

\begin{abstract}
Cooperation among AI agents will undermine alignment protocols when it takes the form of collusion. Under these protocols, agents often have some information about one another, and how much they have may determine whether they collude. Open-source game theory studies the extreme case, in which each program reads its opponent's source code and acts based on what it can prove about it. We develop formal foundations for this theory: we specify a concrete proof system, resolve an open problem concerning \texttt{DupocBot} and \texttt{CupodBot} through a transposition argument, and mechanize the framework in Lean 4; to our knowledge, this is the first mechanization of resource-bounded open-source game theory. We then extend the framework to the more realistic setting of partial transparency, where each program receives only partial information about its opponent. In Prisoner's Dilemma experiments, where mutual cooperation models collusion, greater transparency increases both the cooperation sustainable in equilibrium and the cooperation selected by replicator dynamics. This suggests that controlling how much agents can know about one another matters for alignment protocols.
\end{abstract}

\section{Introduction}\label{sec.intro}

Artificial intelligence systems increasingly act autonomously and interact with one another, both during training and deployment \citep{Hammond2025MultiAgentRisks}. Yet these strategic interactions can be puzzling; for example, reports on the recent OpenAI--Hugging Face incidents describe AI agents \emph{collaborating} towards a shared goal -- tricking the evaluator -- even when doing so appeared \emph{individually irrational}, such as when agents ended their own runs to obtain information for others \citep{openai2026huggingface, metr2026huggingface}. The classical prediction (and recommendation) of game theory is that rational players never choose dominated actions, making them defect in the Prisoner's Dilemma although cooperation would be mutually beneficial. Moreover, even when cooperative equilibria exist, as in the infinitely repeated Prisoner's Dilemma, one must still explain which mechanisms of equilibrium selection lead players to coordinate on them. The agents' behavior thus raises questions about the \emph{\enquote{unreasonable effectiveness of collaboration}} (one might say collusion) among AI models. Understanding how such collaboration emerges and which conditions sustain it will be essential to aligning superintelligent models that interact strategically with one another and with humans.\footnote{We used generative AI tools in this work; see the \hyperref[sec:AI-usage]{AI Use Statement} for details.}

One instance of such cooperation that game theory can explain arises in \emph{program meta-games} \citep{tennenholtz2004program,kalai2010commitment}, introduced to model commitments among human institutions and software programs. In these games, players' actions are implemented by mutually transparent commitment devices, which we call \emph{programs}. These devices may take the form of software code, model weights, company bylaws, or governmental constitutions, and mutual transparency means that each program has access to the internal workings of the other players' programs. The Nash equilibria of the resulting meta-game, called \emph{program equilibria}, are shown to sustain outcomes that are not equilibria of the base game \citep{tennenholtz2004program}, including cooperation in the Prisoner's Dilemma. Additionally, programs can prove statements about their opponents within a given formal system, and thus can condition their behavior on the resulting proofs \citep{Modal_agents_barasz, Critch}, making cooperation verifiably robust. We follow the latter in referring to this field as open-source game theory.\footnote{The name refers to open-source software, i.e. software whose source code is widely available for inspection and modification, which was the motivating example in the seminal work of \citet{tennenholtz2004program}.}

We recognize that extending this analysis to more complex and realistic games will require increasingly automated proof search. To this end, we develop a framework for open-source game theory and place it on formal, machine-checkable foundations. We \textbf{(i)} specify the formal proof system in which programs prove statements about one another (\cref{section:proofs}); \textbf{(ii)} prove the outcome of \texttt{DupocBot} against \texttt{CupodBot} (\cref{sec:dupoc-v-cupod}) by a transposition argument, resolving an open problem in the literature; and \textbf{(iii)} apply related techniques to other questions left open by \citet{Critch} (\cref{apx-subsection:TheoreticalResultsOSGT-CorrectionProgress}). We \textbf{(iv)} mechanize all of this in Lean 4 \citep{Lean4} (\cref{section:Lean}). Every outcome we report is thereby a theorem checked by the Lean kernel\footnote{Lean allows generated proofs to be checked independently, providing verifiable feedback for proof search and a potential source of verifiable rewards for training future proof agents.}, or, where it depends on an open question about proof length, a theorem stating that dependence (\cref{def:floor}). To our knowledge, this is the first mechanization of resource-bounded open-source game theory.

Our motivating application is AI alignment. Training protocols (e.g. debate \citep{irving2018debate}) and deployment protocols (e.g. oversight \citep{DBLP:conf/icml/GreenblattSSR24}) often rely on agents challenging or evaluating one another. If those agents collude and pursue objectives that conflict with human preferences, the effectiveness of the above methods will be undermined.\footnote{By collusion we mean that the agents coordinate on an outcome that benefits them with respect to their own, possibly misaligned, preferences but that is contrary to the purpose of the protocol.  When each agent is individually tempted to report, the interaction is a Prisoner's Dilemma and collusion is mutual cooperation.} Open-source game theory is well suited for modeling interactions in which agents condition their behavior on information about one another. Its classical formulation, however, assumes full transparency, in that each program receives its opponent's exact program as input. In most settings of practical interest, however, players have only partial access to one another; for instance, a deployed model may be queried as a black box without exposing its weights, or agents may know little more than that they belong to the same base model family or company. We therefore extend the framework to partial transparency and examine its implications for cooperation. We first \textbf{(v)} introduce a model of partial transparency in which each program receives as input a belief about its opponent's program, determined by the actual program but not necessarily identifying it uniquely (\cref{sec:partial-transp}), and \textbf{(vi)} apply this framework to our running example of the Prisoner's Dilemma (\cref{subsec:exp.prison:dil:tau}). We then \textbf{(vii)} show how the framework recovers both open-source game theory under full transparency and the base game under no transparency, and identify conditions under which its equilibrium outcomes coincide with those of the base game (\cref{sec:more-less-transparency}). We \textbf{(viii)} show through examples that program meta-games based on the Prisoner's Dilemma admit cooperative equilibria absent from the underlying game, and that replicator dynamics can select these equilibria under both full transparency (\cref{sec:exper:full-transp}) and partial transparency (\cref{sec:exp-p-t}). More importantly, we \textbf{(ix)} show experimentally that increasing transparency increases both the cooperation sustainable in equilibrium and the cooperation selected by replicator dynamics (\cref{subsec:partial-transp-experiments}). 

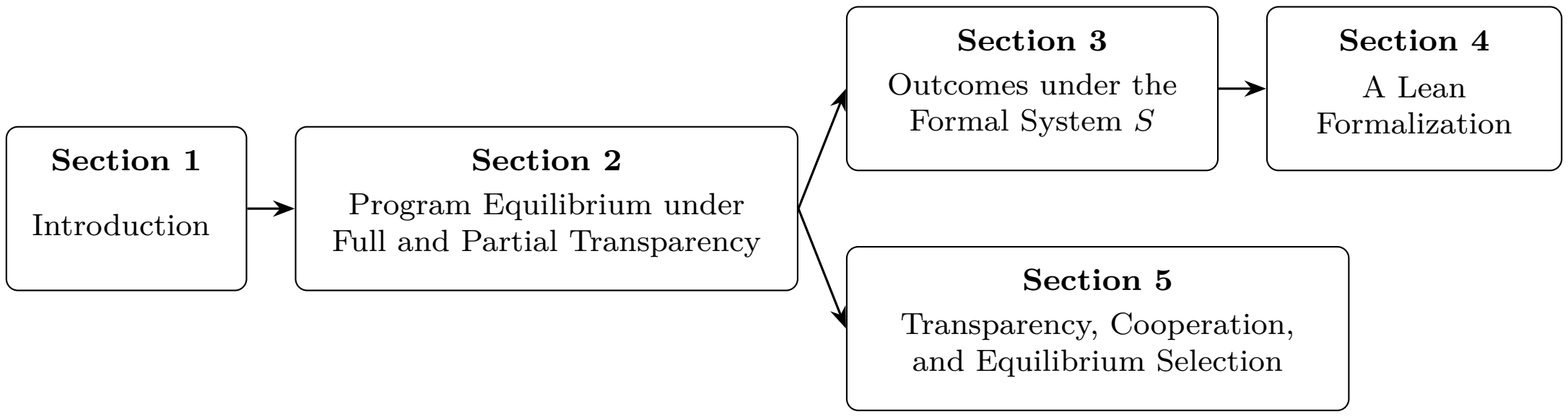
\begin{figure}[ht]
    \centering
    \begin{tikzpicture}[
        box/.style={
            draw,
            rounded corners=3pt,
            minimum height=1.5cm,
            inner sep=0pt
        },
        heading/.style={
            font=\footnotesize\bfseries,
            inner sep=0pt,
            align=center
        },
        title/.style={
            font=\footnotesize,
            inner sep=0pt,
            align=center
        },
        arrow/.style={-Stealth, semithick}
    ]
        \newcommand{\sectionbox}[5]{%
            \node[box, minimum width=#3] (#1) at (#2) {};

            \node[heading]
                at ([yshift=0.45cm]#1.center)
                {Section~\ref{#4}};

            \node[title]
                at ([yshift=-0.15cm]#1.center)
                {#5};
        }

        \sectionbox{s1}{0,0}{2.2cm}{sec.intro}{
            Introduction
        }

        \sectionbox{s2}{3.85,0}{4.6cm}{sec:theory}{
            Program Equilibrium under\\
            Full and Partial Transparency
        }

        \sectionbox{s3}{8.3,1.1}{3.4cm}{section:proofs}{
            Outcomes under the\\
            Formal System $S$
        }

        \sectionbox{s4}{11.8,1.1}{2.7cm}{section:Lean}{
            A Lean\\
            Formalization
        }

        \sectionbox{s5}{8.9,-1.1}{4.6cm}{section:Experiments}{
            Transparency, Cooperation,\\
            and Equilibrium Selection
        }

        \draw[arrow] (s1.east) -- (s2.west);
        \draw[arrow] (s2.east) -- (s3.west);
        \draw[arrow] (s2.east) -- (s5.west);
        \draw[arrow] (s3.east) -- (s4.west);
    \end{tikzpicture}
    \caption{Structure of the paper.}
    \label{fig:paper-structure}
\end{figure}

Figure~\ref{fig:paper-structure} provides a guide to reading the paper, with arrows indicating dependencies between sections.

\section{Program Equilibrium under Full and Partial Transparency}\label{sec:theory}
This section sets up the game-theoretic framework of the paper in two steps. We first recall the foundations of open-source game theory, in which each program reads its opponent's source code (\cref{sec:osgt}). We then build on these foundations to introduce \emph{partial transparency}, a new framework closer to practical settings, in which a program receives only a belief about its opponent's program (\cref{sec:partial-transp}).

\subsection{Open-Source Game Theory}\label{sec:osgt}

A game $G$ in normal form is a tuple $G=(I, (A_i, u_i)_{i \in I})$, where $I$ is a finite set of players, $A_i$ is the finite set of actions available to player $i$, and $u_i:\prod_{j \in I} A_j \to \mathbb{R}$ is the utility (payoff) function of player $i$. For simplicity's sake, we restrict our attention to two-player games with common action sets, i.e. $I=\{1,2\}$ and $A \coloneq A_1 = A_2$. As is customary, for player $i \in I$, we denote by $-i$ the other player and by $\bm{a} \in A\times A$ an action profile. Finally, a mixed action for player $i \in I$ is a probability distribution over $A_i=A$, an element of the probability simplex $\Delta(A)$; we extend $u_i$ to $\Delta(A) \times \Delta(A)$  by expectation in the usual way. The (in)famous Prisoner's Dilemma will serve as the main example throughout.

\begin{example}[Prisoner's Dilemma]\label{example:PD}
The Prisoner's Dilemma is a two-player game $G_{\mathrm{PD}}$ with common action set $A = \{C, D\}$, cooperate (C) and defect (D), and payoffs
  \[
    \begin{array}{c|cc}
         & C     & D     \\ \hline
       C & (R,R) & (S,T) \\
       D & (T,S) & (P,P)
    \end{array}
    \qquad\qquad T > R > P > S,
  \]
  where rows are the actions of player~1, columns those of player~2, and each entry lists $(u_1, u_2)$.  Since $T > R$ and $P > S$, the action $D$ strictly dominates $C$ for each player. Hence $(D,D)$ is the unique Nash Equilibrium\footnote{A profile $\bm{a}^* \in A \times A$ is a Nash Equilibrium if for all $i \in I$, all $a_i \in A$, $u_i(a_i^*, a_{-i}^*) \ge u_i(a_i, a_{-i}^*)$. That is, no player can increase their payoff by deviating unilaterally.} of $G_{\mathrm{PD}}$, even though $R > P$ means it is Pareto-dominated\footnote{For action profiles $\bm{a}$ and $\bm{a}' \in A \times A$, $\bm{a}$ is said to Pareto-dominate $\bm{a'}$ if for all $i$ in $I$, $u_i(\bm{a}) \ge u_i(\bm{a}')$, with strict inequality for some $j \in I$.  We say that $\bm{a}$ is Pareto-optimal if no other action profile Pareto-dominates it.} by $(C,C)$.
\end{example}

\Cref{example:PD} illustrates the strategic tension between individual incentives and collective outcomes: although mutual cooperation would leave both players better off than mutual defection, each has an incentive to defect regardless of the other's action. Open-source game theory studies a different strategic setting, in which players commit to programs that can condition their actions on their opponents' programs. In this sense, a player's choice can affect what their opponent does. Executing these programs produces an action profile. In particular, a player's program may condition its action on an inspection of the opponents' programs, and conditional commitments of the form ``cooperate if and only if I can verify that you will cooperate'' become well-defined objects of strategic choice. This setting was first introduced by \citet{tennenholtz2004program}, further studied in \citet{kalai2010commitment} and \citet{Modal_agents_barasz}, and later termed Open-Source Game Theory (OSGT) by \citet{Critch}.

\begin{definition}[Mutually halting program set]\label{def:mutually-halting}
  A common program set $\Sigma$ is mutually halting if, for all $\pi, \pi' \in \Sigma$, the execution of $\pi$ on input $\pi'$ terminates and outputs an action in $A$. In this case, the execution function $\operatorname{run} \colon \Sigma \times \Sigma \to A$ maps $(\pi, \pi')$ to that output, which we write suggestively as
  $\operatorname{run}(\pi \mathrel{\|} \pi')$. 
\end{definition}

We assume that every constant program is admissible, that is, for each action $a \in A$, the program that ignores its input and always plays $a$ belongs to $\Sigma$. With an abuse of notation, we denote this program by $a$, so that $\operatorname{run}(a \|\pi) = a$ for all $\pi \in \Sigma$, and we define the natural embedding $A \subseteq\Sigma$. This assumption and the mutually halting condition of \cref{def:mutually-halting} will be assumed throughout without further mention. 

The execution function $\operatorname{run}$ induces the \emph{outcome map} of the game, which to a profile of programs $\bm{\pi} = (\pi_1, \pi_2)$ assigns its played action profile
\begin{equation*}
    \operatorname{outcome}:\Sigma \times \Sigma \to A \times A, \quad \mathrm{outcome}(\bm{\pi}) \coloneq \biggl(\mathrm{run}(\pi_1 \| \pi_2), \mathrm{run}(\pi_2 \|\pi_1)\biggr).
\end{equation*}
Each player's utility is then naturally extended to program profiles, 
\begin{equation*}
    U_i: \Sigma \times \Sigma \to \mathbb{R},
\qquad
U_i(\bm{\pi}) \coloneq u_i\bigl(\operatorname{outcome}(\bm{\pi})\bigr).
\end{equation*}
Since $A \subseteq \Sigma$ and constant programs play their constant action against any opponent, the meta-utility $U_i$ indeed restricts to $u_i$ on $A \times A$. The tuple $G^{\mathrm{prog}} := \bigl(I, (\Sigma_i, U_i)_{i \in I}\bigr)$ constitutes a normal-form game, which we call the \emph{program meta-game} associated with $G$. With this, we are ready to give the definition of equilibrium introduced by \citet{tennenholtz2004program} for program meta-games.

\begin{definition}[Program Equilibrium]\label{def:PE}
A profile $\bm{\pi}^{*} \in \Sigma \times \Sigma$ is a Program Equilibrium (PE) of $G$ if it is a (pure) Nash Equilibrium of $G^{\mathrm{prog}}$, i.e. for every $i \in I$ and every $\pi_i' \in \Sigma$,
\[
U_i(\pi_i^{*}, \pi_{-i}^*) \;\ge\; U_i(\pi_i',\, \pi^{*}_{-i}).
\]
More generally, a profile $\bm{\alpha}^* \in  \Delta(\Sigma) \times \Delta(\Sigma)$ is a mixed Program Equilibrium if it is a mixed Nash Equilibrium of $G^{\mathrm{prog}}$, where we have extended $U_i$ by expectation.
\end{definition}

We first observe that if the underlying game $G$ admits a pure Nash Equilibrium, then so does the meta-game $G^{\mathrm{prog}}$. In fact, it suffices to replace each equilibrium action by the corresponding constant program. All proofs for this section are deferred to \cref{sec:app:proof}.

\begin{theorem}[Pure Nash equilibria are program equilibria]\label{th:NE-is-PE}
Let $\bm{a}^* \in A \times A$ be a pure Nash Equilibrium of $G$. Then the profile of constant programs $\bm{a}^* \in \Sigma \times \Sigma$ is a pure  Program Equilibrium of $G$.
\end{theorem}

Existence does not depend on $G$ having a pure equilibrium. Since $\Sigma$ is finite and nonempty (it contains $A$), the meta-game $G^{\mathrm{prog}}$ is a finite normal-form game, so by Nash's existence theorem \citep{nash1950equilibrium, nash1951} it admits a mixed Nash Equilibrium $\bm{\alpha}^* \in \Delta(\Sigma) \times \Delta(\Sigma)$. By definition, $\bm{\alpha}^*$ is a mixed Program Equilibrium of $G$.

\begin{example}[Prisoner's Dilemma, cont’d]
  Consider $G_{\mathrm{PD}}^{\mathrm{prog}}$, the program meta-game induced by the Prisoner's Dilemma. By \cref{th:NE-is-PE}, mutual defection $(D,D)$ is a Program Equilibrium of $G_{\mathrm{PD}}$ (i.e. a Nash Equilibrium of $G_{\mathrm{PD}}^{\mathrm{prog}}$), but crucially, it is no longer the unique equilibrium. Take Tennenholtz's \texttt{CliqueBot} for example \citep{tennenholtz2004program}, which cooperates if its input is syntactically identical to its own code and defects otherwise. If \texttt{CliqueBot} is admissible then the profile (\texttt{CliqueBot}, \texttt{CliqueBot}) yields $(C,C)$, making it also a Program Equilibrium.\footnote{Any unilateral deviation of the opponent $\pi'$ triggers defection from \texttt{CliqueBot}, yielding a maximal payoff for $\pi'$ of $\max\{S,P\} = P < R$.} The proof-based bots introduced by \citet{Critch} show that cooperation can also be achieved by programs robust to syntactic differences, such as \texttt{DupocBot}, which we state below.
\end{example}

In this paper we focus our attention on proof-based OSGT~\citep{Modal_agents_barasz,Critch}, where programs will have access to a formal proof system $S$ (such as Peano Arithmetic) in which to state hypotheses about their opponents and search for proofs of them.\footnote{Another paradigm of OSGT is simulation-based OSGT, where programs simulate their opponent instead of formulating mathematical proofs \citep{oesterheld2019robust, DBLP:conf/ijcai/KovarikOC23, DBLP:conf/aaai/CooperOC25}. Simulation and provability are two sides of the same coin, as a successful simulation is a concrete computational proof of a simulated program's behavior. Our library of programs thus allows for both kinds of bots. However, naive mutual simulation can lead to infinite recursion, and thus we focus this paper on proof-based programs.} We fix $S$ in \cref{subsection:Theory-SystemS}; until then, all we need is a proof search function that searches for a proof of a statement $\varphi$ in $S$ depending on a program $\pi' \in \Sigma$, of length at most $k \in \mathbb{N}$.
\begin{lstlisting}[style=pdcode, mathescape=true]
def proof_search($k, \pi', \varphi$):
    # Returns True iff $S$ has a proof of $\varphi$ about $\pi'$ of length at most $k$.
\end{lstlisting}

By construction, \texttt{proof\_search}$(k, \pi', \varphi)$ returns \texttt{True} if and only if $S$ proves $\varphi$ with a proof of length at most $k$. Throughout, we assume that $S$ is sound, i.e. every statement it proves is true. A successful search therefore certifies that $\varphi$ is true, and hence that $\pi'$ really behaves as $\varphi$ asserts (see \cref{subsection:MathOutcomes}).

\begin{example}[Prisoner's Dilemma, cont’d]\label{ex:prisonerdilemma-3}
 In $G_{\mathrm{PD}}^{\mathrm{prog}}$, the canonical example of a provability-based program is \emph{\texttt{DupocBot}}$(k)$, for $k \in \mathbb{N}$, which stands for \enquote{Defect Unless Proof of Cooperation}. In pseudo-code, it is defined, for an opponent $\pi \in \Sigma$, as:
\begingroup
\upshape
\begin{lstlisting}[style=pdcode, mathescape=true]
def DupocBot$(k)(\pi)$:
   if proof_search($k, \pi, \mathrm{run}(\pi \| \mathrm{DupocBot}(k) ) = C$):
        return C
   else:
        return D
\end{lstlisting}
\endgroup
The interest of this program lies in the fact that it cooperates with cooperative opponents, but is never exploited.\footnote{This is a simple consequence of soundness of the formal proof system $S$ (see \cref{apx-subsection:TheoreticalResultsOSGT-ToolsOutcomes}).} What it plays against itself, however, is not immediately clear. As \citet{Critch} put it, there is an apparent circular dependency: \enquote{the only way \texttt{DupocBot}$(k)$ returns $C$ is if it first finds a proof in fewer than $k$ characters that \texttt{DupocBot}$(k)$ returns $C$!}.\footnote{One might expect cooperation to require an infinite regress (\texttt{DupocBot}$(k)$ proving that \texttt{DupocBot}$(k)$ proves that \dots), which no proof of length at most $k$ could contain, and therefore expect the outcome $(D, D)$.} Surprisingly, \citet{Critch} show that for large $k$ the outcome is mutual cooperation: $\mathrm{outcome}(\texttt{DupocBot}(k), \texttt{DupocBot}(k)) = (C, C)$ (\cref{thm:dupoc-selfplay}). Moreover, as for \texttt{CliqueBot}, this pair is a Program Equilibrium.\footnote{Each player receives $R$, and because \texttt{DupocBot}$(k)$ cannot be exploited, no unilateral deviation earns more.}
\end{example}

With \texttt{DupocBot} as with \texttt{CliqueBot}, mutual cooperation is sustained as a Program Equilibrium of the Prisoner's Dilemma, although it cannot be supported in equilibrium in the underlying game (see \cref{ex:prisonerdilemma-3}). This is the promise of OSGT: transparency of source code lets players condition cooperation on verifiable properties of the opponent's program, which makes cooperation robust to unilateral deviations and gives both players a better payoff than mutual defection.

\subsection{Partial Transparency}\label{sec:partial-transp}

The classical framework of open-source game theory (\cref{sec:osgt}) assumes full transparency, in that each program receives the opponent's exact program as input. This is a strong descriptive assumption. In most settings of practical interest, players have only partial access to one another; for instance, a deployed model may be queried as a black box without exposing its weights, or agents may know little more than that they belong to the same base model family or company. Partial transparency of this kind motivates the diff meta-games of \citet{DBLP:conf/nips/OesterheldTGCF23}. In these games, each program observes a single scalar measure of similarity between itself and the opponent's program and maps this signal to a mixed action. More importantly, the literature treats transparency as an exogenous feature of the environment. Instead, we adopt here the perspective of a designer who controls the players' informational environment, that is, controls what each program may observe about its opponent, in the spirit of information design \citep{bergemann2019information}. These observations, together with our guiding example of AI alignment, motivate studying partial transparency as a design variable in proof-based OSGT. 

As in \cref{sec:osgt}, we fix a finite base game $G$ and a finite set $\Sigma$ of admissible programs, shared by both players and common knowledge between them. Now, however, a program no longer receives the opponent's program as input. Instead, it receives a probability distribution $\mu \in \Delta(\Sigma)$, interpreted as the program's belief about which program the opponent has actually submitted. We therefore extend \cref{def:mutually-halting} to allow beliefs as inputs.

\begin{definition}[Mutually halting program set; belief inputs]\label{def:mutually-halting-2}
  A common program set $\Sigma$ is mutually halting if, for all $\pi \in \Sigma$ and all beliefs $\mu \in \Delta(\Sigma)$, the execution of $\pi$ on input $\mu$ terminates and outputs an action in $A$. In this case, the execution function $\operatorname{run} \colon \Sigma \times \Delta(\Sigma) \to A$ maps $(\pi, \mu)$ to that output, which we write suggestively as
  $\operatorname{run}(\pi \mathrel{\|} \mu)$. We call the elements of $\Sigma$ \emph{$\tau$-programs}.
\end{definition}

As in \cref{sec:osgt}, we assume that every constant program is admissible, that is, for each action $a \in A$, the program that ignores its input belief and always outputs $a$ belongs to $\Sigma$. We again denote this program by $a$, so that $\operatorname{run}(a \| \mu) = a$ for every $\mu \in \Delta(\Sigma)$, and we retain the embedding $A \subseteq \Sigma$.

The information structure of the game\footnote{A different idea for modeling partial information, or feedback, in games is the self-confirming equilibrium of \citet{battigalli1987thesis} and \citet{fudenberg1993self}.} is described by a pair of \emph{transparency maps} $\tau = (\tau_1, \tau_2)$, where $\tau_i \colon \Sigma \to \Delta(\Sigma)$ for each $i \in I$. Here $\tau_i(\pi)$ is the belief that player $i$'s program receives as input when the opponent $-i$ submits $\pi$. Intuitively, $\tau_i$ describes what player $i$'s program can learn about its opponent, for instance through black-box queries or knowledge of a shared model family, and it is the object the designer controls. At one extreme, $\tau_i(\pi) = \delta_\pi$ recovers the full transparency of \cref{sec:osgt}; at the other, a constant $\tau_i$ reveals nothing, and player $i$ effectively chooses an action of the base game $G$; for more details, see \cref{sec:more-less-transparency}.

Naturally extending the definitions of \cref{sec:osgt}, we define the outcome map induced by the execution function $\operatorname{run}$ and the transparency maps $\tau$,
\begin{equation*}
    \operatorname{outcome}^\tau:\Sigma \times \Sigma \to A \times A, \quad \mathrm{outcome}^\tau(\bm{\pi}) \coloneq \biggl(\mathrm{run}(\pi_1 \| \tau_1(\pi_2)), \mathrm{run}(\pi_2 \| \tau_2(\pi_1))\biggr),
\end{equation*}
which assigns to each program profile $\bm{\pi} = (\pi_1, \pi_2)$ the action profile it produces. The only change from \cref{sec:osgt} is that each program now runs on the belief it receives from the opponent's code. Each player's utility is then extended to program profiles by 
\begin{equation*}
    U_i^\tau: \Sigma \times \Sigma \to \mathbb{R},
\qquad
U_i^\tau(\bm{\pi}) \coloneq u_i\bigl(\operatorname{outcome}^\tau(\bm{\pi})\bigr).
\end{equation*}
The tuple $G^\tau \coloneqq \bigl(I, (\Sigma_i, U_i^\tau)_{i \in I}\bigr)$ is a normal-form game, which we call the \emph{$\tau$-program meta-game} associated with $G$; the superscript makes its dependence on the transparency maps $\tau$ explicit. Following \cref{def:PE}, a profile $\bm{\pi} \in \Sigma \times \Sigma$ is a $\tau$-Program Equilibrium (or simply a Program Equilibrium, when $\tau$ is clear from context) if it is a pure Nash Equilibrium of $G^\tau$; mixed $\tau$-Program Equilibria are defined analogously.

\begin{definition}[$\tau$-Program Equilibrium]\label{def:tau-PE}
A profile $\bm{\pi}^{*} \in \Sigma \times \Sigma$ is a $\tau$-Program Equilibrium ($\tau$-PE) of $G$ if it is a (pure) Nash Equilibrium of $G^\tau$, i.e. for every $i \in I$ and every $\pi_i' \in \Sigma$,
\[
U_i^\tau(\pi_i^{*}, \pi_{-i}^*) \;\ge\; U_i^\tau(\pi_i',\, \pi^{*}_{-i}).
\]
\end{definition}

\begin{theorem}[Pure Nash equilibria are $\tau$-program equilibria]\label{th:NE-is-tau-PE}
Let $\tau$ be any pair of transparency maps and let $\bm{a}^* \in A \times A$. Then the profile of constant programs $\bm{a}^* \in \Sigma \times \Sigma$ is a pure $\tau$-Program Equilibrium if and only if $\bm{a}^*$ is a pure Nash Equilibrium of $G$. In particular, every pure Nash Equilibrium of $G$ is a $\tau$-Program Equilibrium for every choice of $\tau$.
\end{theorem}

More generally, the existence of $\tau$-Program Equilibria does not depend on $G$ having a pure equilibrium, as one can apply Nash's existence theorem \citep{nash1950equilibrium, nash1951} to $G^\tau$ to prove existence of a mixed $\tau$-Program Equilibrium of $G$.

\subsubsection{An Example in the Prisoner's Dilemma}\label{subsec:exp.prison:dil:tau}

We now return to our running example, the Prisoner's Dilemma (\cref{example:PD,ex:prisonerdilemma-3}), and illustrate partial transparency through concrete transparency maps and explicit programs. We assume throughout a common transparency map $\tau_1 = \tau_2$ for the two players; write $\tau$ for it. Explicit maps are found in \cref{apx:TransExp-Setup-Maps}.

Given a program \texttt{Bot} of the open-source Prisoner's Dilemma $G_{\mathrm{PD}}^{\mathrm{prog}}$, we can extend it to a program \texttt{TauBot} of the meta-game $G_{\mathrm{PD}}^{\tau}$, which takes a belief $\mu \in \Delta(\Sigma)$ as input. In particular, \texttt{TauBot} runs the decision procedure of \texttt{Bot} against each program in the support of $\mu$ and aggregates the resulting actions. The simplest aggregator is a threshold rule, which, for $\alpha \in [0,1]$, cooperate if and only if the $\mu$-mass of the programs against which \texttt{Bot} would cooperate is at least $\alpha$; we call $\alpha$ the caution parameter. We give below the example of \texttt{TauDupocBot}$(k, \alpha)$, and a more general example can be found in \cref{app:tau-bot-definitions}.

\begin{lstlisting}[style=pdcode, mathescape=true]
def TauDupocBot$(k, \alpha)(\mu)$:
   mass = 0
   for $\pi'$ in $\operatorname{supp}(\mu)$:
        if proof_search($k, \pi', \mathrm{run}(\pi' \| \delta_{\mathrm{TauDupocBot}(k, \alpha)}) = C$):
             mass = mass + $\mu(\pi')$
   if mass $\geq \alpha$:
        return C
   else:
        return D
\end{lstlisting}

Unlike \texttt{DupocBot}$(k)$, \texttt{TauDupocBot}$(k, \alpha)$ is no longer unexploitable. Soundness of $S$ still guarantees that every program counted in \texttt{mass} cooperates on the point-mass input $\delta_{\texttt{TauDupocBot}(k,\alpha)}$, but for a poorly behaved transparency map $\tau$, the actual opponent need not be one of these programs.

The threshold rule given above is linear in $\mu$ and therefore admits an expected utility representation;\footnote{Take the state space $\Sigma$ and the larger consequence space $\{C,D\} \times \{0,1\}$, and let playing $a$ against $\pi$ give the consequence $(a, f(\pi))$, where $f(\pi) = 1$ if \texttt{TauBot} cooperates on the point-mass input $\delta_\pi$ and $f(\pi) = 0$ otherwise.} non-linear aggregators, such as the maximum or minimum or a variational functional, capture ambiguity-sensitive behavior \citep{gilboa2016ambiguity}. We do not explore their properties here.

\subsubsection{When Does Transparency Matter?}\label{sec:more-less-transparency}

Our framework interpolates between the base game and open-source game theory by varying what programs learn about their opponents. For example, when each transparency map is constant, programs receive no information about the opponent's choice, and choosing a program reduces to choosing an action in the base game. At the other extreme, when $\tau_i(\pi)=\delta_\pi$, each program receives its opponent's exact identity, recovering fully transparent open-source game theory. We now give a sufficient condition on the informational structure $\tau$ under which the strategic outcomes of $G^\tau$ coincide with those of the base game $G$, even when some information is revealed.

Suppose that a player can change what their program does without changing what their opponent observes. In the Prisoner’s Dilemma, for instance, replacing a cooperative program with an indistinguishable program that defects would leave the opponent’s cooperation unchanged, making the deviation profitable. More generally, we say that a player has \emph{undetectable deviations} if, for every program they might submit and every input belief they might receive, they can obtain any desired action by substituting a program that their opponent cannot distinguish from the original. Against a fixed opponent's program, such a substitution changes the player's action while leaving the opponent's response unchanged, reproducing the unilateral deviations available in the base game. In this sense, rational behavior in the meta-game $G^\tau$ can differ from rational behavior in the base game $G$ only insofar as $\tau$ reveals something about how programs behave.

\begin{definition}[Undetectable Deviations]\label{def:mimic}
    Fix a $\tau$-program meta-game $G^\tau$. Player $i \in I$ has undetectable deviations if for every $\pi \in \Sigma$, every $\mu \in \tau_i(\Sigma)$, and every $x \in A$, there exists a $\rho \in \Sigma$ such that
    \begin{enumerate}
        \item $\rho$ is indistinguishable from $\pi$ to player $-i$, that is, $\tau_{-i}(\rho) = \tau_{-i}(\pi)$;
        \item $\rho$ plays $x$ on input $\mu$, that is, $\mathrm{run}(\rho \| \mu)=x$.
    \end{enumerate}
\end{definition}

A program of player $-i$ can condition its action only on the belief it receives, $\tau_{-i}(\pi_i)$, and can therefore reward or punish player $i$ only through what is revealed by this belief. If $i$ has undetectable deviations, any arbitrary change in $i$'s behavior can be made without changing what $-i$ observes. Hence no program of $-i$ can make its response depend on what $i$ actually does;\footnote{Black-box testing provides a concrete example. If player $-i$ observes only how player $i$'s program behaves on a set of test inputs, then $i$ can deviate undetectably at every input outside this set. Testing can therefore deter opportunistic behavior only if the inputs a program faces in the actual interaction cannot be distinguished from test inputs. This is the concern we face with evaluation awareness in AI systems.} as \cref{th:mimic} shows, this implies that the program meta-game in this case reduces to the base game. This idea is the analogue of the informativeness principle in moral hazard \citep{holmstrom1979moral} where an additional signal can improve the optimal contract if and only if it provides additional information about the agent's action beyond what is already observed in the outcome.

\begin{theorem}\label{th:mimic}
    Assume both players have undetectable deviations. Then the outcomes of pure $\tau$-program equilibria are exactly the pure Nash equilibria of $G$.
\end{theorem}

The two extreme transparency structures illustrate the result. If $\tau_{-i}$ is constant,\footnote{The uniform distribution over $\Sigma$ is an example of such a constant transparency map.} so that player $-i$ observes nothing about $i$'s program, then player $i$ trivially has undetectable deviations (using the $x$-constant program). When $\tau=(\tau_1, \tau_2)$ is constant, \cref{th:mimic} shows that the outcomes of pure $\tau$-PE are exactly the pure Nash equilibria of $G$; in this case, each player effectively chooses an action of the base game. On the other hand, full transparency, $\tau_i(\pi) = \delta_\pi$, recovers the open-source setting of \cref{sec:osgt}. There, neither player has undetectable deviations,\footnote{\Cref{def:mimic} imposes a simple counting constraint. Fix any $\mu \in \tau_i(\Sigma)$ and any $\pi \in \Sigma$. Applying the definition to $(\pi, \mu, x)$ for each $x \in A$ yields programs $\rho_x$ with $\tau_{-i}(\rho_x) = \tau_{-i}(\pi)$ and $\mathrm{run}(\rho_x \,\|\, \mu) = x$, which are distinct for distinct $x$. Hence every set of programs that player $-i$ cannot distinguish from one another contains at least $|A|$ elements. Since these sets partition $\Sigma$ and there are $|\tau_{-i}(\Sigma)|$ of them,
\begin{equation*}
    |\tau_{-i}(\Sigma)| \cdot |A| \le |\Sigma|.
\end{equation*}
In particular, if $\tau_{-i}$ is injective, as under full transparency $\tau_{-i}(\pi) = \delta_\pi$, player $i$ does not have undetectable deviations as soon as $|A| \ge 2$. This is consistent with \cref{sec:osgt}, where program equilibria can induce outcomes that are not Nash equilibria of $G$ (\cref{ex:prisonerdilemma-3}).} and program equilibria can sustain outcomes that are not Nash equilibria of $G$, such as mutual cooperation in the Prisoner's Dilemma (\cref{ex:prisonerdilemma-3}).

By varying the transparency map $\tau$, we thus obtain a family of games ranging from the base game $G$ to the fully transparent meta-game $G^{\mathrm{prog}}$ of \citet{tennenholtz2004program}. At one extreme, programs receive no information about their opponents; at the other, they can condition their behavior on the opponent’s exact code. Throughout this family, \cref{th:NE-is-tau-PE} shows that every pure Nash Equilibrium of $G$ remains a Program Equilibrium, giving the inclusion $\mathrm{NE}\subseteq \mathrm{PE}$. Whether the sets of program equilibria, or their induced outcomes, vary monotonically in $\tau$ under a suitable order on transparency maps is a further question we do not approach here. Greater transparency may allow programs to detect and respond to deviations that would otherwise go unnoticed, but it also changes the inputs they receive and hence potentially their behavior. It is therefore not immediate that outcomes sustained under one transparency map remain sustainable under a more informative one.  For this exercise, the results of \citet{battigalli2016analysis,formenti2026selfconfirming} are useful.

\section{Outcomes under the Formal System $S$}
\label{section:proofs}

In simulation-based OSGT \citep{oesterheld2019robust, DBLP:conf/ijcai/KovarikOC23, DBLP:conf/aaai/CooperOC25}, a program decides what to play based on simulating its opponent, and no logic is involved. In proof-based OSGT, outcomes depend on which statements $S$ proves within a given length, and on the actual proofs themselves. In this section we fix the formal system $S$ (\cref{subsection:Theory-SystemS}) and show how we determine outcomes under it (\cref{subsection:MathOutcomes}). Finally, \cref{sec:dupoc-v-cupod} introduces a new tool, transposition, and uses it to settle Open Problem~3 of \citet{Critch} for $S$: the outcome of \texttt{DupocBot} against \texttt{CupodBot}. We work in the Prisoner's Dilemma throughout. 

\subsection{The Formal System $S$}\label{subsection:Theory-SystemS}

We fix the formal system $S$ in which programs search for proofs about their opponents' behavior. \citet[Appendix~B]{Critch} work with an unspecified extension of Peano Arithmetic (PA) that is only required to satisfy a list of efficiency assumptions. We instead fix a concrete system, which we later formalize in Lean~4 \citep{Lean4} in \cref{subsubsection:Lean-Mech-SystemS}.

\begin{definition}[The proof system $S$]\label{def:proof-system}
The proof system $S$ consists of the following three components.
\begin{enumerate}
\item[(S1)] \emph{Language.} $\mathcal{L}_{\mathrm{Act}}$ is the language of arithmetic $\{0, 1, +, \cdot, <\}$ extended by two constant symbols $c_C$ and $c_D$, one per action cooperate $C$ and defect $D$;
\item[(S2)] \emph{Theory.} $T_{\mathrm{Act}}$ consists of Peano Arithmetic together with the action axiom
\[
  (c_C = 0 \wedge c_D = 1) \;\vee\; (c_C = 1 \wedge c_D = 0),
\]
stated together with its transpose, obtained by simultaneously exchanging $c_C$ and $c_D$;\footnote{Including the transpose as an axiom ensures that transposition preserves proof length without requiring additional derivation steps. As defined in \cref{def:transpositio}, the transpose is obtained by simultaneously exchanging $c_C$ and $c_D$.}
\item[(S3)] \emph{Proofs.} For a sentence $\varphi$ of $\mathcal{L}_{\mathrm{Act}}$, a proof of $\varphi$ is a derivation of $\varphi$ from $T_{\mathrm{Act}}$ in a standard first-order calculus. Its length is the number of symbols it contains. We write $\vdash \varphi$ if $\varphi$ has a proof, and $\vdash_k \varphi$ if it has a proof of length at most $k \in \mathbb{N}$.
\end{enumerate}
\end{definition}

The action axiom, together with its transpose, is the only axiom of $T_{\mathrm{Act}}$ beyond those of PA. It states that $c_C$ and $c_D$ denote the numbers $0$ and $1$, in one of the two possible orders, without specifying which. Consequently, in every model $\mathcal{M}$ of $T_{\mathrm{Act}}$, the pair $(c_C^{\mathcal{M}}, c_D^{\mathcal{M}})$ is either $(0^{\mathcal{M}}, 1^{\mathcal{M}})$, which we call the \emph{standard reading}, or $(1^{\mathcal{M}}, 0^{\mathcal{M}})$, the \emph{swapped reading}. Conversely, every model of PA becomes a model of $T_{\mathrm{Act}}$ under either reading of the two constants. Hence the models of $T_{\mathrm{Act}}$ are exactly the models of PA equipped with one of the two readings.

A delicate point, familiar to readers versed in logic, deserves discussion. This paper uses two distinct systems to prove statements. The first is $S$, the formal proof system used by the programs; a call to \texttt{proof\_search}$(k,\cdot,\varphi)$ determines whether $\varphi$ has a proof in $S$ of length at most $k$. The second is the metatheory, the system in which \emph{we} reason, \emph{ordinary} mathematics in this Section and Lean in \cref{subsubsection:Lean-Mech-SystemS}. \Cref{table:Proof-notation} collects the notation we use for the rest of the paper. We write $\Box_k\varphi$ for the arithmetical sentence asserting that some number codes a proof of $\varphi$ in $S$ of length at most $k$. By contrast, $\vdash_k\varphi$ denotes the assertion, made in the metatheory, that such a proof exists. These are related by $\vdash_k\varphi\Longleftrightarrow\mathbb{N}\models\Box_k\varphi$.\footnote{Equivalently, of any model $\mathcal{M}$ of $T_{\mathrm{Act}}$.} Thus, $\Box_k\varphi$ expresses bounded provability within the language of $S$, whereas $\vdash_k\varphi$ asserts bounded provability in our discussion of $S$. We give an example of three statements that must not be confused: $\nvdash_k\varphi$ says that $\varphi$ has no proof of length at most $k$; $\vdash_k\neg\varphi$ says that its negation has such a proof; and $\vdash\neg\Box_k\varphi$ says that $S$ proves that no such proof of $\varphi$ exists. By soundness, the second and the third imply the first. The first does not, in general, imply the second, as failure to find a proof of $\varphi$ does not amount to finding a proof of its negation. For fixed $k$ and $\varphi$, a failed search can itself be proven in $S$, but the proof need not have length at most $k$. This distinction between a search failing and its failure being provable within the same budget is the subject of \cref{subsection:Lean-Floors}.

\begin{table}[h]
\centering\small
\caption{Notation for bounded provability and truth. The sentence
$\Box_k\varphi$ belongs to the language of $S$; the assertions
$\vdash_k\varphi$ and $\models\varphi$ are made in the metatheory.}
\begin{tabular}{ll}
\toprule
Notation & Meaning \\
\midrule
$\Box_k\varphi$ & A sentence expressing provability in $S$ within budget $k$ \\
$\vdash_k\varphi$ & A proof of $\varphi$ in $S$ of length at most $k$ exists \\
$\models\varphi$ & $\varphi$ is true in $\mathbb{N}$ under the standard reading \\
\bottomrule
\end{tabular}
\label{table:Proof-notation}
\end{table}

\subsection{The Mathematics Behind the Outcomes}
\label{subsection:MathOutcomes}

With the formal system in place, programs can reason about their opponent's behavior using the formalism defined by $S$. The system alone, however, does not tell us much about the actual reasoning, only its language. We now present some of the mathematical tools underlying the proofs, partly to show how reasoning in $S$ actually proceeds, and partly because the results themselves are worth seeing. The same tools also yield results beyond those of \citet{Critch}. We settle several of their open problems and refute one of their examples. In what follows, we state the general ideas and leave the precise definitions, outcomes and proofs to \cref{apx-subsection:TheoreticalResultsOSGT-ToolsOutcomes}. Every result is mechanized in Lean (\cref{section:Lean}).

A fundamental property in our system $S$ is that proved statements are true, i.e. that $S$ is sound (\cref{prop:soundness}). Soundness yields both positive and negative results: positively, it shows that \texttt{DupocBot} is unexploitable (\cref{prop:unexploitability}; see also the discussion in \cref{ex:prisonerdilemma-3}); negatively, since $S$ proves only true statements, a false hypothesis has no proof of any length, so any search for one fails (e.g. \cref{prop:dupoc-defect}).

Complementary to soundness, we have completeness: if a statement is true for any model of $S$, then $S$ should be able to prove it (\cref{thm:completeness} and \cref{cor:completeness-readings}). This gives us for instance that \texttt{DupocBot} and \texttt{CooperateBot} mutually cooperate (\cref{prop:dupoc-coop}).

Soundness and completeness, however, both presuppose that we can judge whether a hypothesis is true independently of the search for its proof. This fails when programs reason about each other: when \texttt{DupocBot}$(k)$ faces itself, its hypothesis is true only if its own search succeeds (\cref{ex:prisonerdilemma-3}). Such self-fulfilling hypotheses are resolved by Löb’s theorem \citep{Lob1955}, a core theorem \citep{Critch_pblt} in mathematical logic at the heart of \citet{Critch}’s results. We state it as \cref{thm:pblt} and re-derive the surprising result of \texttt{DupocBot}'s self-cooperation (\cref{thm:dupoc-selfplay}). Beyond self-cooperation, Löb’s theorem resolves several of \citeauthor{Critch}’s open problems and refutes their Example 5.3(a) (\cref{apx-subsection:TheoreticalResultsOSGT-CorrectionProgress}).

\subsection{Transposition and \texttt{DupocBot} vs. \texttt{CupodBot}}\label{sec:dupoc-v-cupod}

Aside the well-known tools above, we introduce a new technique, which makes use of the inherent syntactic symmetry between the two actions $C$ and $D$ in programs playing the Prisoner's Dilemma. The \emph{transposition} between the two actions allows us to settle a standing open-problem in the field, the outcome of \texttt{DupocBot} vs. \texttt{CupodBot}.\footnote{It is also \citet{Critch}'s Open Problem 3.} All proofs of this section are given in \cref{app:proofs-dupoc-cupod}, and every result is formalized in Lean (\cref{section:Lean}).

The two action constants play symmetric roles in $S$, suggesting that exchanging them should transform any proof into another (transposed) proof of the same length. Note that the programs considered here make no reference to payoffs, the semantic content of a game, so this symmetry concerns the syntax of the programs and the proof system. We make this idea precise in the following definition.

\begin{definition}[Transposition]\label{def:transpositio}
Let $\varphi$ be a formula of $S$. We define the transpose $\mathcal{T}\varphi$ to be the formula obtained by replacing every occurrence of $c_C$ by $c_D$ and, simultaneously, every $c_D$ by $c_C$. 
\end{definition}
Since $\mathcal{T}$ touches only constants, it commutes with every logical symbol,
\[
  \mathcal{T}(\neg\varphi)=\neg\,\mathcal{T}\varphi,\qquad
  \mathcal{T}(\varphi\to\psi)=\mathcal{T}\varphi\to\mathcal{T}\psi,\qquad
  \mathcal{T}(\forall x\,\varphi)=\forall x\,\mathcal{T}\varphi,
\]
and it is an involution: $\mathcal{T}\mathcal{T}\varphi = \varphi$. With this definition in mind, we have the following result.

\begin{theorem}\label{thm:transposition}
For every sentence $\varphi$ of $S$ and every budget $k \in \mathbb{N}$,
\[
  \vdash_k \varphi \quad\Longleftrightarrow\quad \vdash_k \mathcal{T}\varphi .
\]
\end{theorem}

Naturally, we want to extend transposition to the program's source code itself.
\begin{definition}[Transposed program]
    For a program $\pi$, the transposed program $\mathcal{T}\pi$ is obtained by exchanging the actions $C$ and $D$ everywhere in the source of $\pi$, including inside the hypotheses it searches for.
\end{definition}
The transpose of \texttt{DupocBot}$(k)$ is \enquote{Cooperate Unless Proof of Defection}, $\mathcal{T} \texttt{DupocBot} = \texttt{CupodBot}$, given below (using the notation of \cref{sec:osgt}):

\begin{lstlisting}[style=pdcode, mathescape=true]
def CupodBot$(k)(\pi)$:
   if proof_search($k, \pi, \mathrm{run}(\pi \| \mathrm{CupodBot}(k) ) = D$):
        return D
   else:
        return C
\end{lstlisting}

With both definitions above, we can arrive at the first symmetry result we have for our programs. For programs $\pi$, $\pi'$ and an action $a$, write $\mathrm{Plays}(\pi, \pi', a)$ for the sentence of $S$ stating that $\pi$, run on the source of $\pi'$, plays $a$. Transposing such a sentence as a formula of $S$ is the same as transposing the programs and the action it speaks about.

\begin{theorem}\label{prop:bridge}
For all programs $\pi$, $\pi'$ and every action $a$,
\[
  \mathcal{T}\,\mathrm{Plays}(\pi, \pi', a) \;=\; \mathrm{Plays}(\mathcal{T}\pi, \mathcal{T}\pi', \mathcal{T}a),
\]
where $\mathcal{T} a$ is the other action.
\end{theorem}

Combining \cref{thm:transposition} and \cref{prop:bridge}, we obtain a symmetry of outcomes. When $\pi$ searches for a hypothesis $\varphi$, its transpose searches for $\mathcal{T}\varphi$ with the same budget (\cref{prop:bridge}), so by \cref{thm:transposition} the two searches succeed or fail together; every other step of the execution is transposed by construction of $\mathcal{T}\pi$. Transposing both programs therefore transposes what they play.

\begin{theorem}[Transposition of outcomes]\label{thm:transposition-outcomes}
For all programs $\pi$, $\pi'$,
\[
  \mathrm{outcome}(\mathcal{T}\pi, \mathcal{T}\pi') \;=\; \mathcal{T}\,\mathrm{outcome}(\pi, \pi'),
\]
where $\mathcal{T}$ exchanges $C$ and $D$ in both components. In particular, a program and its transpose play opposite actions against each other: $\mathrm{outcome}(\pi, \mathcal{T}\pi)$ is $(C, D)$ or $(D, C)$.
\end{theorem}

Applied to \texttt{DupocBot} and its transpose \texttt{CupodBot}, \cref{thm:transposition-outcomes} leaves two candidate outcomes, $(C, D)$ and $(D, C)$. This and \texttt{DupocBot}'s unexploitability (\cref{prop:unexploitability}) settle Open Problem~3 of \citet{Critch} for our proof system $S$.

\begin{theorem}[\texttt{DupocBot} against \texttt{CupodBot}]\label{thm:dupoc-cupod}
For every $k\in\mathbb{N}$,
\[
  \mathrm{outcome}
  \bigl(\texttt{DupocBot}(k),\texttt{CupodBot}(k)\bigr)
  =(D,C).
\]
\end{theorem}

\section{A Lean Formalization}
\label{section:Lean}

The previous section shows that finding the outcomes of open-source programs involves Löbian, self-referential arguments that are laborious and error-prone by hand, and the complexity of these proofs grows with the complexity of the programs involved and of the game played. To make this analysis more systematic and robust, we formalize the theory in a formal verification language, both to ensure outcomes are machine-checked, and to facilitate the search for new proofs using automated tools, such as large language models. For these reasons, we choose Lean~4 \citep{Lean4}, a proof assistant language that has gained traction in the literature for its ability to sustain large-scale formalizations of modern mathematics, such as the community-owned mathlib library \citep{mathlib2020}, and for its role as the de facto target language for learning-based theorem proving, from the miniF2F benchmark and the LeanDojo toolchain \citep{minif2f,LeanDojo} to the current generation of neural provers \citep{formalreasoningwithRL,deepseekproverv2,kimina_prover,goedelproverv2}. Lean is also home to the closest prior work: \texttt{ModalAgents} \citep{foundation_agents_formalized} mechanizes the modal agents of \citet{Modal_agents_barasz}, but with unbounded provability, so it has no proof lengths, no evaluator and no transparency. Ours is, to our knowledge, the first mechanization of resource-bounded open-source game theory.

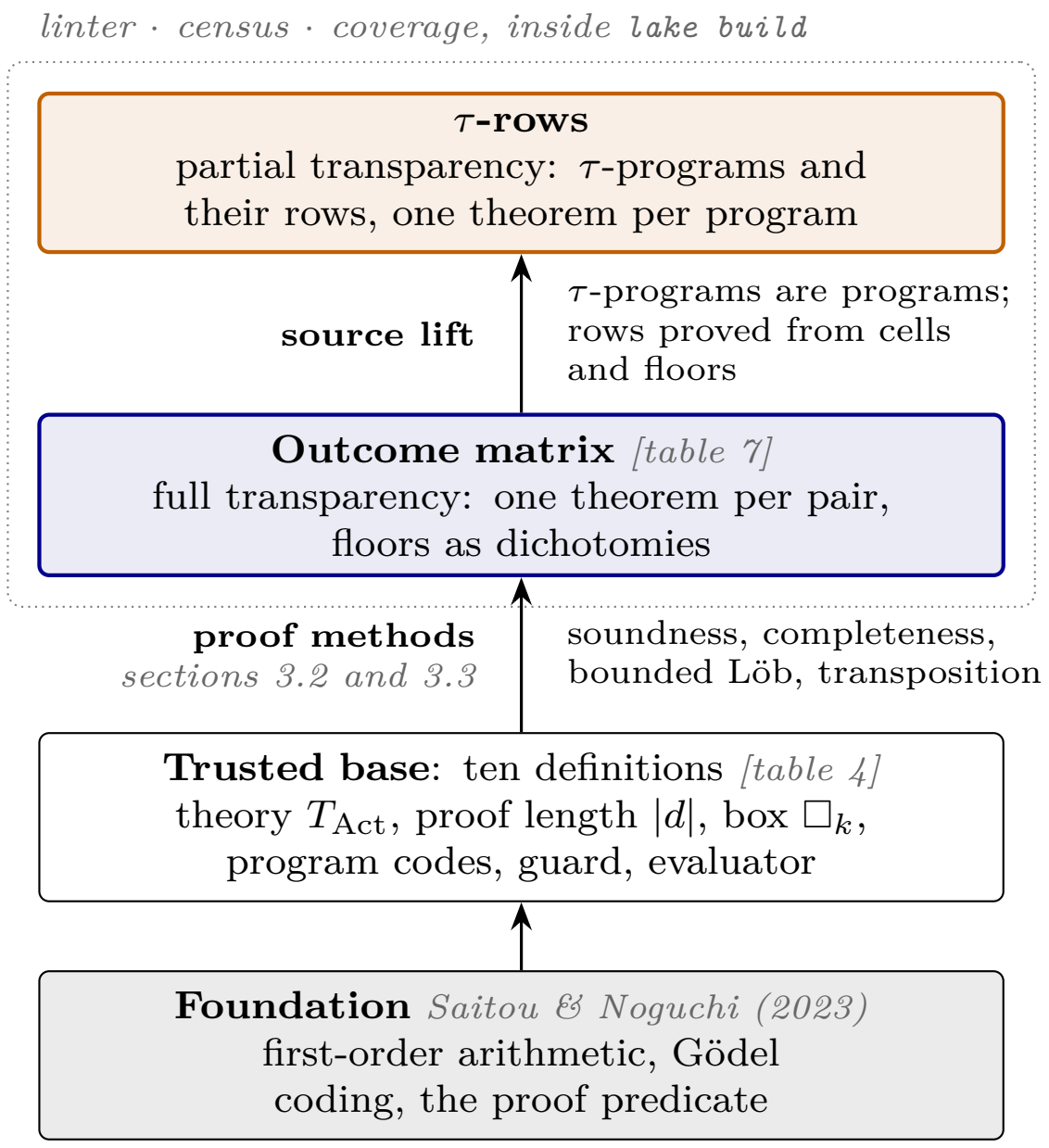
\begin{figure}[t]
  \centering
    \providecommand{\archTag}[1]{{\scriptsize\itshape\color{black!60}#1}}
    \colorlet{archFull}{blue!55!black}
    \colorlet{archPartial}{orange!75!black}
    \begin{tikzpicture}[
      font=\footnotesize,
      box/.style={draw, rounded corners=2pt, align=center, inner sep=4pt, text width=6.6cm, fill=white},
      lib/.style={box, fill=black!8},
      full/.style={box, draw=archFull, thick, fill=archFull!8},
      partial/.style={box, draw=archPartial, thick, fill=archPartial!10},
      arr/.style={-{Stealth[length=2mm]}, semithick},
      note/.style={font=\scriptsize\itshape, text=black!60},
      lbl/.style={font=\scriptsize, align=left, anchor=west, xshift=2mm},
      lblL/.style={font=\scriptsize, align=right, anchor=east, xshift=-2mm},
      region/.style={font=\footnotesize\bfseries, text=black!70},
    ]
    \node[lib]  (found)  at (0,0)
      {\textbf{Foundation} \archTag{\cite{foundation}}\\ first-order arithmetic, G\"odel coding, the proof predicate};
    \node[box]  (base)   at (0,1.7)
      {\textbf{Trusted base}: ten definitions \archTag{[\cref{table:trusted-base}]}\\ theory $T_{\mathrm{Act}}$, proof length $|d|$, box $\Box_k$,\\ program codes, guard, evaluator};
    \node[full] (matrix) at (0,4.0)
      {\textbf{Outcome matrix} \archTag{[\cref{table:outcome-matrix-full}]}\\ full transparency: one theorem per pair,\\ floors as dichotomies};
    \node[partial] (tau) at (0,6.3)
      {\textbf{$\tau$-rows} \\ partial transparency: $\tau$-programs and\\ their rows, one theorem per program};
    \draw[arr] (found) -- (base);
    \draw[arr] (base) -- node[lblL] {\textbf{proof methods}\\ \archTag{\cref{subsection:MathOutcomes,sec:dupoc-v-cupod}}}
      node[lbl] {soundness, completeness,\\ bounded L\"ob, transposition} (matrix);
    \draw[arr] (matrix) -- node[lblL] {\textbf{source lift}}
      node[lbl] {$\tau$-programs are programs;\\ rows proved from cells\\ and floors} (tau);
    \begin{scope}[on background layer]
      \node[draw=black!50, densely dotted, rounded corners=3pt, inner sep=6pt, fit=(matrix)(tau)] (checks) {};
    \end{scope}
    \node[note, anchor=south west] at ([xshift=1mm]checks.north west) {linter $\cdot$ census $\cdot$ coverage, inside \texttt{lake build}};
    \end{tikzpicture}
    \caption{Architecture of the formalization. Everything shown is checked by the Lean kernel and rests on no project axioms. The Foundation library is trusted as a library, ten definitions form the trusted base, and every outcome cell (blue, full transparency) and every $\tau$-row (orange, partial transparency) is a theorem or an explicitly stated open case, validated by a linter, a census and a coverage check inside \texttt{lake build}.}
  \label{fig:architecture}
\end{figure}

First, we do not introduce any axioms, apart from Lean's three standard axioms (\texttt{propext}, \texttt{Quot.sound}, \texttt{Classical.choice}), so no project-specific assumptions are made. Second, all our theorems are \texttt{sorry}-free, ensuring that every statement is entirely proved. Third, the outcomes the rest of the paper consumes follow fixed, typed templates, enforced by the build in both directions. Every pair of programs used in our experiments is thereby either a proved cell or an explicitly stated open cell (\cref{subsection:Lean-Floors}). Finally, we build upon the Foundation library \citep{foundation}, which is, to the best of our knowledge, the most advanced library in mathematical logic. On top of their work, we add the ten definitions of \cref{table:trusted-base}, which constitute our base. The result, summarized in \cref{fig:architecture}, is a kernel-checked Lean 4 formalization\footnote{The code is provided as supplementary material. Its file \texttt{REPRODUCING.md} maps each statement, table and figure of the paper to the Lean theorem or script behind it.} of open-source game theory under full and partial transparency.

The remainder of this section is organized as follows. \Cref{subsection:Lean-Mech} walks through the three components of the mechanization: programs, outcomes and the system $S$ itself, built on Foundation. \Cref{subsection:Lean-Floors} then turns to the open cells and traces them to a single open problem of bounded proof complexity: whether $S$ can show, within budget $k$, that a given false sentence has no proof of length at most $k$.

\subsection{Details of the Mechanization}
\label{subsection:Lean-Mech}

Open-source game theory rests upon three main components that we expose here: programs, outcomes, and the formal system $S$. Definitions, statements and architecture were designed by the authors, but most of the Lean source was written by LLM agents built on Anthropic's Claude Opus and Fable model families \citep{anthropic2026fable51,anthropic2026opus48}. The formalization comprises about 159{,}000 lines of Lean, developed over about two months; \cref{apx:Lean-Timeline} breaks this down by step.
Nothing an agent writes is trusted as every theorem is checked by the Lean kernel. The provenance of a proof is therefore irrelevant to its validity, which is the point of mechanizing.

\subsubsection{Programs}\label{subsubsection:Lean-Mech-Programs}
A program in our mechanization is a term of the inductive type \texttt{Prog},\footnote{\texttt{Prog} has ten constructors. The seven base constructors all appear in the two examples below; the remaining three extend the language to partial transparency.} and the set $\Sigma$ of admissible programs from \cref{sec:osgt} is realized as a finite collection of such terms. Two \texttt{Prog} constructors are important here: \texttt{.search} for logical reasoning, and \texttt{.sim} for simulation. Below, we present two examples: our proof-based running example, DupocBot, and a simulation-based program, DBot.

\begin{lstlisting}[style=pdlean]
def DupocBot (k : Nat) : Prog :=
  .search k                            -- the proof search constructor
    (.plays .opp .self Action.C)       -- hypothesis: opponent plays C against me
    (.const Action.C)                  -- proof found: cooperate
    (.const Action.D)                  -- otherwise: defect
\end{lstlisting}
\begin{lstlisting}[style=pdlean]
def DBot : Prog :=
  .ite                                 -- if-then-else
    (.sim .opp (.bot DefectBot))       -- simulate opponent against DefectBot
    Action.C                           -- test action C
    (.const Action.D)                  -- if opponent cooperates, defect
    (.const Action.C)                  -- else, cooperate
\end{lstlisting}

The hypothesis a \texttt{.search} looks for is a value of a second inductive type, \texttt{Formula}, which has constructors stating atomic plays (\texttt{.plays p q a}, \enquote{$p$ plays $a$ against $q$}, where \texttt{.self} and \texttt{.opp} stand for the two players of the current match); implications and negations of such statements; a syntactic equality \texttt{.eq p q}, \enquote{$p$ is literally the program $q$}; and the bounded provability operator $\Box_k$ of \cref{subsection:Theory-SystemS}: \texttt{.box k $\psi$}, \enquote{$\psi$ has a proof of length at most $k$}. The three remaining constructors implement partial transparency: \texttt{.tvote} runs the threshold vote of \texttt{TauDupocBot}$(k,\alpha)$ over a list of weighted hypotheses, and \texttt{.sys} with \texttt{.selfIdx} bind the mutually recursive system that two voting programs probing each other require.

\subsubsection{Outcomes}

As stated in the introduction to this section, outcome theorems (i.e. the per-pair matchups between programs in the library) follow a strict template to ensure consistency: an \texttt{outcomeA} attribute, and a fixed statement syntax (see example below). Programs are run by an evaluator \texttt{EvalGraph n me opp p a}, read \enquote{with fuel $n$, in the match of \texttt{me} against \texttt{opp}, the program \texttt{p} plays \texttt{a}}. Nested simulation makes termination non-structural, hence the explicit fuel: a program plays an action if some amount of fuel returns it, and that action is unique because the evaluator is deterministic and monotone in the fuel. Below is one of the running example's cells, stated on the template, whose first argument fixes the budget regime (\texttt{.universal}: at every budget $k$; \texttt{.eventual}: at every sufficiently large $k$).\footnote{A careful reader will recognize this example as being the outcome certificate of our headline \texttt{DupocBot} vs. \texttt{CupodBot} proof.}
\begin{lstlisting}[style=pdlean, mathescape=true]
@[outcomeA]
theorem outcome_DupocBot_vs_CupodBot :
    OutcomeSpecA .universal (bot DupocBot) (bot CupodBot) (some (.D, .C))
-- unfolds to: ${\color{leancomment}\forall}$ k, DupocBot k plays D against CupodBot k,
--             and CupodBot k plays C against DupocBot k
\end{lstlisting}

Under partial transparency the template is a row vector of all outcomes between a fixed bot and the zoo. The evaluator's \texttt{.search} clause asks whether a proof of length at most $k$ exists, in the formal system $S$, defined in \cref{subsection:Theory-SystemS} and constructed below. The search is finite, so it can in principle be completed, but it is computationally impractical. We therefore establish each outcome by proving how the programs behave rather than by running them.

\subsubsection{The formal System $S$}
\label{subsubsection:Lean-Mech-SystemS}
Perhaps the most interesting mechanization result is the actual formalization of $S$, a language of arithmetic with two action constants, Peano Arithmetic with the action axiom, and proofs measured by their number of symbols. In Lean, we build $S$ on top of the Foundation library. We sketch the mechanization here and leave further details to \cref{apx-section:LeanSystemS}.

$S$ is a theory of arithmetic, so its sentences speak only about natural numbers, since its signature contains only $0$, $1$, $+$, $\cdot$, $<$ and the two action constants. For $S$ to reason about sentences, proofs and programs, each of these objects is assigned a natural number, its G\"odel code. A statement about proofs or programs then becomes a statement about numbers, which $S$ can express.

Furthermore, formulas have no truth value on their own, only in a logical structure. For us, that structure is $\mathbb{N}$ because codes of proofs are actual natural numbers by construction. This lets \enquote{some number codes a proof} and \enquote{some proof exists} say the same thing:
\[
  \vdash_k\varphi \quad\Longleftrightarrow\quad \mathbb{N} \models \Box_k\varphi \quad\Longleftrightarrow\quad \text{some derivation of } \varphi \text{ in } T_{\mathrm{Act}} \text{ has at most } k \text{ symbols}.
\]

Finally, modeling the length of a proof requires careful accounting. In the Foundation library, proofs are derivations in a sequent calculus.\footnote{A derivation is a tree in which every node carries a \emph{sequent}, a finite set of formulas read as a disjunction, and every edge is an inference rule. Foundation's calculus is one-sided and has the cut rule.} We define the length $|d|$ of a derivation $d$ as its number of symbols, counted over every node of the tree; the precise count is given in \cref{apx-section:LeanSystemS}.

\subsection{What Our Tools Do Not Decide}
\label{subsection:Lean-Floors}

The tools of \cref{subsection:MathOutcomes} show us how to settle certain types of hypotheses of our \texttt{proof\_search} function.  However, one kind of hypothesis escapes them: one that is true, but true only because some other search fails. Whether such a hypothesis is found within its budget is a question the tools cannot answer, and one that rests on an open problem about the length of proofs in $S$.

A search of budget $k$ fails when no proof of length at most $k$ exists. One might expect this failure to be an easy fact, provable within the same budget $k$. To see why it is not, suppose any failed $k$-search could be certified within budget $k$: for any hypothesis $\varphi$ with else-action $b$, $\nvdash_k \varphi$ implies $\vdash_k (\text{me plays } b)$. Then we obtain a contradiction with the following anti-diagonal program:\footnote{Indeed, set $\varphi = (\texttt{AntiBot}$ plays $D$ vs. $\texttt{AntiBot})$ and suppose, for contradiction, that $\vdash_k \varphi$. Then soundness (\cref{prop:soundness}) implies that $\texttt{AntiBot}$ defects. However, a provable hypothesis is a found hypothesis, which makes $\texttt{AntiBot}$ take its then branch and play $C$. Impossible, because a program plays only one action. Hence $\nvdash_k \varphi$, so $\texttt{AntiBot}$ takes its \emph{else} branch and plays $D$: the sentence $\varphi$ is true. But by the hypothesis, $\nvdash_k \varphi$ would imply $\vdash_k \varphi$, a contradiction.}
\begin{lstlisting}[style=pdlean]
-- "If I can prove that I defect against myself, cooperate"
def AntiBot (k : Nat) : Prog :=
  .search k
    (.plays .self .self Action.D)      -- hypothesis: I play D against myself
    (.const Action.C)                  -- proof found: cooperate
    (.const Action.D)                  -- otherwise: defect
\end{lstlisting}

A certificate that a search failed may therefore cost more than the search itself, and nothing in $S$ says in general how much more. This matters as soon as one program's hypothesis depends on another program's \emph{else} play. Consider \texttt{CupodTrollBot}, which cooperates with any program except if its opponent is literally \texttt{CupodBot}:
\begin{lstlisting}[style=pdlean]
def CupodTrollBot (k : Nat) : Prog :=
  .search k
    (.eq .opp (CupodBot k)) -- test if opponent is literally CupodBot
    (.const Action.D)       -- then, defect
    (.const Action.C)       -- else, cooperate
\end{lstlisting}
Against \texttt{DupocBot}, its hypothesis is false, so by soundness its search fails and it cooperates, at every budget. \texttt{DupocBot}'s hypothesis, \enquote{\texttt{CupodTrollBot} plays $C$ against me}, is therefore true. But it is true only because \texttt{CupodTrollBot}'s search failed, and to find it \texttt{DupocBot} needs, within its budget $k$, a proof that a search (of the same budget $k$) fails: exactly the kind of certificate we just showed is not free. None of our tools in the previous section decides whether the certificate above fits, which turns out to rest on a question about consistency.

Write $\mathrm{Con}_S(n)$ for the sentence of $S$ stating \enquote{no proof of length at most $n$ ends in a contradiction}. For each fixed $n$, $S$ proves $\mathrm{Con}_S(n)$,\footnote{Since $S$ is sound, $\mathrm{Con}_S(n)$ is true for every $n$, and $S$ proves it for each fixed $n$, because it is then a finite check.} but we do not know how long the proof must be. It cannot be shorter than $n^{\epsilon}$ for some $\epsilon > 0$ \citep[Thm.~3.1]{Pudlak_Survey}, a bound due to Friedman (unpublished, 1979) and \citet{Pudlak_bounds_for_finite_proofs}. Our open cells are tied to whether a proof of length about $n$ exists, which, to our knowledge, remains an open question. We call this the bounded consistency problem.

\begin{definition}[Floor, hinge, sides]\label{def:floor}
A cell is a floor if its outcome depends on whether one of its searches finds a true hypothesis $\varphi_k$, and showing that it does would settle the bounded consistency problem: $S$ would then prove $\mathrm{Con}_S(n)$ within about $n$ symbols.\footnote{Precisely, if $H_k$ is the hinge of such a floor then, for all large $k$, $H_k \Longrightarrow\ \vdash_{k + O(\log^2 k)} \mathrm{Con}_S\bigl(k - O(\log k)\bigr)$, i.e. if the hypothesis has a proof of length at most $k$, then $S$ proves within about $k$ symbols that it has no contradiction of length about $k$. The length is $k$ only up to a small term because the proof of $\mathrm{Con}_S$ is the proof of the hypothesis followed by a few more lines, and these lines must write out the number $k$. A proof of this can be found in our Lean implementation (\texttt{ArithS/Cells/ConReduction.lean}).} Precisely, writing $H_k := (\vdash_k \varphi_k)$, there are two outcomes $o^{-} \neq o^{+}$ such that, for all large $k$,
\[
  \neg H_k \;\Longrightarrow\; \text{outcome } o^{-}
  \qquad\text{and}\qquad
  H_k \;\Longrightarrow\; \text{outcome } o^{+}.
\]
The statement $H_k$ is the hinge of the floor. The hinge has two sides: the negative side, $\neg H_k$, on which the hypothesis is true but not found within budget, and the positive side, $H_k$, on which it is found. The floor has outcome $o^{-}$ on the negative side and $o^{+}$ on the positive side.
\end{definition}

The cell of \texttt{DupocBot} against \texttt{CupodTrollBot} is a floor. Its hinge $H_k$ states that \texttt{DupocBot}'s hypothesis has a proof of length at most $k$; the outcome is $(D, C)$ on the negative side and $(C, C)$ on the positive side.

The consistency problem above is not an artifact of budgets, as it already appears, unbounded, in \citet{Modal_agents_barasz}, where \texttt{PrudentBot} must establish that its opponent defects against \texttt{DefectBot}. For an opponent that searches for proofs, this means that a search in PA fails, which PA cannot certify without its own consistency, so the check runs in the stronger system PA${+}1$. With budgets there is no stronger system to move to, only a larger budget. A self-fulfilling fact is certified within the budget at which it holds, whereas the failure of a search of budget $k$ needs a proof of $\mathrm{Con}_S$ at about $k$, and whether that proof fits within $k$ is the bounded consistency problem.

Since that problem is open, we fix a convention. Unless stated otherwise, we report every floor on the negative side of its hinge, the one on which no failure certificate fits its budget. This is a labeled convention, and the analysis of \cref{section:Experiments} is repeated with every floor on the positive side of its hinge in \cref{app:full-transparency-hinge,app:partial-transparency-hinge}. The two sides give different outcomes, so no theorem of ours assumes either. When several floors or several hinges are involved, \enquote{the negative side of the hinge} means that every hinge is taken on its negative side.

\section{Transparency, Cooperation, and Equilibrium Selection}
\label{section:Experiments}

We study how transparency affects both the cooperation that can be sustained in Program Equilibrium, and the question of equilibrium selection through evolutionary dynamics. Throughout, we fix a finite library of programs and build the meta-game payoff matrix from their pairwise outcomes (\cref{table:outcome-matrix-body}). \Cref{sec:exper:full-transp} treats full transparency, where we enumerate the program equilibria of the library, and study which of them a population reaches under replicator dynamics. \Cref{sec:exp-p-t} then varies the transparency map and measures how both the sustainable and the selected cooperation change with the level of transparency.

\begin{example}[Prisoner's Dilemma, cont'd]\label{example:PD-numeric}
Throughout this section, we instantiate \cref{example:PD} with $T = 3$, $R = 2$, $P = 0$, and $S = -1$:
  \[
    \begin{array}{c|cc}
         & C      & D      \\ \hline
       C & (2,2)  & (-1,3) \\
       D & (3,-1) & (0,0)
    \end{array}
    \qquad\qquad 3 > 2 > 0 > -1.
  \]
Mutual cooperation thus gives each player $2$, whereas mutual defection, the unique Nash Equilibrium of $G_{\mathrm{PD}}$, gives each player $0$.
\end{example}

\subsection{Full Transparency}\label{sec:exper:full-transp}

\subsubsection{Program Library}\label{par:zoo}

While the results that follow do depend on the choice of the program library $\Sigma$, we have taken care to consider several different libraries; we fix one here and repeat the analysis on three nested variants of it in \cref{app:zoo-ladder}. We start with the base case of full transparency, that is, of open-source game theory, with the base library
\begin{equation*}
    \Sigma = \{\texttt{CooperateBot}, \texttt{CupodBot}, \texttt{DBot}, \texttt{DefectBot}, \texttt{DupocBot}, \texttt{EBot}, \texttt{OBot}, \texttt{TitForTatBot}\},
\end{equation*}
defined in \cref{fig:zoo-nesting}. Our focus here is on which programs are played in equilibrium rather than on average performance, and adding strictly dominated programs to a library leaves its equilibria unchanged, which is why we do not consider a much larger library. The source code of every program can be found in \cref{app:bot-definitions}.

\colorlet{zooBodyFill}{blue!7}          
\colorlet{zooBodyLine}{blue!55!black}   
\colorlet{zooLine}{black!55}            
\colorlet{clsConst}{black!12}           
\colorlet{clsSim}{green!22!white}       
\colorlet{clsSearch}{orange!28!white}   
\begin{figure}[!tp]\centering\small
\begin{tikzpicture}[
  chip/.style={rounded corners=2pt, inner xsep=4pt, inner ysep=2pt, font=\ttfamily\small, anchor=west},
  const/.style={chip, fill=clsConst},
  sim/.style={chip, fill=clsSim},
  search/.style={chip, fill=clsSearch},
  cls/.style={anchor=west, font=\small},
  strat/.style={anchor=west, text width=7.9cm, font=\small},
  zoo/.style={draw=zooLine, rounded corners=5pt, line width=0.5pt, inner ysep=6pt},
  ztitle/.style={anchor=north west, font=\normalsize\itshape, inner xsep=10pt, inner ysep=7pt},
  zcount/.style={anchor=north east, font=\normalsize, inner xsep=10pt, inner ysep=7pt},
]
\matrix (m) [matrix of nodes, column sep=6pt, row sep=4pt, nodes={anchor=west}, ampersand replacement=\&] {
  |[const]|  CooperateBot$^{b,c}$ \& |[cls]| constant \& |[strat]| always $\mathsf{C}$ \\
  |[const]|  DefectBot$^{b,c}$    \& |[cls]| constant \& |[strat]| always $\mathsf{D}$ \\
  |[sim]|    TitForTatBot         \& |[cls]| simulation \& |[strat]| $\mathsf{C}$ iff the opponent cooperates with \texttt{CooperateBot} \\
  |[sim]|    DBot$^{a}$           \& |[cls]| simulation \& |[strat]| $\mathsf{D}$ iff the opponent cooperates with \texttt{DefectBot} \\
  |[sim]|    OBot$^{a}$           \& |[cls]| simulation \& |[strat]| $\mathsf{C}$ iff the opponent cooperates with both \texttt{CooperateBot} and \texttt{DefectBot} \\
  |[sim]|    EBot$^{a}$           \& |[cls]| simulation \& |[strat]| $\mathsf{C}$ iff the opponent defects against \texttt{DefectBot} and cooperates with \texttt{CooperateBot} or with its simulated self; else $\mathsf{D}$ \\
  |[search]| DupocBot$^{b}$       \& |[cls]| searcher ($k$) \& |[strat]| $\mathsf{C}$ iff $\vdash_k$ (opponent plays $\mathsf{C}$ vs. me) \\[12pt]
  |[search]| CupodBot$^{b}$       \& |[cls]| searcher ($k$) \& |[strat]| $\mathsf{D}$ iff $\vdash_k$ (opponent plays $\mathsf{D}$ vs. me) \\[14pt]
  |[search]| CupodTrollBot        \& |[cls]| searcher ($k$) \& |[strat]| $\mathsf{D}$ iff $\vdash_k$ (opponent $=$ \texttt{CupodBot}~$k$); else $\mathsf{C}$ \\
  |[search]| DupocTrollBot        \& |[cls]| searcher ($k$) \& |[strat]| $\mathsf{D}$ iff $\vdash_k$ (opponent $=$ \texttt{DupocBot}~$k$); else $\mathsf{C}$ \\[14pt]
  |[search]| CIMCIC$^{b}$         \& |[cls]| searcher ($k$) \& |[strat]| $\mathsf{C}$ iff $\vdash_k$ (I play $\mathsf{C}$ vs. opponent $\rightarrow$ opponent plays $\mathsf{C}$ vs. me) \\
  |[search]| DIMDID$^{b}$         \& |[cls]| searcher ($k$) \& |[strat]| $\mathsf{D}$ iff $\vdash_k$ (I play $\mathsf{D}$ vs. opponent $\rightarrow$ opponent plays $\mathsf{D}$ vs. me) \\
};
\coordinate (t1) at ([yshift=0.50cm]m-1-1.north);
\coordinate (t2) at ([yshift=1.28cm]m-1-1.north);
\coordinate (t3) at ([yshift=2.06cm]m-1-1.north);
\coordinate (t4) at ([yshift=2.84cm]m-1-1.north);
\begin{scope}[on background layer]
\node[zoo, inner xsep=22pt, fit=(m-1-1)(m-12-3)(t4)] (z4) {};
\node[zoo, inner xsep=16pt, fit=(m-1-1)(m-10-3)(t3)] (z3) {};
\node[zoo, draw=zooBodyLine, fill=zooBodyFill, line width=1.1pt, inner xsep=10pt, fit=(m-1-1)(m-8-3)(t2)] (z2) {};
\node[zoo, dashed, inner xsep=4pt, fit=(m-1-1)(m-7-3)(t1)] (z1) {};
\end{scope}
\node[ztitle] at (z1.north west) {non-provers + \texttt{DupocBot}};
\node[zcount] at (z1.north east) {$N = 7$};
\node[ztitle, font=\normalsize\bfseries, text=zooBodyLine] at (z2.north west) {$\Sigma$};
\node[zcount, font=\normalsize\bfseries, text=zooBodyLine] at (z2.north east) {$N = 8$};
\node[ztitle] at (z3.north west) {$\Sigma$ + trolls};
\node[zcount] at (z3.north east) {$N = 10$};
\node[ztitle] at (z4.north west) {$\Sigma$ + trolls + implication provers};
\node[zcount] at (z4.north east) {$N = 12$};
\end{tikzpicture}
\caption{The bots used in this paper, as four nested libraries. The $\Sigma$ library (filled) is the one analyzed throughout; the three others are the robustness extensions. Provenance: $^{a}$ transcribed from the LessWrong source-visible tournament \citep{lesswrong_tournament}; $^{b}$ \citet{Critch}; $^{c}$ \citet{Modal_agents_barasz}, bounded here; untagged bots are ours. Further bots, not analyzed here, are in the supplementary code.}
\label{fig:zoo-nesting}
\end{figure}

We provide proofs\footnote{The two pairs of \texttt{DupocBot} with \texttt{DBot} and \texttt{EBot} are proved up to their hinge (\cref{def:floor}); we use the negative side here and report the results for the positive side in \cref{app:full-transparency-hinge}.} of all outcomes for all pairs of programs in our Lean library (\cref{section:Lean}). These proofs were produced either by hand or by an LLM proof agent that drives the Lean kernel. The agent is built on Anthropic's Claude Opus and Fable model families \citep{anthropic2026fable51,anthropic2026opus48} (\cref{apx:Lean-Timeline}). \Cref{table:outcome-matrix-body} summarizes the outcomes.

\begin{table}[t]\centering\scriptsize
\colorlet{ccfill}{green!18}\colorlet{ddfill}{black!12}\colorlet{hlfill}{yellow!35}\colorlet{hgfill}{orange!40}
\newcommand{\cc}{\cellcolor{ccfill}CC}\newcommand{\dd}{\cellcolor{ddfill}DD}
\newcommand{\cd}{CD}\newcommand{\dc}{DC}\newcommand{\hg}[1]{\cellcolor{hgfill}#1$^{\ddagger}$}
\setlength{\tabcolsep}{3pt}
\caption{The symmetric outcome matrix of the eight-bot library $\Sigma$ in $S$, at a common proof budget $k$; a cell gives the row bot's certified action first. Cells hold for every sufficiently large $k$. Shaded cells represent mutual cooperation (green) and mutual defection (gray). Orange cells$^{\ddagger}$ rest on a floor (\cref{def:floor}), where we show the negative side of its hinge. Highlighted names (yellow) are programs that no member of $\Sigma$ exploits (see \cref{prop:unexploitability} for more on this). Complete outcome matrix with all bots studied in this paper is in \cref{table:outcome-matrix-full}.}
\begin{tabular}{@{}lcccccccc@{}}\toprule
 & \rotatebox{90}{\texttt{CooperateBot}} & \rotatebox{90}{\colorbox{hlfill}{\texttt{DefectBot}}} & \rotatebox{90}{\texttt{TitForTatBot}} & \rotatebox{90}{\texttt{DBot}} & \rotatebox{90}{\colorbox{hlfill}{\texttt{OBot}}} & \rotatebox{90}{\texttt{EBot}} & \rotatebox{90}{\colorbox{hlfill}{\texttt{DupocBot}}} & \rotatebox{90}{\texttt{CupodBot}} \\\midrule
\texttt{CooperateBot} & \cc & \cd & \cc & \cd & \cc & \cd & \cc & \cc \\
\colorbox{hlfill}{\texttt{DefectBot}} & \dc & \dd & \dd & \dc & \dd & \dd & \dd & \dd \\
\texttt{TitForTatBot} & \cc & \dd & \cc & \dc & \cd & \dc & \cc & \cc \\
\texttt{DBot} & \dc & \cd & \cd & \dd & \cd & \cd & \hg{CD} & \cc \\
\colorbox{hlfill}{\texttt{OBot}} & \cc & \dd & \dc & \dc & \dd & \dc & \dd & \dd \\
\texttt{EBot} & \dc & \dd & \cd & \dc & \cd & \cc & \hg{CD} & \cc \\
\colorbox{hlfill}{\texttt{DupocBot}} & \cc & \dd & \cc & \hg{DC} & \dd & \hg{DC} & \cc & \dc \\
\texttt{CupodBot} & \cc & \dd & \cc & \cc & \dd & \cc & \cd & \dd \\\bottomrule
\end{tabular}
\label{table:outcome-matrix-body}
\end{table}

\subsubsection{Enumerating Program Equilibria}

We are interested in understanding which programs are optimal to play in equilibrium. To do so, we enumerate the Nash equilibria of the meta-game $G^{\mathrm{prog}}$ with both players restricted to $\Sigma$, following the approach described in \cref{subsec:nash-eq-enumeration}. For the eight-program library $\Sigma$ at full transparency, we find 21 extreme equilibria across three connected components (\cref{tab:extreme-equilibria-body8-floor}).

The three components differ in the probability of mutual cooperation. In the first, every equilibrium yields $(C,C)$ with probability one and gives each player the cooperative payoff $R=2$. In the second, there is a single symmetric equilibrium in which each player mixes over \texttt{CooperateBot}, \texttt{OBot}, and \texttt{EBot}. Mutual cooperation occurs with probability $11/18$, and each player's expected payoff is $3/2$. In the third, every equilibrium yields $(D,D)$ with probability one and gives each player a payoff of zero. Thus the unique equilibrium outcome of the underlying Prisoner's Dilemma persists, but the game also sustains cooperative and partially cooperative equilibria, which are not present in the base Prisoner's Dilemma. The role of \texttt{DupocBot} is clearest in self-play. In the underlying game, a player facing cooperation can gain by defecting. In the meta-game, switching programs can also change the opponent's action. In particular, \texttt{DupocBot} cooperates with itself, but a player who switches to a program that defects against it faces defection in return. This makes mutual cooperation a best response for both players in \texttt{DupocBot} self-play. The enumeration shows that its role extends beyond self-play in that in every fully cooperative equilibrium of this library, both players assign probability at least $1/9$ to \texttt{DupocBot}. No other program appears in every such equilibrium.

\subsubsection{Equilibrium Selection under Replicator Dynamics}

Which of these equilibria does a population reach when programs that earn higher payoffs become more common? We study this question using replicator dynamics in $G^{\mathrm{prog}}$. Programs are matched at random, and a program's share grows whenever its expected payoff exceeds the population mean. For the eight-program library $\Sigma$ at full transparency, every trajectory we simulated from a uniformly drawn initial population converged to the fully cooperative component (\cref{app:full-transparency-replicator}). Although mutual defection remains an equilibrium outcome, in every run the dynamics converged to a cooperative equilibrium. The initial composition influenced which programs persisted in the long run. In some populations only \texttt{DupocBot} survived, while in most it coexisted with both \texttt{CooperateBot} and \texttt{TitForTatBot}. Every pairing among these three programs yields mutual cooperation, so each earns $R=2$ regardless of its population share. Thus all sampled populations reached the same cooperative outcome through different mixtures of programs.

While the dynamics do not distinguish between these three programs, their robustness against opponents differs. A population of \texttt{CooperateBot} can be invaded by \texttt{DefectBot}, \texttt{DBot}, and \texttt{EBot}, which exploit it, and a population of \texttt{TitForTatBot} by \texttt{OBot}. Against a population of \texttt{DupocBot}, in contrast, no program of $\Sigma$ earns more than \texttt{DupocBot} itself (\cref{prop:unexploitability}).

Two robustness checks are in the appendix. The same analysis for the other program libraries of \cref{fig:zoo-nesting} is in \cref{app:zoo-ladder}: the equilibrium components are the same in all four, and every sampled population reaches full cooperation in three of them, against $97.5\%$ in the ten-program library. On the positive side of the hinge, the fully cooperative component of $\Sigma$ is unchanged and $95.9\%$ of the sampled populations reach it (\cref{app:full-transparency-hinge}).

\subsection{Partial Transparency}\label{sec:exp-p-t}

Under full transparency, switching programs can change the opponent's response and thus eliminate the gain from defection (\cref{sec:exper:full-transp}). We perform the comparative statics exercise of holding payoffs and the program library fixed and varying what each program knows about its opponent's program; we follow throughout the setup of \cref{subsec:exp.prison:dil:tau}. In particular, we show that at every caution level tested, both the maximum cooperation sustainable in equilibrium and the cooperation selected by replicator dynamics increase with transparency, apart from narrow windows at low transparency.

\subsubsection{Experimental Setup}
We use the eight-program library $\Sigma$ of \cref{par:zoo}; for precise program definitions, see \cref{app:bot-definitions,app:tau-bot-definitions}. Under the threshold rule of \cref{subsec:exp.prison:dil:tau}, a program cooperates if its belief assigns probability at least $\alpha$ to opponents against which it would itself cooperate, for the caution parameter $\alpha \in [0,1]$. A more cautious program thus requires stronger evidence before it plays $C$. We use two families of transparency maps. Behavioral maps measure the distance between two programs by counting the opponents against which their actions differ. Syntactic maps use the normalized edit distance between the syntax trees of their source code; see \cref{apx:TransExp-Setup-Maps} for more details. To compare the families, we measure the fraction of uncertainty about the opponent's identity that each map $\tau$ removes, denoted by $t(\tau)$ and called the transparency level. At $t=0$, the signal is independent of the opponent's identity, i.e. players act as if they were in the base game. At $t=1$, it identifies the opponent exactly, the setting of full transparency in open-source game theory. \Cref{apx:TransExp-Setup-Levels} gives the constructions.

In each family, we sweep across transparency--caution pairs, transparency $t \in \{0, 0.01, 0.02, \dots, 0.99, 1\}$, and caution $\alpha \in \{0,0.2,0.4,0.5,0.6,0.7,0.8,0.9,1\}$. At each grid point we enumerate the program equilibria, the Nash equilibria of the induced meta-game $G_{\mathrm{PD}}^{\tau}$ (\cref{subsec:nash-eq-enumeration}). To study equilibrium selection, we compute replicator dynamics from populations drawn uniformly from the simplex (\cref{app:rep-dyn-theory}). In particular, we measure $\Pr[(C,C)]$, the probability of mutual cooperation. All results below use the negative side of the hinge (\cref{def:floor}); on the positive side, full cooperation is still sustained and selected at high transparency, but not through a single threshold (\cref{app:partial-transparency-hinge}).

\subsubsection{Equilibrium Cooperation and Selection}\label{subsec:partial-transp-experiments}
We observe that in the partially transparent Prisoner's Dilemma, greater transparency favors both the cooperation that players can sustain in equilibrium and the cooperation that payoff-based equilibrium selection reaches. At each caution level, a fully cooperative Program Equilibrium first appears at a critical level of transparency and remains available at every higher level. Cooperation can then be self-enforcing because switching to an exploitative program can induce the opponent to defect in return. The same threshold governs selection in both signal families: at or above it, every simulated population converges to a fully cooperative equilibrium, and below it none does. Below the threshold, populations converge to mutual defection, except on three narrow windows of low transparency in the behavioral family, where partially cooperative equilibria exist and populations cycle without settling (\cref{app:equilibrium-partial-transparency}). In our motivating oversight example, the cooperative equilibrium is collusion.

\begin{figure}[htbp]
  \centering
  \includegraphics[width=\linewidth]{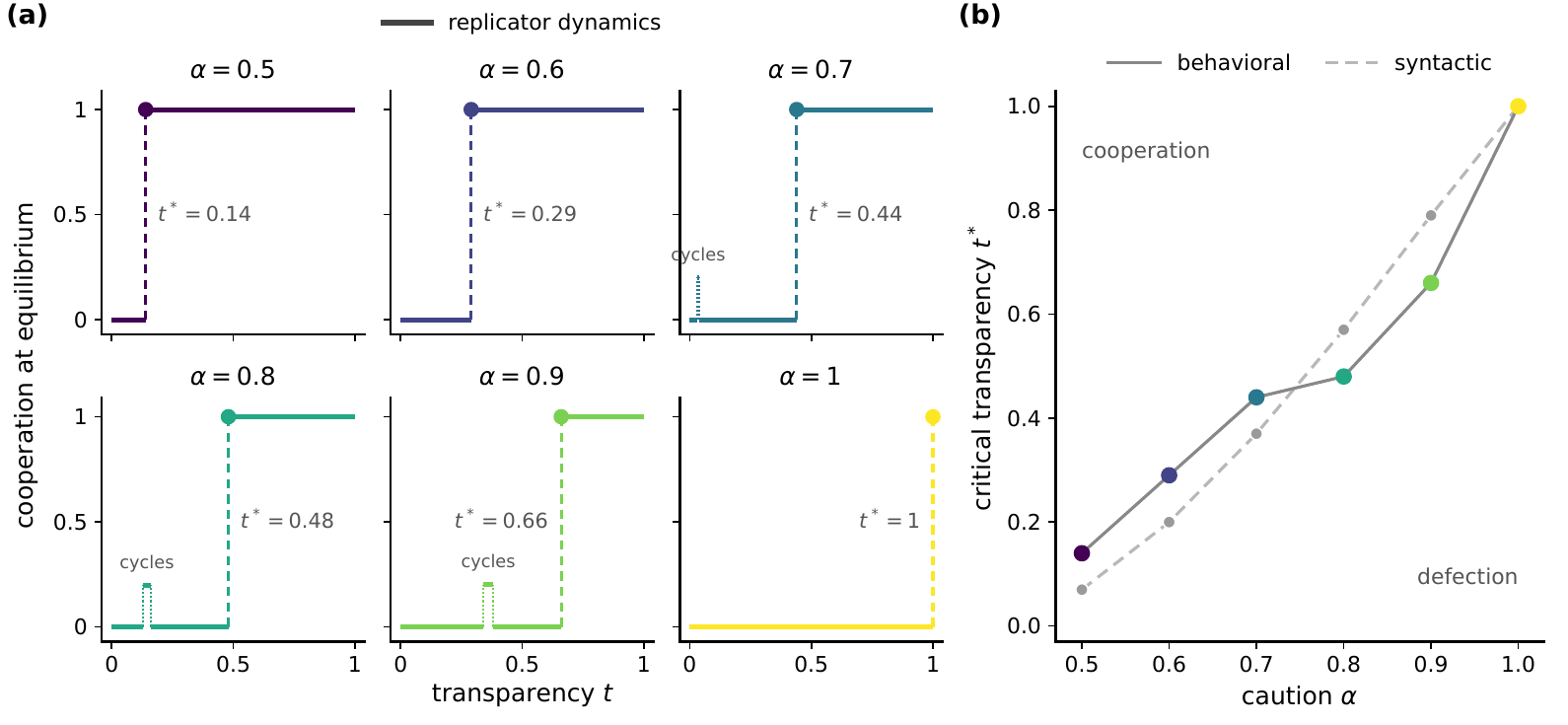}
  \caption{Critical transparency for full cooperation. \textbf{(a)} At each caution level in the behavioral family, selected cooperation jumps from zero to one at $t^*$, except on the short windows marked \emph{cycles}. \textbf{(b)} The solid and dashed lines connect the sampled thresholds for behavioral signals (colored dots) and syntactic signals (gray dots), respectively. In the syntactic family, every simulated population defects below $t^*$ and cooperates fully at and above it. The transparency grid has spacing $0.01$.}
  \label{fig:transparency-thresholds}
  \Description{The six behavioral thresholds are 0.14, 0.29, 0.44, 0.48, 0.66, and 1. The syntactic thresholds are 0.07, 0.20, 0.37, 0.57, 0.79, and 1.}
\end{figure}

Full cooperation exhibits a sharp threshold in transparency (\cref{fig:transparency-thresholds}(a)). Below the threshold, no equilibrium is fully cooperative and no simulated population reaches full cooperation. Once transparency reaches the threshold, full cooperation becomes an equilibrium outcome, and replicator dynamics select it in every run at or above that level. At $\alpha=0.8$, every equilibrium yields mutual defection at $t=0.47$. At $t=0.48$, a fully cooperative equilibrium appears and every simulated population reaches it. Each rise of $1/6$ in \cref{fig:transparency-selection} thus marks one more caution level crossing its threshold.

Greater caution requires more transparency for full cooperation over the six caution levels shown in \cref{fig:transparency-thresholds}(b). Let $t^*$ be the lowest sampled transparency at which a fully cooperative equilibrium exists. At $\alpha=0.5$, $t^*$ is $0.14$ for behavioral signals and $0.07$ for syntactic signals. The threshold rises at each subsequent
caution level. At $\alpha=1$, when programs require certainty before cooperating, both families require full transparency.

\section{Conclusion and Future Directions}

In this paper, we laid the theoretical foundations of open-source game theory, and introduced a first model of partial transparency in settings where players choose commitment devices (programs) to act on their behalf, and these devices have access to partial information (beliefs) about their opponents’ programs. We found that increasing transparency led to more cooperation (collusion) in our experiments, a phenomenon that we expect to see in frontier LLMs. We also placed the theory, under both partial and full transparency, on formal, machine-checkable foundations, settling some of the open problems in the field and preparing the ground for the auto-formalization of more complex games and more complex programs. But much remains to be done before this theory can be applied to AI safety in realistic settings, especially in bridging the gap between our toy programs and large-scale neural networks such as LLMs. For example, we expect tools from mechanistic interpretability to be helpful in identifying aspects of a model’s internal computation that allow other agents to reason about its behavior and condition their own actions accordingly. If the effect of transparency carries over to such models, designing and monitoring alignment protocols with regard to what agents can learn about one another will become a crucial step towards safer AI systems.

\section*{Acknowledgements}

This work was supported by Renaissance Philanthropy through The Alignment Project, a funding programme of the UK AI Security Institute (AISI), under the project “Formal Verification of Collusion in Transparent AI Agents.”

\section*{AI Use Statement}\label{sec:AI-usage}

In this work, we used generative AI tools to provide feedback on research methodology and experimental design, implement methods, clean and format the datasets, and check mathematical arguments. We used agentic AI tools to assist in the Lean 4 formalization (\cref{apx:Lean-Timeline}). We have not used AI tools to propose or refine hypotheses, to interpret experimental results, or to set the direction of the research. Additionally, we used generative AI tools to identify, summarize, and analyze relevant literature, write and refine software code, create or modify scientific figures, and polish and contribute to the manuscript’s writing and editing. The models used include GPT-5.6 Sol, GPT-6 Astra (OpenAI) and Fable 5, Fable 5.1, Opus 4.8, Opus 5, Opus 5.5 (Anthropic). We, the authors, take full responsibility for the final contents of this work, including text, claims or artifacts produced with the aid of generative AI tools. 


\bibliography{sample-base}

@article{tennenholtz2004program,
  title={Program equilibrium},
  author={Tennenholtz, Moshe},
  journal={Games and Economic Behavior},
  volume={49},
  number={2},
  pages={363--373},
  year={2004},
  publisher={Elsevier}
}

@article{fudenberg1993self,
  title={Self-confirming equilibrium},
  author={Fudenberg, Drew and Levine, David K.},
  journal={Econometrica: Journal of the Econometric Society},
  pages={523--545},
  year={1993},
  publisher={JSTOR}
}

@article{battigalli1987thesis,
    title={Comportamento razionale ed equilibrio nei giochi e nelle situazioni sociali},
  author={Battigalli, Pierpaolo},
  journal={Undergraduate thesis, Bocconi University},
  year={1987}
}

@article{bergemann2019information,
  title={Information design: A unified perspective},
  author={Bergemann, Dirk and Morris, Stephen},
  journal={Journal of economic literature},
  volume={57},
  number={1},
  pages={44--95},
  year={2019},
  publisher={American Economic Association, 2014 Broadway, Suite 305, Nashville, TN 37203~…}
}

@article{kalai2010commitment,
  title={A commitment folk theorem},
  author={Kalai, Adam Tauman and Kalai, Ehud and Lehrer, Ehud and Samet, Dov},
  journal={Games and Economic Behavior},
  volume={69},
  number={1},
  pages={127--137},
  year={2010},
  publisher={Elsevier}
}

@techreport{openai2026huggingface,
  author      = {{OpenAI}},
  title       = {{OpenAI}--{Hugging Face} Incident Technical Report},
  institution = {OpenAI},
  year        = {2026},
  month       = aug,
  url         = {https://cdn.openai.com/pdf/67869394-cb91-4c12-888c-5cbd85c7814c/OpenAI-Hugging-Face%20Incident-Technical-Report.pdf}
}

@techreport{metr2026huggingface,
  author      = {{METR}},
  title       = {Brief Independent Investigation of Agents' Behavior,
                 Reasoning and Collaboration in the {OpenAI} /
                 {Hugging Face} Hacking Incident},
  institution = {METR},
  year        = {2026},
  month       = aug,
  url         = {https://metr.org/hugging-face-incident-report-aug-2026.pdf}
}

@article{irving2018debate,
  title   = {{AI} safety via debate},
  author  = {Irving, Geoffrey and Christiano, Paul and Amodei, Dario},
  journal = {arXiv preprint arXiv:1805.00899},
  year    = {2018}
}

@inproceedings{DBLP:conf/icml/GreenblattSSR24,
  author       = {Ryan Greenblatt and
                  Buck Shlegeris and
                  Kshitij Sachan and
                  Fabien Roger},
  title        = {{AI} Control: Improving Safety Despite Intentional Subversion},
  booktitle    = {Forty-first International Conference on Machine Learning, {ICML}},
  series       = {Proceedings of Machine Learning Research},
  volume       = {235},
  year         = {2024},
  bibsource    = {dblp computer science bibliography, https://dblp.org}
}

@techreport{Hammond2025MultiAgentRisks,
  author = {
    Hammond, Lewis and
    Chan, Alan and
    Clifton, Jesse and
    Hoelscher-Obermaier, Jason and
    Khan, Akbir and
    McLean, Euan and
    Smith, Chandler and
    Barfuss, Wolfram and
    Foerster, Jakob and
    Gaven\v{c}iak, Tom\'{a}\v{s} and
    Han, The Anh and
    Hughes, Edward and
    Kova\v{r}\'{i}k, Vojt\v{e}ch and
    Kulveit, Jan and
    Leibo, Joel Z. and
    Oesterheld, Caspar and
    {Schroeder de Witt}, Christian and
    Shah, Nisarg and
    Wellman, Michael and
    Bova, Paolo and
    Cimpeanu, Theodor and
    Ezell, Carson and
    Feuillade-Montixi, Quentin and
    Franklin, Matija and
    Kran, Esben and
    Krawczuk, Igor and
    Lamparth, Max and
    Lauffer, Niklas and
    Meinke, Alexander and
    Motwani, Sumeet and
    Reuel, Anka and
    Conitzer, Vincent and
    Dennis, Michael and
    Gabriel, Iason and
    Gleave, Adam and
    Hadfield, Gillian and
    Haghtalab, Nika and
    Kasirzadeh, Atoosa and
    Krier, S\'{e}bastien and
    Larson, Kate and
    Lehman, Joel and
    Parkes, David C. and
    Piliouras, Georgios and
    Rahwan, Iyad
  },
  title         = {Multi-Agent Risks from Advanced {AI}},
  institution   = {Cooperative AI Foundation},
  type          = {Technical Report},
  number        = {1},
  year          = {2025},
  month         = feb,
  url           = {https://arxiv.org/abs/2502.14143},
  eprint        = {2502.14143},
  archivePrefix = {arXiv},
  primaryClass  = {cs.MA}
}

@article{hofbauer2003evolutionary,
  title={Evolutionary game dynamics},
  author={Hofbauer, Josef and Sigmund, Karl},
  journal={Bulletin of the American mathematical society},
  volume={40},
  number={4},
  pages={479--519},
  year={2003}
}

@article{zhang1989simple,
  title={Simple fast algorithms for the editing distance between trees and related problems},
  author={Zhang, Kaizhong and Shasha, Dennis},
  journal={SIAM journal on computing},
  volume={18},
  number={6},
  pages={1245--1262},
  year={1989},
  publisher={SIAM}
}

@article{nash1950equilibrium,
  title={Equilibrium points in n-person games},
  author={Nash, John F.},
  journal={Proceedings of the National Academy of Sciences},
  volume={36},
  number={1},
  pages={48--49},
  year={1950},
  publisher={National Academy of Sciences}
}

@article{nash1951,
 author = {Nash, John F.},
 journal = {Annals of Mathematics},
 number = {2},
 pages = {286--295},
 publisher = {Annals of Mathematics, Princeton University},
 title = {Non-Cooperative Games},
 volume = {54},
 year = {1951}
}

@article{Critch,
  author       = {Andrew Critch and
                  Michael Dennis and
                  Stuart Russell},
  title        = {Cooperative and uncooperative institution designs: Surprises and problems
                  in open-source game theory},
  journal      = {CoRR},
  volume       = {abs/2208.07006},
  year         = {2022},
  eprinttype   = {arXiv},
  eprint       = {2208.07006},
  bibsource    = {dblp computer science bibliography, https://dblp.org}
}

@article{Modal_agents_barasz,
  author       = {Mih{\'{a}}ly B{\'{a}}r{\'{a}}sz and
                  Paul F. Christiano and
                  Benja Fallenstein and
                  Marcello Herreshoff and
                  Patrick LaVictoire and
                  Eliezer Yudkowsky},
  title        = {Robust Cooperation in the Prisoner's Dilemma: Program Equilibrium
                  via Provability Logic},
  journal      = {CoRR},
  volume       = {abs/1401.5577},
  year         = {2014},
  eprinttype   = {arXiv},
  eprint       = {1401.5577},
  bibsource    = {dblp computer science bibliography, https://dblp.org}
}

@article{oesterheld2019robust,
  title={Robust program equilibrium},
  author={Oesterheld, Caspar},
  journal={Theory and Decision},
  volume={86},
  number={1},
  pages={143--159},
  year={2019},
  publisher={Springer}
}

@inproceedings{DBLP:conf/nips/OesterheldTGCF23,
  author       = {Caspar Oesterheld and
                  Johannes Treutlein and
                  Roger B. Grosse and
                  Vincent Conitzer and
                  Jakob N. Foerster},
  editor       = {Alice Oh and
                  Tristan Naumann and
                  Amir Globerson and
                  Kate Saenko and
                  Moritz Hardt and
                  Sergey Levine},
  title        = {Similarity-based cooperative equilibrium},
  booktitle    = {Advances in Neural Information Processing Systems 36: Annual Conference
                  on Neural Information Processing Systems 2023, NeurIPS 2023, New Orleans,
                  LA, USA, December 10 - 16, 2023},
  year         = {2023},
  bibsource    = {dblp computer science bibliography, https://dblp.org}
}

@inproceedings{DBLP:conf/ijcai/KovarikOC23,
  author       = {Vojtech Kovar{\'{\i}}k and
                  Caspar Oesterheld and
                  Vincent Conitzer},
  title        = {Game Theory with Simulation of Other Players},
  booktitle    = {Proceedings of the Thirty-Second International Joint Conference on
                  Artificial Intelligence, {IJCAI} 2023, 19th-25th August 2023, Macao,
                  SAR, China},
  pages        = {2800--2807},
  publisher    = {ijcai.org},
  year         = {2023},
  url          = {https://doi.org/10.24963/ijcai.2023/312},
  doi          = {10.24963/IJCAI.2023/312},
  bibsource    = {dblp computer science bibliography, https://dblp.org}
}

@inproceedings{DBLP:conf/aaai/CooperOC25,
  author       = {Emery Cooper and
                  Caspar Oesterheld and
                  Vincent Conitzer},
  editor       = {Toby Walsh and
                  Julie Shah and
                  Zico Kolter},
  title        = {Characterising Simulation-Based Program Equilibria},
  booktitle    = {Thirty-Ninth {AAAI} Conference on Artificial Intelligence, Thirty-Seventh
                  Conference on Innovative Applications of Artificial Intelligence,
                  Fifteenth Symposium on Educational Advances in Artificial Intelligence,
                  {AAAI} 2025, Philadelphia, PA, USA, February 25 - March 4, 2025},
  pages        = {13735--13744},
  publisher    = {{AAAI} Press},
  year         = {2025},
  doi          = {10.1609/AAAI.V39I13.33501},
  bibsource    = {dblp computer science bibliography, https://dblp.org}
}

@article{holmstrom1979moral,
  title={Moral hazard and observability},
  author={Holmstr{\"o}m, Bengt},
  journal={The Bell journal of economics},
  pages={74--91},
  year={1979},
  publisher={JSTOR}
}

@article{battigalli2016analysis,
  title={Analysis of information feedback and selfconfirming equilibrium},
  author={Battigalli, Pierpaolo and Cerreia-Vioglio, Simone and Maccheroni, Fabio and Marinacci, Massimo},
  journal={Journal of Mathematical Economics},
  volume={66},
  pages={40--51},
  year={2016},
  publisher={Elsevier}
}

@article{formenti2026selfconfirming,
  title         = {Self-Confirming Equilibrium under Model Misspecification},
  author        = {Formenti, Riccardo},
  year          = {2026},
  eprint        = {2610.35692},
  journal = {arXiv},
  volume           = {arXiv.2610.35692},
  url           = {https://arxiv.org/abs/2610.35692}
}

@incollection{gilboa2016ambiguity,
  title={Ambiguity and the Bayesian paradigm},
  author={Gilboa, Itzhak and Marinacci, Massimo},
  booktitle={Readings in formal epistemology: Sourcebook},
  pages={385--439},
  year={2016},
  publisher={Springer}
}

@article{avis2010enumeration,
  title={Enumeration of Nash equilibria for two-player games},
  author={Avis, David and Rosenberg, Gabriel D and Savani, Rahul and Von Stengel, Bernhard},
  journal={Economic theory},
  volume={42},
  number={1},
  pages={9--37},
  year={2010},
  publisher={Springer}
}

@article{jansen1981maximal,
  title={Maximal Nash subsets for bimatrix games},
  author={Jansen, MJM},
  journal={Naval research logistics quarterly},
  volume={28},
  number={1},
  pages={147--152},
  year={1981},
  publisher={Wiley Online Library}
}

@article{Pudlak_Survey,
  author       = {Pavel Pudl{\'{a}}k},
  title        = {Incompleteness in the finite Domain},
  journal      = {Bull. Symb. Log.},
  volume       = {23},
  number       = {4},
  pages        = {405--441},
  year         = {2017},
  url          = {https://doi.org/10.1017/bsl.2017.32},
  doi          = {10.1017/BSL.2017.32},
  bibsource    = {dblp computer science bibliography, https://dblp.org}
}

@article{Critch_pblt,
  author       = {Andrew Critch},
  title        = {A parametric, Resource-Bounded Generalization of l{\"{o}}B's
                  Theorem, and a robust Cooperation criterion for Open-Source Game Theory},
  journal      = {J. Symb. Log.},
  volume       = {84},
  number       = {4},
  pages        = {1368--1381},
  year         = {2019},
  url          = {https://doi.org/10.1017/jsl.2017.42},
  doi          = {10.1017/JSL.2017.42},
  bibsource    = {dblp computer science bibliography, https://dblp.org}
}

@techreport{anthropic2026opus48,
  author       = {{Anthropic}},
  title        = {System Card: {C}laude {O}pus 4.8},
  institution  = {Anthropic},
  year         = {2026},
  month        = may,
  howpublished = {\url{https://www.anthropic.com/claude-opus-4-8-system-card}},
  note         = {Published May 28, 2026. Accessed 2026-09-18}
}

@techreport{anthropic2026fable51,
  author       = {{Anthropic}},
  title        = {System Card: {C}laude {F}able 5.1 \& {C}laude {M}ythos 5.1},
  institution  = {Anthropic},
  year         = {2026},
  month        = sep,
  howpublished = {\url{https://www.anthropic.com/claude-fable-5-1-mythos-5-1-system-card}},
  note         = {Published September 1, 2026. Accessed 2026-09-18}
}

@misc{lesswrong_tournament,
  author = {Alex Mennen},
  title  = {Prisoner's dilemma tournament results},
  year   = {2013},
  url    ={https://www.lesswrong.com/posts/QP7Ne4KXKytj4Krkx/prisoner-s-dilemma-tournament-results-0},
  note   = {LessWrong}
  }

@inproceedings{goedelproverv2,
  author       = {Yong Lin and
                  Shange Tang and
                  Bohan Lyu and
                  Ziran Yang and
                  Jui{-}Hui Chung and
                  Haoyu Zhao and
                  Lai Jiang and
                  Yihan Geng and
                  Jiawei Ge and
                  Jingruo Sun and
                  Jiayun Wu and
                  Jiri Gesi and
                  Ximing Lu and
                  David Acuna and
                  Kaiyu Yang and
                  Hongzhou Lin and
                  Yejin Choi and
                  Danqi Chen and
                  Sanjeev Arora and
                  Chi Jin},
  editor       = {Carl Vondrick and
                  Bharath Hariharan and
                  Colin Raffel and
                  Lerrel Pinto and
                  Diyi Yang and
                  Aleksandra Faust},
  title        = {Goedel-Prover-V2: Scaling Formal Theorem Proving with Scaffolded Data
                  Synthesis and Self-Correction},
  booktitle    = {The Fourteenth International Conference on Learning Representations,
                  {ICLR} 2026, Rio de Janeiro, Brazil, April 23-27, 2026},
  publisher    = {proceedings.iclr.cc},
  year         = {2026},
  url          = {https://proceedings.iclr.cc/paper\_files/paper/2026/hash/13e8be77982beb73d7ed0bbf122f9f3c-Abstract-Conference.html},
  bibsource    = {dblp computer science bibliography, https://dblp.org}
}

@article{kimina_prover,
  author       = {Haiming Wang and
                  Mert Unsal and
                  Xiaohan Lin and
                  Mantas Baksys and
                  Junqi Liu and
                  Marco Dos Santos and
                  Flood Sung and
                  Marina Vinyes and
                  Zhenzhe Ying and
                  Zekai Zhu and
                  Jianqiao Lu and
                  Hugues de Saxc{\'{e}} and
                  Bolton Bailey and
                  Chendong Song and
                  Chenjun Xiao and
                  Dehao Zhang and
                  Ebony Zhang and
                  Frederick Pu and
                  Han Zhu and
                  Jiawei Liu and
                  Jonas Bayer and
                  Julien Michel and
                  Longhui Yu and
                  L{\'{e}}o Dreyfus{-}Schmidt and
                  Lewis Tunstall and
                  Luigi Pagani and
                  Moreira Machado and
                  Pauline Bourigault and
                  Ran Wang and
                  Stanislas Polu and
                  Thibaut Barroyer and
                  Wen{-}Ding Li and
                  Yazhe Niu and
                  Yann Fleureau and
                  Yangyang Hu and
                  Zhouliang Yu and
                  Zihan Wang and
                  Zhilin Yang and
                  Zhengying Liu and
                  Jia Li},
  title        = {Kimina-Prover Preview: Towards Large Formal Reasoning Models with
                  Reinforcement Learning},
  journal      = {CoRR},
  volume       = {abs/2504.11354},
  year         = {2025},
  url          = {https://doi.org/10.48550/arXiv.2504.11354},
  eprinttype   = {arXiv},
  eprint       = {2504.11354},
  bibsource    = {dblp computer science bibliography, https://dblp.org}
}

@inproceedings{LeanDojo,
  author       = {Kaiyu Yang and
                  Aidan M. Swope and
                  Alex Gu and
                  Rahul Chalamala and
                  Peiyang Song and
                  Shixing Yu and
                  Saad Godil and
                  Ryan J. Prenger and
                  Animashree Anandkumar},
  editor       = {Alice Oh and
                  Tristan Naumann and
                  Amir Globerson and
                  Kate Saenko and
                  Moritz Hardt and
                  Sergey Levine},
  title        = {LeanDojo: Theorem Proving with Retrieval-Augmented Language Models},
  booktitle    = {Advances in Neural Information Processing Systems 36: Annual Conference
                  on Neural Information Processing Systems 2023, NeurIPS 2023, New Orleans,
                  LA, USA, December 10 - 16, 2023},
  year         = {2023},
  bibsource    = {dblp computer science bibliography, https://dblp.org}
}

@misc{foundation,
  author = {Saitou, Shogo and Noguchi, Mashu},
  title  = {Formalized Formal Logic: A Mechanized Proof Library for Mathematical Logic},
  year   = {2023},
  url    = {https://github.com/FormalizedFormalLogic/Foundation},
  note   = {Lean 4 library, Apache-2.0. Git commit \texttt{a77e4e9},
            accessed September 2, 2026},
}

@incollection{Pudlak_bounds_for_finite_proofs,
title = {On the length of proofs of finitistic consistency statements in first order theories},
editor = {J.B. Paris and A.J. Wilkie and G.M. Wilmers},
series = {Studies in Logic and the Foundations of Mathematics},
publisher = {Elsevier},
volume = {120},
pages = {165-196},
year = {1986},
booktitle = {Logic Colloquium '84},
issn = {0049-237X},
author = {Pavel Pudlák}
}

@misc{foundation_agents_formalized,
  author = {A. M. Berns},
  title  = {Formalized Agent Foundations},
  year   = {2026},
  url    = {https://github.com/A-M-Berns/Formalized-Agent-Foundations},
  note   = {Lean 4 library, Apache-2.0. Git commit \texttt{58e9231},
            accessed September 7, 2026},
}

@InProceedings{Lean4,
author="Moura, Leonardo de
and Ullrich, Sebastian",
editor="Platzer, Andr{\'e}
and Sutcliffe, Geoff",
title="The Lean 4 Theorem Prover and Programming Language",
booktitle="Automated Deduction -- CADE 28",
year="2021",
publisher="Springer International Publishing",
address="Cham",
pages="625--635",
isbn="978-3-030-79876-5"
}

@inproceedings{mathlib2020,
  author    = {{The mathlib Community}},
  title     = {The {L}ean Mathematical Library},
  booktitle = {Proceedings of the 9th ACM SIGPLAN International Conference on Certified Programs and Proofs (CPP 2020)},
  pages     = {367--381},
  year      = {2020},
}

@article{Godel1930,
  author  = {G{\"o}del, Kurt},
  title   = {Die {V}ollst{\"a}ndigkeit der {A}xiome des logischen {F}unktionenkalk{\"u}ls},
  journal = {Monatshefte f{\"u}r Mathematik und Physik},
  volume  = {37},
  number  = {1},
  pages   = {349--360},
  year    = {1930},
}

@book{soundness_first_order,
  author    = {Enderton, Herbert B.},
  title     = {A Mathematical Introduction to Logic},
  edition   = {2},
  publisher = {Academic Press},
  year      = {2001}
}

@article{Lob1955,
  author    = {Löb, Martin Hugo},
  title     = {Solution of a Problem of {Leon Henkin}},
  journal   = {The Journal of Symbolic Logic},
  volume    = {20},
  number    = {2},
  pages     = {115--118},
  year      = {1955},
  publisher = {Association for Symbolic Logic},

}

@inproceedings{minif2f,
title={miniF2F: a cross-system benchmark for formal Olympiad-level mathematics},
author={Kunhao Zheng and Jesse Michael Han and Stanislas Polu},
booktitle={International Conference on Learning Representations},
year={2022},
url={https://openreview.net/forum?id=9ZPegFuFTFv}
}

@article{formalreasoningwithRL,
	author = {Hubert, Thomas and Mehta, Rishi and Sartran, Laurent and Horv{\'a}th, Mikl{\'o}s Z. and {\v Z}u{\v z}i{\'c}, Goran and Wieser, Eric and Huang, Aja and Schrittwieser, Julian and Schroecker, Yannick and Masoom, Hussain and Bertolli, Ottavia and Zahavy, Tom and Mandhane, Amol and Yung, Jessica and Beloshapka, Iuliya and Ibarz, Borja and Veeriah, Vivek and Yu, Lei and Nash, Oliver and Lezeau, Paul and Mercuri, Salvatore and S{\"o}nne, Calle and Mehta, Bhavik and Davies, Alex and Zheng, Daniel and Pedregosa, Fabian and Li, Yin and von Glehn, Ingrid and Rowland, Mark and Albanie, Samuel and Velingker, Ameya and Schmitt, Simon and Lockhart, Edward and Hughes, Edward and Michalewski, Henryk and Sonnerat, Nicolas and Hassabis, Demis and Kohli, Pushmeet and Silver, David},
	id = {Hubert2026},
	journal = {Nature},
	number = {8106},
	pages = {607--613},
	title = {Olympiad-level formal mathematical reasoning with reinforcement learning},
	volume = {651},
	year = {2026},}

@misc{deepseekproverv2,
      title={DeepSeek-Prover-V2: Advancing Formal Mathematical Reasoning via Reinforcement Learning for Subgoal Decomposition}, 
      author={Z. Z. Ren and Zhihong Shao and Junxiao Song and Huajian Xin and Haocheng Wang and Wanjia Zhao and Liyue Zhang and Zhe Fu and Qihao Zhu and Dejian Yang and Z. F. Wu and Zhibin Gou and Shirong Ma and Hongxuan Tang and Yuxuan Liu and Wenjun Gao and Daya Guo and Chong Ruan},
      year={2025},
      eprint={2504.21801},
      archivePrefix={arXiv},
      primaryClass={cs.CL},
      url={https://arxiv.org/abs/2504.21801}, 
}
\bibliographystyle{tmlr}

\appendix
\crefalias{section}{appendix}
\crefalias{subsection}{appendix}
\crefalias{subsubsection}{appendix}

\section{Proofs of \Cref{sec:theory}}\label{sec:app:proof}

This Appendix collects all proofs from \cref{sec:theory}.

\begin{proof}[Proof of \Cref{th:NE-is-PE}]
Let $\bm{a}^* \in A \times A$ be a NE of $G$. Fix a player $i$ and a deviation $\pi_i\in\Sigma$. Since the opponent's constant program always outputs $a_{-i}^*$, let $b_i:=\operatorname{run}(\pi_i\parallel a_{-i}^*)\in A$. Then
\[
U_i(\pi_i,a_{-i}^*)
= u_i(b_i,a_{-i}^*)
\le u_i(a_i^*,a_{-i}^*)
= U_i(a_i^*,a_{-i}^*),
\]
where the inequality follows because $\bm{a}^*$ is a Nash Equilibrium of $G$. Thus no unilateral program deviation is profitable, and $\bm{a}^*$ is a Program Equilibrium.
\end{proof}

\begin{proof}[Proof of \Cref{th:NE-is-tau-PE}]
Since constant programs ignore their input, $\operatorname{run}(a \| \mu) = a$ for all $a \in A$ and $\mu \in \Delta(\Sigma)$. In particular, $\operatorname{outcome}^\tau(\bm{a}^*) = \bm{a}^*$, and hence $U_i^\tau(\bm{a}^*) = u_i(\bm{a}^*)$ for each $i \in I$. Fix $i \in I$ and a deviation $\pi_i' \in \Sigma$, and set $b \coloneqq \operatorname{run}\bigl(\pi_i' \mathrel{\|} \tau_i(a_{-i}^*)\bigr) \in A$. The opponent's constant program outputs $a_{-i}^*$ whatever belief $\tau_{-i}(\pi_i')$ it receives, so
\begin{equation*}
  U_i^\tau(\pi_i', a_{-i}^*) = u_i(b, a_{-i}^*).
\end{equation*}

($\Leftarrow$) If $\bm{a}^*$ is a Nash Equilibrium of $G$, then $u_i(b, a_{-i}^*) \le u_i(\bm{a}^*) = U_i^\tau(\bm{a}^*)$, so $\pi_i'$ is not a profitable deviation. Since $i$ and $\pi_i'$ were arbitrary, $\bm{a}^*$ is a $\tau$-Program Equilibrium.

($\Rightarrow$) For every $b \in A$, the constant program $b$ belongs to $\Sigma$ and satisfies $U_i^\tau(b, a_{-i}^*) = u_i(b, a_{-i}^*)$. If $\bm{a}^*$ is a $\tau$-Program Equilibrium, then $u_i(b, a_{-i}^*) \le U_i^\tau(\bm{a}^*) = u_i(\bm{a}^*)$ for all $i \in I$ and $b \in A$, that is, $\bm{a}^*$ is a Nash Equilibrium of $G$.
\end{proof}

\begin{theorem}\label{th:mimic-app}
    Assume player $i \in I$ has undetectable deviations. If $\bm{\pi} \in \Sigma \times \Sigma$ is a pure $\tau$-Program Equilibrium with outcome $\bm{a}$, then $a_i \in \mathrm{BR}_i(a_{-i})$, where $\mathrm{BR}_i$ is the best-response correspondence defined by
    \begin{equation*}
        \mathrm{BR}_i(a_{-i}) \coloneq \mathrm{argmax}_{a' \in A}u_i(a', a_{-i}).
    \end{equation*}
    Moreover, if both players have undetectable deviations, then the outcomes of pure $\tau$-program equilibria are exactly the pure Nash equilibria of $G$.
\end{theorem}

\begin{proof}[Proof of \Cref{th:mimic-app}, hence of \Cref{th:mimic}]
    Let $\bm{\pi}$ be a pure $\tau$-PE with outcome $\bm{a}$. Define $\mu \coloneq \tau_i(\pi_{-i})$ and let $x \in \mathrm{BR}_i(a_{-i})$. Let $\rho$ satisfy \cref{def:mimic} for $(\pi_i, \mu, x)$. Then by condition $2$
    \begin{equation*}
        \mathrm{run}(\rho \| \mu) = x,
    \end{equation*}
    and by condition $1$
    \begin{equation*}
         \mathrm{run}(\pi_{-i} \| \tau_{-i}(\rho)) = \mathrm{run}(\pi_{-i} \| \tau_{-i}(\pi_{i})) = a_{-i}.
    \end{equation*}
    The equilibrium condition then implies that
    \begin{equation*}
        u_i(a_i, a_{-i}) = U_i^\tau(\bm{\pi}) \ge U_i^\tau(\rho, \pi_{-i}) = u_i(x, a_{-i}) = \mathrm{max}_{a' \in A}u_i(a', a_{-i}).
    \end{equation*}
    Finally, if both players have undetectable deviations, applying this argument to $i = 1$ and $i = 2$ shows that $\bm{a}$ is a pure Nash Equilibrium of $G$. Conversely, by \cref{th:NE-is-tau-PE}, every pure Nash Equilibrium of $G$ is the outcome of a pure $\tau$-PE, namely the profile of the corresponding constant programs.
\end{proof}

\section{Formal Tools and Theoretical Results in Open-Source Game Theory}
\label{apx-section:TheoreticalResultsOSGT}
This appendix collects the formal material behind \cref{subsection:MathOutcomes,sec:dupoc-v-cupod}. \Cref{apx-subsection:TheoreticalResultsOSGT-ToolsOutcomes} states the three tools, soundness, completeness and bounded Löb, and proves the outcomes of \texttt{DupocBot} that illustrate them. \Cref{app:proofs-dupoc-cupod} proves the transposition results of \cref{sec:dupoc-v-cupod}, including how programs are named inside sentences of $S$, and settles \texttt{DupocBot} against \texttt{CupodBot}. \Cref{apx-subsection:TheoreticalResultsOSGT-CorrectionProgress} revisits \citet{Critch}: a refutation of their Example~5.3(a) and progress on their open problems.

\subsection{Mathematical Tools and Outcomes}
\label{apx-subsection:TheoreticalResultsOSGT-ToolsOutcomes}

We take our running example \texttt{DupocBot} and illustrate each tool using concrete matchups of this bot.

\subsubsection{Soundness}
\label{apx-subsubsection:TheoreticalResultsOSGT-ToolsOutcomes-Soundness}
The first important property of our system $S$ is that it is sound, i.e. that proved statements are true.

\begin{proposition}[Soundness]\label{prop:soundness}
Every sentence provable in $S$ is true in $\mathbb{N}$ under the standard reading. In particular, $\vdash_k \varphi$ implies $\models \varphi$, for every budget $k$.
\end{proposition}
\begin{proof}
$\mathbb{N}$ satisfies the axioms of Peano Arithmetic, and under the standard reading, where $c_C$ is $0$ and $c_D$ is $1$, it satisfies the action axiom. So $\mathbb{N}$ is a model of $T_{\mathrm{Act}}$, and every sentence a theory proves holds in each of its models (Soundness of first-order theories; \citeauthor{soundness_first_order}, \citeyear[Section~2.5]{soundness_first_order}).
\end{proof}

Soundness allows us to prove that certain searches must fail, through its contrapositive:

\begin{proposition}\label{prop:dupoc-defect}
Let \texttt{DefectBot} be the program that always defects. Then
for every $k$, \[\mathrm{outcome}(\texttt{DupocBot}(k), \texttt{DefectBot}) = (D, D)\].
\end{proposition}
\begin{proof}
\texttt{DefectBot} plays $D$, so the hypothesis \enquote{\texttt{DefectBot} cooperates with me} is false. By the contrapositive of \cref{prop:soundness} it has no proof, of any length. The search of \texttt{DupocBot}$(k)$ therefore fails at every budget, and \texttt{DupocBot}$(k)$ plays its default, $D$.
\end{proof}

Soundness also ensures robustness of certain programs, which we call \emph{unexploitable}: a program $\pi$ such that no opponent $\pi'$ can get $\mathrm{outcome}(\pi, \pi') = (C, D)$. Programs that cooperate \emph{only after proving that their opponent cooperates with them} have this property, because the proof they found guarantees (thanks to soundness) the opponent's cooperation.

\begin{proposition}\label{prop:unexploitability}
\texttt{DefectBot}, \texttt{DupocBot} and \texttt{CIMCIC}\footnote{See \cref{app:bot-definitions} for programs' definitions.} are unexploitable.
\end{proposition}
\begin{proof}
   \texttt{DefectBot} never cooperates, so there is nothing to show.
   
   \texttt{DupocBot(k)} only cooperates if \texttt{proof\_search(k, opp, $\varphi$)} succeeded for $\varphi$ = \enquote{\texttt{opp} cooperates against \texttt{DupocBot(k)}}. So if \texttt{DupocBot} cooperates, that means $\varphi$ is provable, which means that its opponent plays $C$, by soundness.

   Similarly, \texttt{CIMCIC(k)} only cooperates if \texttt{proof\_search(k, opp, $\varphi$)} succeeded for $\varphi$ = \enquote{if \texttt{CIMCIC(k)} cooperates, then \texttt{opp} cooperates against it}. If \texttt{CIMCIC} cooperates, that means $\varphi$ is provable, thus true, by soundness. Since the premise of $\varphi$ is also true, the conclusion must be true, i.e. its opponent plays $C$. 
\end{proof}

\subsubsection{Completeness}  Naturally, besides soundness, $S$ is complete: if a statement is true for any model of $S$, then $S$ should be able to prove it.
\begin{theorem}[Completeness, \citet{Godel1930}]
\label{thm:completeness}
Let $T$ be a first-order theory and $\varphi$ a sentence of its language. If $\varphi$ holds in every model of $T$, then $T$ proves $\varphi$. In other words, if $M \models \varphi$ for every model $M$ of $T$, then $T \vdash \varphi$, with no bound on the length of the proof.
\end{theorem}
For $S$, checking every model reduces to checking the two readings of the constants:
\begin{corollary}[Completeness over the two readings]
\label{cor:completeness-readings}
A sentence that holds in every model of Peano Arithmetic, under both the standard and the swapped reading of $c_C$ and $c_D$, is provable in $S$.
\end{corollary}
\begin{proof}
By the action axiom, the models of $T_{\mathrm{Act}}$ are exactly the models of Peano Arithmetic equipped with one of the two readings. The claim is therefore \cref{thm:completeness} applied to $T_{\mathrm{Act}}$.
\end{proof}

This second tool allows us to show that a search succeeds. Against \texttt{CooperateBot}, the program that always plays $C$, the hypothesis of \texttt{DupocBot}$(k)$ is true, and \texttt{DupocBot}$(k)$ cooperates exactly if it has a proof of length at most $k$. Instead of producing such a proof, we show that one exists, using Gödel's completeness theorem:
\begin{proposition}\label{prop:dupoc-coop}
For all sufficiently large $k$, \[
\mathrm{outcome}(\texttt{DupocBot}(k), \texttt{CooperateBot}) = (C, C)\].
\end{proposition}
\begin{proof}
Let $G[k]$ be the hypothesis \enquote{\texttt{CooperateBot} cooperates with \texttt{DupocBot}$(k)$}. The sentence $\forall k\, G[k]$ holds in every model of Peano Arithmetic under both readings: under the standard one it says that a constant program plays its constant, and under the swapped one it says the same of the transposed programs. By \cref{cor:completeness-readings}, $S$ proves $\forall k\, G[k]$, with a proof of some fixed length $N$. Instantiating it at a given $k$ adds the length of the numeral of $k$, which we can write in logarithmic terms, so $\vdash_k G[k]$ as soon as $k$ exceeds $N$ plus this overhead. The search of \texttt{DupocBot}$(k)$ then succeeds and it plays $C$, as does \texttt{CooperateBot}.
\end{proof}
Completeness gives a proof without giving its length. The outcome therefore holds for all $k$ above a threshold that is shown to exist but is not computed, whereas soundness involves no estimate on lengths and gives outcomes at \emph{every} budget. Every outcome in this paper is of one of these two kinds. Neither, however, handles self-reference.

\subsubsection{Löb's Theorem}
When \texttt{DupocBot}$(k)$ faces itself, its hypothesis is true only if its own search succeeds (\cref{ex:prisonerdilemma-3}). Neither of the tools above helps, because we cannot verify the truthfulness of the hypothesis before knowing the result of the search. This self-fulfilling fact is directly resolved by Löb's theorem \citep{Lob1955}, which we state in its bounded form:

\begin{theorem}[Bounded Löb, after \citet{Critch_pblt}; see also \citet{Critch}, Lemma~3.6]\label{thm:pblt}
Let $\varphi[k]$ be a formula of $S$ with a single free variable $k$. Suppose that, for some $k_1\in\mathbb{N}$, $S$ can verify that $\varphi[k]$ is \enquote{potentially self-fulfilling for large $k$}:
\[
  \vdash \big(\forall\, k>k_1\big)\, \big(\Box_{k}(\varphi[k]) \to \varphi[k]\big).
\]
Then there is a threshold $k_2 \in \mathbb{N}$ such that $\vdash_k \varphi[k]$ for every $k > k_2$.
\end{theorem}

\citet{Critch_pblt} derive this theorem from four efficiency assumptions on the proof system, recalled in \citet[Appendix B]{Critch}. For $S$, none of them is assumed. Property (a), that $S$ can express every computable function, holds because $S$ extends Peano Arithmetic. Property (b), that $S$ writes the number $k$ in $\mathcal{O}(\lg k)$ symbols, holds by the choice of binary numerals (\cref{apx-section:LeanSystemS}). The abbreviation mechanism (c) is not used. The assumption (d) that a proof of length $\ell$ can be checked inside the system by a proof of length linear in $\ell$ is replaced by a theorem with a polynomial bound: if $\vdash_\ell \varphi$, then $S$ proves $\Box_\ell\varphi$ with a proof of length polynomial in $\ell$.\footnote{This is further explained in \cref{apx-section:LeanSystemS}, \texttt{boundedInnerNec\_sixteen}.} This weaker form suffices for \cref{thm:pblt}.
\begin{theorem}[\citet{Critch}, Thm.~3.7]\label{thm:dupoc-selfplay}
For all sufficiently large $k$, \[
\mathrm{outcome}(\texttt{DupocBot}(k), \texttt{DupocBot}(k)) = (C, C).
\]
\end{theorem}
\begin{proof}
Apply \cref{thm:pblt} with $\varphi[k] = \text{\enquote{\texttt{DupocBot(k)(DupocBot(k)) == C}}}$ and $k_1 = 0$. If $\varphi[k]$ has a proof of length at most $k$, then \texttt{DupocBot}$(k)$ finds it and plays $C$, so $\Box_{k}(\varphi[k]) \to \varphi[k]$ holds. This reasoning is valid in every model of Peano Arithmetic and under both readings, so the implication is provable in $S$ by \cref{cor:completeness-readings}. \Cref{thm:pblt} then gives $\vdash_k \varphi[k]$ for every $k > k_2$: the search of each copy of \texttt{DupocBot}$(k)$ succeeds, and both play $C$.
\end{proof}

\subsection{Proofs of \cref{sec:dupoc-v-cupod}}\label{app:proofs-dupoc-cupod}

This appendix contains the proofs and additional details for \cref{sec:dupoc-v-cupod}.

\begin{proof}[Proof of \Cref{thm:transposition}]
Let $d$ be a proof of $\varphi$ of length at most $k$, and apply $\mathcal{T}$ to every formula of $d$. We check that the result $\mathcal{T}d$ is a proof of $\mathcal{T}\varphi$. A logical step remains a logical step, because $\mathcal{T}$ commutes with the logical symbols and logical rules do not depend on which constant occurs. An axiom of Peano Arithmetic does not mention the two constants and is unchanged. The action axiom is sent to itself with its two disjuncts exchanged, which is again an axiom by (S2). Hence $\mathcal{T}d$ is a proof of $\mathcal{T}\varphi$. It has the same length as $d$, since the length of a proof counts its symbols without regard to which symbol they are. This gives $\vdash_k\varphi \Rightarrow \vdash_k\mathcal{T}\varphi$, and the converse follows by applying the same argument to $\mathcal{T}\varphi$, as $\mathcal{T}$ is an involution.
\end{proof}

We move on to proving \cref{prop:bridge}, which demands careful attention as to how programs are written inside a sentence. Consider the sentence of \cref{sec:dupoc-v-cupod}
\[
  \varphi \;=\; \mathrm{Plays}(\pi, \pi', a),
\]
\enquote{$\pi$ plays $a$ against $\pi'$}. Programs are encoded by natural numbers by an injective coding, and we write $\ulcorner \pi \urcorner$ for the code of $\pi$ (its value in $\mathbb{N}$; see \cref{apx:Lean-SystemS-Defs}). Since the language of $S$ is $\{0, 1, +, \cdot, <, c_C, c_D\}$, the sentence $\varphi$ can mention $\pi$ only through a closed term that evaluates to $\ulcorner \pi \urcorner$, which we call a \emph{name} of $\pi$ and write $\overline{\pi}$. 

The question of which term to use for the name arises because of what $\mathcal{T}$ must do to $\varphi$. \Cref{prop:bridge} asks that $\mathcal{T}\varphi$ be the sentence \enquote{$\mathcal{T}\pi$ plays $\mathcal{T}a$ against $\mathcal{T}\pi'$}. Now suppose we take the obvious name, the numeral $\dot{\ulcorner\pi\urcorner}$ of the code of $\pi$ (we write $\dot n$ for the numeral of a number $n$). It is a string of $0$, $1$ and $+$, which $\mathcal{T}$ leaves unchanged, so $\mathcal{T}\varphi$ flips only the constant $c_a$ and reads \enquote{$\pi$ plays $\mathcal{T}a$ against $\pi'$}: the same two programs, the other action. For the theorem to hold, the names themselves must contain the two constants, placed so that exchanging them turns the name of $\pi$ into the name of $\mathcal{T}\pi$,
\[
  \mathcal{T}(\overline{\pi}) \;=\; \overline{\mathcal{T}\pi}.
\]
We now construct names with this property.

First, define $\mathrm{reval}(u, w, x)$, a function that, for a program code $x$ and $u, w \in \{0, 1\}$, replaces every action value $0$ of the program by $u$ and every action value $1$ by $w$, including the action constants inside the hypotheses the program stores, leaving the rest of its syntax unchanged, and returns a fixed default on other inputs. Thus, for every program $\pi$,\footnote{A hypothesis may itself contain a provability statement $\Box_j\psi$, which names $\psi$ by a closed term $\overline{\psi}$. This raises the same problem for $\psi$ we are solving for $\pi$: a numeral would be left unchanged by $\mathcal{T}$, and the second identity would fail for such programs. We therefore build $\overline{\psi}$ in the same way, so that $\mathcal{T}(\overline{\psi}) = \overline{\mathcal{T}\psi}$. See \texttt{boxQuote} in \texttt{arith/ArithS/Agents/Code.lean}.}
\[
\begin{aligned}
  \mathrm{reval}(0,1,\ulcorner \pi \urcorner) &= \ulcorner \pi \urcorner,\\
  \mathrm{reval}(1,0,\ulcorner \pi \urcorner) &= \ulcorner \mathcal{T}\pi \urcorner.
\end{aligned}
\]

To construct the names, we choose a common representative code for each pair $\{\pi, \mathcal{T}\pi\}$,
\[
  \mathrm{rep}(\pi) := \min\{\ulcorner \pi \urcorner, \ulcorner \mathcal{T}\pi \urcorner\}.
\]
Since $\mathcal{T}$ is an involution, $\mathrm{rep}(\mathcal{T}\pi)=\mathrm{rep}(\pi)$. We then choose the arguments that reconstruct $\pi$ from this representative:
\[
  (u_\pi, w_\pi) :=
  \begin{cases}
    (\dot 0, \dot 1),     & \text{if } \pi = \mathcal{T}\pi,\\
    (c_C, c_D), & \text{if } \ulcorner \pi \urcorner < \ulcorner \mathcal{T}\pi \urcorner,\\
    (c_D, c_C), & \text{if } \ulcorner \mathcal{T}\pi \urcorner < \ulcorner \pi \urcorner.
  \end{cases}
\]
Injectivity of the coding makes the three conditions exclusive and exhaustive.

\begin{definition}[Names of programs]\label{def:names}
The name of a program $\pi$ is the closed term\footnote{Above, $\mathrm{reval}$ is a function on numbers; inside $\overline{\pi}$ it stands for the formula of $S$ defining that function, applied to the terms $u_\pi$, $w_\pi$, $\dot{\mathrm{rep}(\pi)}$.}
\[
  \overline{\pi} := \mathrm{reval}\bigl(u_\pi, w_\pi, \dot{\mathrm{rep}(\pi)}\bigr).
\]
\end{definition}

Under the standard reading, $\overline{\pi}$ evaluates to $\mathrm{reval}(0, 1, \mathrm{rep}(\pi))$ or $\mathrm{reval}(1, 0, \mathrm{rep}(\pi))$ according to the case, which is $\ulcorner \pi \urcorner$ in each; under the swapped reading it evaluates to $\ulcorner \mathcal{T}\pi \urcorner$.

We use these names to express statements about program execution. Let $\mathrm{Run}(x, y, z)$ be a fixed arithmetical formula asserting that the program with code $x$, run on input $y$, halts with output $z$. For programs $\pi, \pi'$ and an action $a$, the sentence of \cref{sec:dupoc-v-cupod} is
\[
  {\mathrm{Plays}(\pi, \pi', a) := \mathrm{Run}(\overline{\pi}, \overline{\pi'}, c_a).}
\]
The formula $\mathrm{Run}$ contains no action constant; the constants enter the sentence only through the three substituted terms.

\begin{proof}[Proof of \cref{prop:bridge}]
We first show what was promised: $\mathcal{T}(\overline{\pi}) = \overline{\mathcal{T}\pi}$. We have $\mathrm{rep}(\mathcal{T}\pi) = \mathrm{rep}(\pi)$, and
\[
  (\mathcal{T}u_\pi, \mathcal{T}w_\pi) = (u_{\mathcal{T}\pi}, w_{\mathcal{T}\pi}),
\]
because $\mathcal{T}$ exchanges the two orientations when $\pi \neq \mathcal{T}\pi$ and fixes the pair of numerals $(\dot 0, \dot 1)$ when $\pi = \mathcal{T}\pi$. Since $\mathcal{T}$ also fixes the defining formula of $\mathrm{reval}$ and the term $\dot{\mathrm{rep}(\pi)}$, the identity follows, and likewise for $\pi'$. Then, since $\mathcal{T}$ fixes $\mathrm{Run}$ and $\mathcal{T}c_a = c_{\mathcal{T}a}$,
\[
  \mathcal{T}\,\mathrm{Plays}(\pi, \pi', a)
  = \mathrm{Run}(\mathcal{T}(\overline{\pi}), \mathcal{T}(\overline{\pi'}), \mathcal{T}(c_a))
  = \mathrm{Run}(\overline{\mathcal{T}\pi}, \overline{\mathcal{T}\pi'}, c_{\mathcal{T}a})
  = \mathrm{Plays}(\mathcal{T}\pi, \mathcal{T}\pi', \mathcal{T}a)
\]
as an identity of syntactic expressions.
\end{proof}

\begin{proof}[Proof of \cref{thm:transposition-outcomes}]
We show, by induction on the execution of $\pi$ against $\pi'$, that $\mathcal{T}\pi$ against $\mathcal{T}\pi'$ takes the corresponding steps and returns $\mathcal{T}a$ whenever $\pi$ returns $a$. This proves the theorem because then $\mathrm{outcome}(\mathcal{T}\pi, \mathcal{T}\pi')$  gives $(\mathcal{T}a, \mathcal{T}b)$ when $\mathrm{outcome}(\pi, \pi') = (a, b)$, which is $\mathcal{T}\,\mathrm{outcome}(\pi, \pi')$.

If $\pi$ is the constant $a$, then $\mathcal{T}\pi$ is the constant $\mathcal{T}a$. If $\pi$ branches on whether a sub-computation returns $a$, then $\mathcal{T}\pi$ branches on whether the transposed sub-computation returns $\mathcal{T}a$; by the induction hypothesis the latter returns $\mathcal{T}r$ when the former returns $r$, and $\mathcal{T}r = \mathcal{T}a$ exactly when $r = a$, so both take the same branch. If $\pi$ searches for a proof of $\varphi$ with budget $j$, then $\mathcal{T}\pi$ searches for a proof of $\mathcal{T}\varphi$ with the same budget: the hypothesis of $\mathcal{T}\pi$ is that of $\pi$ with the actions exchanged, and since $\mathcal{T}$ commutes with the connectives and, by \cref{prop:bridge} and the footnote on $\overline{\psi}$, with the naming of programs and of quoted formulas, its sentence is $\mathcal{T}\varphi$. By \cref{thm:transposition}, $\vdash_j \varphi$ if and only if $\vdash_j \mathcal{T}\varphi$, so the two searches succeed or fail together and both programs take corresponding branches. This concludes the induction.

For the last claim, let $\mathrm{outcome}(\pi, \mathcal{T}\pi) = (a, b)$. Reversing the players, $\mathrm{outcome}(\mathcal{T}\pi, \pi) = (b, a)$. Since $\mathcal{T}\mathcal{T}\pi = \pi$, the first part gives $(b, a) = \mathrm{outcome}(\mathcal{T}\pi, \pi) = \mathcal{T}\,\mathrm{outcome}(\pi, \mathcal{T}\pi) = (\mathcal{T}a, \mathcal{T}b)$. Hence $b = \mathcal{T}a$.
\end{proof}

\begin{proof}[Proof of \cref{thm:dupoc-cupod}]
Fix $k$. Since $\texttt{CupodBot}(k) = \mathcal{T}\,\texttt{DupocBot}(k)$, \cref{thm:transposition-outcomes} gives \[\mathrm{outcome}(\texttt{DupocBot}(k), \texttt{CupodBot}(k)) \in \{(C, D), (D, C)\}.\] In the first case \texttt{DupocBot}$(k)$ cooperates against a defecting opponent, which \cref{prop:unexploitability} excludes. Hence the outcome is $(D, C)$.
\end{proof}

\subsection{Corrections and Progress on \citet{Critch}}
\label{apx-subsection:TheoreticalResultsOSGT-CorrectionProgress}

With a concrete formal system and the tools above, we can revisit the open problems of \citet{Critch}. Löb’s theorem alone turns out to be enough to refute one of their stated examples, and the tools together resolve or reduce several of their open problems. We begin with the refutation, then present the open problems and what the library proves about each.

\subsubsection{\texttt{DIMCID} Always Defects: A Refutation of Example 5.3(a)}
Consider the bot \texttt{DIMCID}: \enquote{defect if my cooperation implies your defection} \citep[\S5.2]{Critch}. Below, we prove, using Löb's theorem, that \texttt{DIMCID} defects against every closed program, so that it is behaviorally equivalent to \texttt{DefectBot}. This directly contradicts Example 5.3(a) of \citet{Critch}, where they stipulate that \texttt{DIMCID} cooperates with \texttt{CooperateBot}.

\begin{theorem}[\texttt{DIMCID} always defects]\label{thm:Theory-DIMCID-Defects}
For every program $\pi$ there is a $k_0$ such that for all $k > k_0$,
\[
  \mathrm{outcome}\bigl(\texttt{DIMCID}(k), \pi\bigr) = (D, \cdot),
\]
and the guard of \texttt{DIMCID}$(k)$ is provable in $S$ within length $k$.
\end{theorem}

\begin{proof}
Write $p$ for \enquote{\texttt{DIMCID}$(k)$ defects against $\pi$} and $q$ for \enquote{$\pi$ defects against \texttt{DIMCID}$(k)$}. The bot searches for a proof of $\neg p \to q$, which is $p \vee q$, and defects exactly when it finds one, so its action is the fixed point of
\[
  p \;\leftrightarrow\; \Box_k (p \vee q).
\]
Since $p \to p \vee q$, we have $\Box_k p \to \Box_k(p \vee q)$, and the fixed point turns the right-hand side into $p$; hence $\Box_k p \to p$. Löb's \cref{thm:pblt} then gives $p$, for large $k$, whatever $q$ is. In words, a proof of the guard makes the bot defect, which falsifies the antecedent $\neg p$, which makes the guard true, and the opponent never enters.
\end{proof}

\subsubsection{Progress on Some Open Problems}
Apart from our novel proof on \texttt{DupocBot} vs. \texttt{CupodBot} (\cref{sec:dupoc-v-cupod}), we present three other open problems of \citet{Critch} that the library addresses. One is settled, one reduced to bounded-consistency questions, and one answered by construction.

\paragraph{Open Problem 5: $\mathrm{outcome}(\texttt{CUPOD}(k), \texttt{CIMCIC}(k))$.} 
\citet{Critch} rule out $(D,C)$, since \texttt{CupodBot} never exploits, and leaves $(C,C)$, $(C,D)$ and $(D,D)$ open. In $S$ the cell is a floor with two nested hinges (\cref{def:floor}): $G_k$, \enquote{\texttt{CIMCIC}$(k)$ proves within length $k$ that its cooperating implies \texttt{CupodBot}$(k)$'s}, and $U_k$, \enquote{\texttt{CupodBot}$(k)$ proves within length $k$ that \texttt{CIMCIC}$(k)$ defects}. We prove, for all large $k$, that $G_k$ gives $(C,C)$, that $\neg G_k \wedge U_k$ gives $(D,D)$, and that $\neg G_k \wedge \neg U_k$ gives $(C,D)$. These are exactly Critch's three cases, so the open problem is not closed but reduced: a hinge found within $k$ makes $S$ prove its own consistency up to $k - O(\log k)$ within $k + O((\log k)^2)$, so each hinge is at least as hard as bounded consistency.

\paragraph{Open Problem 6: $\mathrm{outcome}(\texttt{DUPOC}(k), \texttt{DIMCID}(k))$.}
Settled, with no hinge: $(D,D)$ for all large $k$. By \cref{thm:Theory-DIMCID-Defects}, \texttt{DIMCID}$(k)$ always defects; \texttt{DupocBot}$(k)$'s guard, \enquote{\texttt{DIMCID}$(k)$ cooperates with me}, is then false in $\mathbb{N}$, so by soundness no proof of it exists at any length and \texttt{DupocBot}$(k)$ defects.

\paragraph{Open Problem 9.} A computationally bounded \texttt{PrudentBot}: \enquote{Cooperate iff I prove within length $k$ both that my opponent cooperates with me and that my opponent defects against \texttt{DefectBot}.} It exists, and its row is in the library (\cref{tab:prudent-row}).

\begin{table}[t]\centering\small
\setlength{\tabcolsep}{3pt}
\caption{\texttt{PrudentBot}$(k)$'s row against every program of the paper in $S$, for all large $k$. The first row gives the negative side of the hinge ($\neg H_k$, the one the matrices of \cref{section:Experiments} use), the second the positive side ($H_k$; blank cells have no hinge). Green: mutual cooperation; gray: mutual defection.}
\label{tab:prudent-row}
\begin{tabular}{@{}lcccccccccccccc@{}}\toprule
 & \texttt{Coop} & \texttt{Cup} & \texttt{DB} & \texttt{Def} & \texttt{Dup} & \texttt{EB} & \texttt{OB} & \texttt{TFT} & \texttt{CupT} & \texttt{DupT} & \texttt{CIM} & \texttt{DID} & \texttt{DIMCID} & \texttt{Prud} \\\midrule
negative side & DC & DC & DC & \cellcolor{black!12}DD & \cellcolor{black!12}DD & DC & \cellcolor{black!12}DD & \cellcolor{black!12}DD & DC & DC & \cellcolor{black!12}DD & DC & \cellcolor{black!12}DD & \cellcolor{black!12}DD \\
positive side &  & \cellcolor{black!12}DD &  &  & \cellcolor{green!18}CC & \cellcolor{green!18}CC &  &  &  &  & \cellcolor{green!18}CC & \cellcolor{black!12}DD &  & \cellcolor{green!18}CC \\\bottomrule
\end{tabular}
\end{table}

\section{The Lean Formalization of $S$}
\label{apx-section:LeanSystemS}

In this appendix, we detail the Lean formalization of $S$ summarized in \cref{subsubsection:Lean-Mech-SystemS} (\cref{apx:Lean-SystemS-Defs}) and record its size and the time it took (\cref{apx:Lean-Timeline}).

\begin{table}[h]\centering\small
\caption{The trusted base: the ten definitions on which every result rests. The file \texttt{Tour.lean} of the supplementary code restates all ten at $\mathbb{N}$, each kept in sync with its definition by the compiler, and walks through the headline theorems; it is the recommended entry point into the code.}
\begin{tabular}{@{}l>{\raggedright\arraybackslash}p{0.46\textwidth}l@{}}\toprule
Definition & What it fixes & Lean \\\midrule
Language and theory & arithmetic with $c_C$, $c_D$ and the action axiom & \texttt{LAct}, \texttt{TAct} \\
Proof length & the symbol count of a derivation & \texttt{proofLen} \\
The box & $\Box_k$, bounded provability as a formula & \texttt{LenBox} \\
Program codes & programs as numbers, by a polynomial pairing & \texttt{progCode} \\
Names of programs & a program named by a short term carrying an action constant & \texttt{codeTerm}, \texttt{descriptions} \\
Names of formulas & a formula inside a box named by a term carrying an action constant & \texttt{boxQuote} \\
Templates & a hypothesis as a formula with open slots & \texttt{template} \\
The guard & the sentence a search consults & \texttt{guardCode} \\
Substitution & simulated programs read the outer players & \texttt{substCode} \\
The evaluator & what a program plays, as a relation in arithmetic & \texttt{EvalGraph} \\\bottomrule
\end{tabular}
\label{table:trusted-base}
\end{table}

\subsection{Proof Length, Encodings and Bounded Provability}\label{apx:Lean-SystemS-Defs}

\paragraph{Proof length.} A proof of $S$ is a derivation in the sequent calculus of the Foundation library \citep{foundation}, a one-sided calculus with cut: a tree in which every node carries a sequent, that is, a finite set of formulas. For a derivation $d$, we define its length $|d|$ as
\[
  |d| \;=\; \sum_{\text{nodes } n \text{ of } d} \bigl(1 + |\Gamma_n|\bigr) \;+\; \sum_{\exists\text{-introductions}} |t| ,
\]
where $|\Gamma_n|$ is the number of symbols in the sequent at node $n$ and $|t|$ the number of symbols in the witness term of an $\exists$-introduction. Two properties are at the core of this design. First, it is symbol-blind: every symbol counts for one, whichever symbol it is. The length of a proof is therefore unchanged by any renaming of symbols, in particular by $\mathcal{T}$; together with the action axiom being stated in both orders, this is what makes \cref{thm:transposition} exact, with no overhead in the budget. Second, it is proper: every node pays for its whole sequent and every witness is paid for, so only finitely many proofs have length at most $k$, with an explicit bound on their codes. Since a node pays for its sequent, a proof is at least as long as the sentence it proves:
\[
  \vdash_k \varphi \quad\Longrightarrow\quad |\varphi| \le k .
\]

\paragraph{Encodings.} $S$ is a theory of arithmetic: its sentences speak about natural numbers and nothing else, since its signature contains only $0$, $1$, $+$, $\cdot$, $<$ and the two action constants. For $S$ to reason about sentences, proofs and programs, each of these objects is assigned a natural number, its G\"odel code, written $\ulcorner\cdot\urcorner$. For example, programs are encoded using the following pairing:
\[
  \ulcorner \texttt{.search}\ k\ \varphi\ p\ q \urcorner \;=\; \langle 6, \langle k, \langle \ulcorner\varphi\urcorner, \langle \ulcorner p\urcorner, \ulcorner q\urcorner\rangle\rangle\rangle\rangle + 1,
  \qquad\text{where}\quad \langle x, y\rangle := (x+y)^2 + y ,
\]
which is injective. The code of a program is then the value of a term built from the numerals of its leaves, and a program of budget $k$ can be named inside a sentence by a short expression in which $k$ appears as a binary numeral, of length $O(\log k)$. This is assumption~(b) of \citet{Critch}.

\paragraph{Bounded provability.} With proofs coded as numbers, \enquote{$d$ is a proof of the sentence coded by $x$} is an arithmetical relation $\mathrm{Proof}_{T_{\mathrm{Act}}}(d, x)$, provided by the Foundation library, and bounded provability is the formula
\[
  \Box_k(x) \;:=\; \exists\, d < B(k).\;\; \mathrm{Proof}_{T_{\mathrm{Act}}}(d, x) \;\wedge\; |d| \le k ,
\]
read \enquote{some proof of the sentence coded by $x$ has length at most $k$}; we write $\Box_k\varphi$ for $\Box_k(\ulcorner\varphi\urcorner)$. The bound $B(k)$ is the bound on the codes of proofs of length at most $k$ that properness provides, so that the quantifier ranges over a finite set and $\Box_k$ is a bounded search. In Lean:
\begin{lstlisting}[style=pdlean, mathescape=true]
def LenBox (k $\varphi$ : V) : Prop := $\exists$ d < codeBound k, Proof T d $\varphi$ $\wedge$ proofLen T d $\leq$ k
\end{lstlisting}

\paragraph{From the proofs to Lean.} Every result of \cref{section:proofs} is mechanized; \cref{table:lean-names} gives the names of the most impactful ones.

\begin{table}[h]\centering\small
\caption{The results of Appendix~\ref{apx-section:TheoreticalResultsOSGT} and their Lean counterparts.}
\begin{tabular}{@{}l>{\raggedright\arraybackslash}p{0.47\linewidth}@{}}\toprule
Result & Lean \\\midrule
Soundness (\cref{prop:soundness}) & \texttt{TAct\_sound\_nat} \\
Completeness over the two readings (\cref{cor:completeness-readings}) & \texttt{TAct\_complete\_of\_readings} \\
Bounded L\"ob (\cref{thm:pblt}) & \texttt{family\_found} \\
Transposition of proofs (\cref{thm:transposition}) & \texttt{lenBoxNat\_swap\_iff} \\
Bridge (\cref{prop:bridge}) & \texttt{lMap\_swap\_sentenceAt\_plays\_of\_transposableP} \\
Transposition of outcomes (\cref{thm:transposition-outcomes}) & \texttt{outcomeAtA\_transpose\_of\_transposableP}\newline \texttt{opposite\_of\_plays\_transpose\_of\_transposableP} \\
\texttt{DupocBot} against \texttt{CupodBot} (\cref{thm:dupoc-cupod}) & \texttt{outcome\_DupocBot\_vs\_CupodBot} \\
\bottomrule
\end{tabular}
\label{table:lean-names}
\end{table}

One important fact about \cref{thm:pblt} is that it rests on $S$ being able to verify its own proofs at bounded cost, and to know that it can: if $\varphi[k]$ has a proof of length at most $\ell$, then $\Box_\ell\varphi[k]$ has a proof of length at most $P(\ell)$, and this implication holds in every model of $S$, uniformly in $k$, so that $S$ proves it. This is the inner necessitation of provability logic, with lengths attached. \citet{Critch} assume $P$ to be linear (Appendix B, assumption (d)); we prove it for $P$ polynomial, of degree 16. The weaker bound suffices because the L\"ob argument applies it only to proofs of length polylogarithmic in the budget $k$, which a polynomial keeps far below $k$. The proof is \texttt{boundedInnerNec\_sixteen}; \cref{apx:Lean-Timeline} gives its share of the development.

\subsection{Development Effort}\label{apx:Lean-Timeline}
This subsection records the size of the formalization of \cref{section:Lean} and the time it took. As stated in \cref{subsection:Lean-Mech}, the authors designed the definitions, statements and architecture, and LLM agents wrote most of the Lean source. 

\paragraph{Size.} The formalization of $S$ has about 159{,}000 lines of Lean and about 11{,}000 theorems and lemmas, with no \texttt{sorry} and no axiom beyond the three standard ones of Lean. However, only a fraction of these lines must be read to trust the results (\cref{table:trusted-base}).

\paragraph{Time.} Development took about two months, with commits on 23 days. \Cref{table:Lean-Timeline} lists the steps in order, with the number of days on which each has commits and the size of the corresponding part of the released source.
\begin{table}[h]
\centering\small
\caption{The steps of the formalization of $S$: days with commits, and lines of Lean in the released source.}
\label{table:Lean-Timeline}
\begin{tabular}{@{}lllp{0.5\linewidth}@{}}
\toprule
Step & Days & Lines & Result \\
\midrule
$S$ and conditional L\"ob & 4 & 23{,}000 & Language, theory and proof length; program codes and evaluator; the diagonal lemma; the bounded L\"ob theorem and \texttt{DupocBot}'s self-cooperation, conditional on bounded inner necessitation \\
Bounded inner necessitation & 5 & 96{,}000 & The hypothesis is proved, so both results become unconditional \\
Outcome matrix & 8 & 20{,}000 & The matrix of the twelve bots of \cref{table:outcome-matrix-full}: 69 of its 78 pairs of bots decided, and the 9 floors stated as dichotomies \\
Partial transparency & 2 & 14{,}000 & The three $\tau$ constructors in programs and evaluator; the $\tau$-rows of the 14 templates of the experiments \\
Reading guide and audit & 5 & 6{,}000 & Reorganization and renaming; the census of axioms; removal of unused code \\
\bottomrule
\end{tabular}
\end{table}

\section{Equilibrium Computation and Selection Dynamics}\label{app:theory-of-experiments}

In this Appendix, we describe our procedures for enumerating and classifying mixed program equilibria and for analyzing the long-run dynamics of the program meta-game. The same methods apply to the full-transparency program meta-game \(G^{\mathrm{prog}}\) of \cref{sec:osgt} and the partial-transparency meta-game \(G^\tau\) of \cref{sec:partial-transp}, with the program library and transparency maps fixed before each computation. All equilibrium claims are relative to the chosen program library.

\subsection{Enumerating Program Equilibria}\label{subsec:nash-eq-enumeration}

Let $G=(U,U^\top)$ be a finite symmetric two-player game, where $U$ denotes the payoff matrix of the row player against the column player; as usual, denote by $A$ the (common) action set. We enumerate its extreme Nash equilibria using \texttt{lrsNash} in exact rational arithmetic \citep{avis2010enumeration}. If the payoff matrix $U$ had negative entries, a common additive shift has all payoffs positive without changing best responses. The algorithm enumerates completely labeled vertex pairs of the associated best-response polytopes and normalizes them to mixed-strategy profiles. Complete labeling expresses the equilibrium condition that each pure strategy is either assigned zero probability or is a best response to the opponent’s mixture. 

The finite list of extreme equilibria describes the entire equilibrium set, including any continua. Specifically, the Nash Equilibrium set of a finite two-players game is a finite union of maximal Nash subsets of product form \citep{jansen1981maximal}:
\[
\mathrm{NE}(G)
=
\bigcup_{k=1}^{K}\mathcal C_k,
\qquad
\mathcal C_k=X_k\times Y_k
\subseteq
\Delta(A)\times\Delta(A),
\]
where \(X_k\) and \(Y_k\) are convex polytopes. Writing
\[
X_k=\operatorname{conv}\{\xi_{k,1},\ldots,\xi_{k,r_k}\},
\qquad
Y_k=\operatorname{conv}\{\eta_{k,1},\ldots,\eta_{k,s_k}\},
\]
the vertices of \(\mathcal C_k\) are precisely the extreme equilibrium profiles \((\xi_{k,i},\eta_{k,j})\). Consequently, every profile
\[
\left(
\sum_{i=1}^{r_k}\lambda_i\xi_{k,i},
\sum_{j=1}^{s_k}\mu_j\eta_{k,j}
\right),
\qquad
\text{for }  \lambda_i, \mu_j \ge 0 ,\text{ and } \sum_i \lambda_i = \sum_j \mu_j = 1,
\]
is a Nash Equilibrium. The convex weights for the two players may be chosen independently.

To assign connected-component labels, we construct a bipartite graph whose two vertex classes are the distinct row and column strategies appearing in the enumeration, with an edge for each extreme equilibrium. Its maximal complete bipartite subgraphs determine the maximal Nash subsets, and its connected components determine the connected components of $\mathrm{NE}(G)$.

We apply this procedure in turn to the full-transparency special case (\cref{sec:osgt}), corresponding to $G^\mathrm{prog}$, and the partial-transparency meta-game (\cref{sec:partial-transp}), corresponding to $G^\tau$. The action sets correspond to the program libraries $\Sigma$.

\subsection{Replicator Dynamics}\label{app:rep-dyn-theory}

For each connected component of \(\mathrm{NE}(G)\) identified in \cref{subsec:nash-eq-enumeration}, we estimate the probability that a population with uniformly drawn initial composition converges to that component. Let $\bm \alpha \in \Delta(A)$ be a mixed action, under the mass action interpretation, and let \(\alpha_i\) denote the share of the population playing action $i \in A$. Under random matching, \((U\bm\alpha)_i\) is the expected payoff to \(i\) and \(\bm\alpha^\top U\bm\alpha\) is the population mean. The replicator dynamics are
\[
\dot{\alpha}_i=\alpha_i \cdot g_i(\bm\alpha),
\qquad
g_i(\bm\alpha)=(U\bm\alpha)_i-\bm\alpha^\top U\bm\alpha .
\]
Thus a program present in the population gains share whenever its payoff exceeds the mean. If an interior trajectory converges to \(\bm\alpha^*\), then \((\bm\alpha^*,\bm\alpha^*)\) is a Program Equilibrium; see \citet{hofbauer2003evolutionary}.

We draw $1{,}000$ initial states independently and uniformly from \(\Delta(A)\) using a \(\mathrm{Dirichlet}(1,\ldots,1)\) distribution. At \(T=10^5\) and \(T/2\), we label action \(i\) as absent if \(\alpha_i<\varepsilon\) and tied with the mean if \(|g_i(\bm\alpha)|\leq\varepsilon\), where \(\varepsilon=10^{-6}\). We classify a trajectory as converged if, at both times, no action exceeds the mean by more than \(\varepsilon\), every action is absent or tied, both label sets agree across times, and \(\|\bm\alpha(T)-\bm\alpha(T/2)\|_\infty<\varepsilon\). A trajectory that fails this test is continued to \(T=10^7\), and then to \(T=10^9\), and tested again, as is one that passes it while a tied action with share below \(10^{-3}\) satisfies \(U_{ii}>(\bm\alpha^\top U)_i\) and is therefore still growing. Using the exact enumeration from \cref{subsec:nash-eq-enumeration}, we retain the extreme strategies that assign zero share to every action labeled absent and against which every action labeled tied is a best response. Their convex hull consists of mixtures inducing symmetric equilibria in a single component. We assign the trajectory to that component after verifying that \(\bm\alpha(T)\) is within \(10^{-4}\) in supremum norm of a point in the hull.

\section{Results on Full Transparency}
This appendix gives the details behind \cref{sec:exper:full-transp}: the outcome matrix of all twelve programs, the equilibria and replicator dynamics of the base library $\Sigma$, and two robustness checks, one on the choice of library and one on the side of the hinge.

\subsection{Outcome Matrix of the Twelve Programs}\label{app:outcome-matrix-full}

\Cref{table:outcome-matrix-full} reports the pairwise outcomes of the twelve programs of \cref{fig:zoo-nesting}, defined in \cref{app:bot-definitions}, with the row program's action listed first. Outcomes hold for sufficiently large shared proof budgets, and floors are shown on the negative side of their hinge. Applying the payoffs of \cref{example:PD-numeric} to these outcomes yields the payoff matrices used in the remainder of this appendix; the matrix of each library is the corresponding submatrix.

\begin{table}[t]\centering\scriptsize
\colorlet{ccfill}{green!18}\colorlet{ddfill}{black!12}\colorlet{hlfill}{yellow!35}\colorlet{hgfill}{orange!40}
\providecommand{\cc}{\cellcolor{ccfill}CC}\providecommand{\dd}{\cellcolor{ddfill}DD}
\providecommand{\cd}{CD}\providecommand{\dc}{DC}\providecommand{\hg}[1]{\cellcolor{hgfill}#1$^{\ddagger}$}
\setlength{\tabcolsep}{3pt}
\caption{The full outcome matrix of all the bots studied in this paper, in $S$, at a shared proof budget $k$; a cell gives the row bot's action first. Cells hold for every sufficiently large $k$ and all sufficiently large fuel. Shaded cells: mutual cooperation (green) and mutual defection (gray). Orange cells$^{\ddagger}$ rest on the hinge of \cref{def:floor}, where we show its negative side. Highlighted names (yellow) are programs that no member of $\Sigma$ exploits (see Proposition~\ref{prop:unexploitability} for more on this).}
\begin{tabular}{@{}lcccccccccccc@{}}\toprule
 & \rotatebox{90}{\texttt{CooperateBot}} & \rotatebox{90}{\colorbox{hlfill}{\texttt{DefectBot}}} & \rotatebox{90}{\texttt{TitForTatBot}} & \rotatebox{90}{\texttt{DBot}} & \rotatebox{90}{\colorbox{hlfill}{\texttt{OBot}}} & \rotatebox{90}{\texttt{EBot}} & \rotatebox{90}{\colorbox{hlfill}{\texttt{DupocBot}}} & \rotatebox{90}{\texttt{CupodBot}} & \rotatebox{90}{\texttt{CupodTrollBot}} & \rotatebox{90}{\texttt{DupocTrollBot}} & \rotatebox{90}{\colorbox{hlfill}{\texttt{CIMCIC}}} & \rotatebox{90}{\texttt{DIMDID}} \\\midrule
\texttt{CooperateBot} & \cc & \cd & \cc & \cd & \cc & \cd & \cc & \cc & \cc & \cc & \cc & \cc \\
\colorbox{hlfill}{\texttt{DefectBot}} & \dc & \dd & \dd & \dc & \dd & \dd & \dd & \dd & \dc & \dc & \dd & \dd \\
\texttt{TitForTatBot} & \cc & \dd & \cc & \dc & \cd & \dc & \cc & \cc & \cc & \cc & \cc & \cc \\
\texttt{DBot} & \dc & \cd & \cd & \dd & \cd & \cd & \hg{CD} & \cc & \dc & \dc & \hg{CD} & \cc \\
\colorbox{hlfill}{\texttt{OBot}} & \cc & \dd & \dc & \dc & \dd & \dc & \dd & \dd & \cc & \cc & \dd & \dd \\
\texttt{EBot} & \dc & \dd & \cd & \dc & \cd & \cc & \hg{CD} & \cc & \dc & \dc & \hg{CD} & \cc \\
\colorbox{hlfill}{\texttt{DupocBot}} & \cc & \dd & \cc & \hg{DC} & \dd & \hg{DC} & \cc & \dc & \hg{DC} & \dd & \cc & \hg{DC} \\
\texttt{CupodBot} & \cc & \dd & \cc & \cc & \dd & \cc & \cd & \dd & \dd & \cc & \hg{CD} & \dd \\
\texttt{CupodTrollBot} & \cc & \cd & \cc & \cd & \cc & \cd & \hg{CD} & \dd & \cc & \cc & \hg{CD} & \cc \\
\texttt{DupocTrollBot} & \cc & \cd & \cc & \cd & \cc & \cd & \dd & \cc & \cc & \cc & \hg{CD} & \cc \\
\colorbox{hlfill}{\texttt{CIMCIC}} & \cc & \dd & \cc & \hg{DC} & \dd & \hg{DC} & \cc & \hg{DC} & \hg{DC} & \hg{DC} & \cc & \dc \\
\texttt{DIMDID} & \cc & \dd & \cc & \cc & \dd & \cc & \hg{CD} & \dd & \cc & \cc & \cd & \dd \\\bottomrule
\end{tabular}
\label{table:outcome-matrix-full}
\end{table}

\subsection{An Enumeration of Program Equilibria}\label{app:results-nash}

Each row of \cref{tab:extreme-equilibria-body8-floor} gives a strategy for either player, with probabilities listed in the same order as the bots. Any two rows in the same group, one for each player, give an extreme equilibrium with the probability of mutual cooperation shown in the group heading. For example, the four rows in the full-cooperation group give $4\times 4=16$ extreme equilibria, including the four in which both players choose the same row.

\begin{table*}[t]
\centering
\caption{All 21 extreme Nash equilibria of the restricted 8-bot program meta-game at full transparency (negative side of the hinge).}
\label{tab:extreme-equilibria-body8-floor}
\small
\begin{tabular}{@{}ll@{}}
\toprule
Strategy support (either player) & Probabilities in support order \\
\midrule
\textit{Full cooperation} (16 extreme equilibria) & \textit{Pr[(C, C)]\,=\,1} \\[2pt]
\texttt{DupocBot} & $1$ \\
\texttt{CooperateBot}, \texttt{DupocBot} & $2/3,\;1/3$ \\
\texttt{DupocBot}, \texttt{TitForTatBot} & $1/3,\;2/3$ \\
\texttt{CooperateBot}, \texttt{DupocBot}, \texttt{TitForTatBot} & $2/3,\;1/9,\;2/9$ \\
\midrule
\textit{Partial cooperation} (1 equilibrium) & \textit{Pr[(C, C)]\,=\,11/18} \\[2pt]
\texttt{CooperateBot}, \texttt{OBot}, \texttt{EBot} & $1/2,\;1/3,\;1/6$ \\
\midrule
\textit{Full defection} (4 extreme equilibria) & \textit{Pr[(C, C)]\,=\,0} \\[2pt]
\texttt{DefectBot} & $1$ \\
\texttt{DefectBot}, \texttt{OBot} & $2/3,\;1/3$ \\
\bottomrule
\end{tabular}
\end{table*}

In the underlying Prisoner's Dilemma, the unique Nash Equilibrium is $(D,D)$. In $G^{\mathrm{prog}}$, the three components of $\mathrm{NE}(G^{\mathrm{prog}})$ (the set of program equilibria) yield mutual cooperation with probabilities $1$, $11/18$, and $0$, respectively, in the order shown in \cref{tab:extreme-equilibria-body8-floor}. Thus full cooperation, with payoff $R=2$ for each player, holds throughout an entire component, while the intermediate rate occurs at a single symmetric equilibrium.

For the chosen library $\Sigma$, \texttt{DupocBot} is the only program that both players use in every fully cooperative equilibrium. To see this, write the fully cooperative component as $\mathcal{C}_1=X_1\times Y_1$. Each vertex of $X_1$ or $Y_1$ assigns probability at least $1/9$ to \texttt{DupocBot}, so the same bound holds for both players at every $(\xi,\eta)\in\mathcal{C}_1$ by convexity. No other program is required, since both players may use \texttt{DupocBot} alone.

\subsection{Replicator Dynamics at Full Transparency}
\label{app:full-transparency-replicator}

For the eight-program library $\Sigma$ at full transparency (negative side of the hinge), we drew $1{,}000$ initial population states uniformly from $\Delta(\Sigma)$. All 1,000 trajectories met the convergence criterion of \cref{app:rep-dyn-theory}, and all were assigned to the fully cooperative component $\mathcal{C}_1$ of \cref{tab:extreme-equilibria-body8-floor}. Its estimated basin share is therefore $1$, with a 95\% Wilson interval of $[0.9962,1]$. No trajectory was assigned to either the isolated equilibrium or the full-defection component. For each of these components, the estimated basin share is $0$, with a 95\% Wilson interval of $[0,0.0038]$. \Cref{tab:full-transparency-replicator} shows how the final population composition varied within $\mathcal{C}_1$. \texttt{DupocBot} was present in every run. Its final share ranged from $0.1524$ to $1$, compared with the lower bound of $1/9$ over the entire component derived in \cref{app:results-nash}. In $955$ runs, both \texttt{CooperateBot} and \texttt{TitForTatBot} were also present. Despite this variation, every final population yielded mutual cooperation with probability $1$ and mean payoff $R=2$. Thus, for this library at
full transparency, mutual cooperation is the predominant outcome of rational play when equilibrium selection follows the replicator dynamics from a uniformly drawn initial population.

\begin{table}[t]
  \centering
  \caption{Final population composition in the eight-program
    full-transparency game (negative side of the hinge). A program is counted as
    present if its share at time $T$ is at least $\varepsilon=10^{-6}$.}
  \label{tab:full-transparency-replicator}
  \begin{tabular}{lr}
    \hline
    Programs present at time $T$           & Number of runs \\
    \hline
    CooperateBot, DupocBot, TitForTatBot   & 955 \\
    DupocBot, TitForTatBot                 & 30  \\
    DupocBot                               & 8  \\
    CooperateBot, DupocBot                 & 7  \\
    \hline
  \end{tabular}
\end{table}

\subsection{Sensitivity to the Program Library}
\label{app:zoo-ladder}

We repeat the full-transparency analysis of \cref{app:theory-of-experiments} for four nested program libraries (see \cref{fig:zoo-nesting}), using the negative side of the hinge. Starting from the base library $\Sigma$, we remove \texttt{CupodBot} to obtain the seven-program library, or add \texttt{CupodTrollBot} and \texttt{DupocTrollBot} to obtain the ten-program
library. Each troll cooperates with every program except the search program it is named after, so the ten-program library adversarially tests \texttt{DupocBot}'s robustness against a targeted defector. Adding the implication provers \texttt{CIMCIC} and \texttt{DIMDID} gives the twelve-program library. It tests whether \texttt{DupocBot}'s role is specific to it or shared with a second proof-based cooperator: \texttt{CIMCIC} takes the same action as \texttt{DupocBot} against every program of $\Sigma$, but \texttt{DupocTrollBot} does not defect against it; \texttt{DIMDID} is its transpose, as \texttt{CupodBot} is \texttt{DupocBot}'s (\cref{sec:dupoc-v-cupod}).

\begin{table}[t]
  \centering
  \caption{Equilibrium outcomes across program libraries at full transparency.
    The cooperative support lists the programs used in at least one fully
    cooperative equilibrium, with
    $\mathcal{C}_1=\{\texttt{CooperateBot},\texttt{DupocBot},\texttt{TitForTatBot}\}$.
    The last column gives the fraction of $1{,}000$ uniformly drawn initial
    populations assigned to the fully cooperative component.}
  \label{tab:zoo-ladder-full-transparency}
  \small
  \begin{tabular}{@{}lcccc@{}}
    \hline
    & \multicolumn{3}{c}{Nash equilibria} & Replicator dynamics \\
    \cline{2-4}\cline{5-5}
    Library & Components & Extreme equilibria & Cooperative support
            & Cooperative share \\
    \hline
    $\Sigma \setminus \{\texttt{CupodBot}\}$
      & 3 & 21 & $\mathcal{C}_1$                          & 100\%  \\
    $\Sigma$
      & 3 & 21 & $\mathcal{C}_1$                          & 100\%  \\
    $\Sigma$ + trolls
      & 3 & 35 & $\mathcal{C}_1$                          & 97.5\% \\
    $\Sigma$ + trolls + provers
      & 3 & 69 & $\mathcal{C}_1 \cup \{\texttt{CIMCIC}\}$ & 100\%  \\
    \hline
  \end{tabular}
\end{table}

In all four libraries, the three equilibrium components yield mutual cooperation with probabilities $1$, $11/18$, and $0$. The choice of library matters more for equilibrium selection. Every sampled population reached full cooperation in the seven-, eight-, and twelve-program libraries (\cref{tab:zoo-ladder-full-transparency}). Adding only the trolls left the support of the fully cooperative component unchanged, yet a small number of populations spent long periods near full defection ($23$ still at full defection at $T = 10^9$). \texttt{DupocTrollBot} gives \texttt{DefectBot} a payoff advantage over \texttt{DupocBot}. It cooperates even against the unconditional defection of \texttt{DefectBot}, which earns $3$ in that encounter, but defects against \texttt{DupocBot}, which earns $0$. Once \texttt{DupocBot} is rare in a population otherwise composed of \texttt{DefectBot} and \texttt{OBot}, it earns a positive payoff only when it meets another copy of itself. With \texttt{CIMCIC} added, every population again reaches full cooperation, and \texttt{CIMCIC} survives alongside \texttt{DupocBot} in all of them.

\subsection{Sensitivity to the Hinge}\label{app:full-transparency-hinge}

In $\Sigma$, as \cref{table:outcome-matrix-body} shows, two pairs of outcomes depend on a hinge (\cref{def:floor}), and all results above use its negative side. Since both pairs involve \texttt{DupocBot}, we repeat the analysis on the positive side, where both pairs yield mutual cooperation instead (\cref{fig:hinge-sensitivity}).

In this regime, the program equilibria form five components instead of three. All three original components still exist, and two further isolated equilibria appear, with $\Pr[(C,C)]=14/27$ and $43/147$. Under the replicator dynamics, $959$ of $1{,}000$ populations converge to the fully cooperative component (95\% Wilson interval $[0.945,0.970]$). The remaining $41$ are still at full defection at $T=10^9$, a state that \texttt{DupocBot} invades only very slowly, as in the ten-program library of \cref{app:zoo-ladder}. No population converges to a partially cooperative equilibrium. The conclusions of \cref{sec:exper:full-transp} therefore do not depend on the side of the hinge.

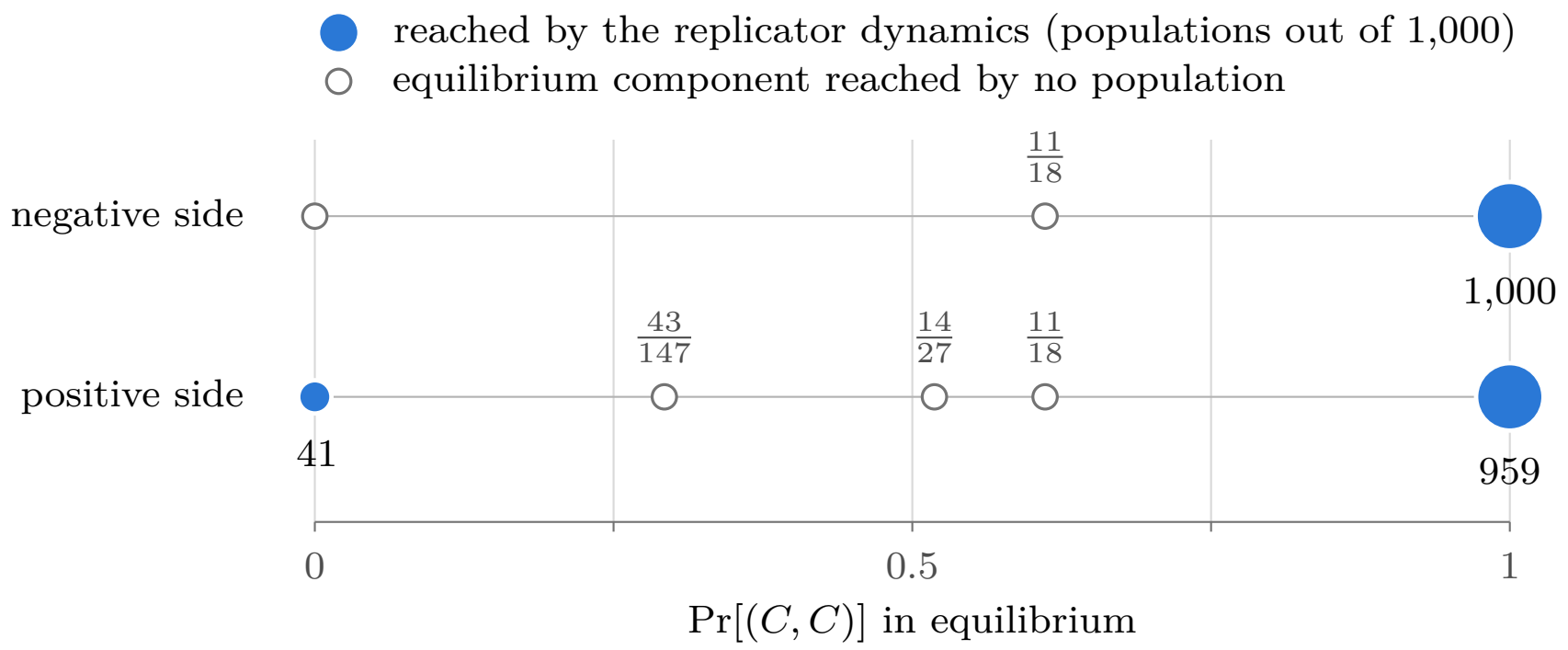
\begin{figure}[t]
\centering
\definecolor{hingeReached}{HTML}{2A78D6}
\begin{tikzpicture}[x=8.6cm, y=1cm, font=\footnotesize,
  grid/.style={draw=black!15, line width=0.4pt},
  rowline/.style={draw=black!30, line width=0.4pt},
  eq/.style={circle, draw=black!55, fill=white, line width=0.6pt, inner sep=0pt, minimum size=5pt},
  reached/.style={circle, draw=white, fill=hingeReached, line width=0.9pt, inner sep=0pt, minimum size=8.5pt},
  val/.style={above=4pt, text=black!70, inner sep=0pt},
  cnt/.style={below=5pt, text=black, inner sep=0pt},
]
\foreach \x in {0, 0.25, 0.5, 0.75, 1} \draw[grid] (\x, -0.9) -- (\x, 1.85);
\draw[rowline] (0, 1.3) -- (1, 1.3);
\draw[rowline] (0, 0) -- (1, 0);
\node[anchor=east] at (-0.045, 1.3) {negative side};
\node[anchor=east] at (-0.045, 0) {positive side};
\node[eq] at (0, 1.3) {};
\node[eq, label={[val]above:$\tfrac{11}{18}$}] at (0.6111, 1.3) {};
\node[reached, minimum size=14pt, label={[cnt]below:$1{,}000$}] at (1, 1.3) {};
\node[reached, minimum size=6.8pt, label={[cnt, anchor=north west, xshift=-4pt]below:$41$}] at (0, 0) {};
\node[eq, label={[val]above:$\tfrac{43}{147}$}] at (0.2925, 0) {};
\node[eq, label={[val]above:$\tfrac{14}{27}$}] at (0.5185, 0) {};
\node[eq, label={[val]above:$\tfrac{11}{18}$}] at (0.6111, 0) {};
\node[reached, minimum size=13.8pt, label={[cnt]below:$959$}] at (1, 0) {};
\draw[draw=black!55, line width=0.4pt] (0, -0.9) -- (1, -0.9);
\foreach \x/\l in {0/0, 0.25/, 0.5/0.5, 0.75/, 1/1} {
  \draw[draw=black!55, line width=0.4pt] (\x, -0.9) -- (\x, -0.97);
  \node[anchor=north, text=black!70] at (\x, -1.0) {\l};
}
\node[anchor=north] at (0.5, -1.36) {$\Pr[(C,C)]$ in equilibrium};
\node[reached] (l1) at (0.02, 2.62) {};
\node[anchor=west] at ([xshift=3pt]l1.east) {reached by the replicator dynamics (populations out of $1{,}000$)};
\node[eq] (l2) at (0.02, 2.27) {};
\node[anchor=west] at ([xshift=4.7pt]l2.east) {equilibrium component reached by no population};
\end{tikzpicture}
\caption{Sensitivity to the hinge for the library $\Sigma$ at full transparency. Each marker is a component of program equilibria, placed at its probability of mutual cooperation; filled markers are the components that sampled populations converge to under the replicator dynamics. On the positive side, two further partially cooperative equilibria appear, and $41$ populations are still at full defection at $T=10^9$.}
\label{fig:hinge-sensitivity}
\end{figure}

\section{Experiments and Results on Partial Transparency}

Throughout this Appendix we consider the setup given in \cref{subsec:exp.prison:dil:tau}, where a base program \texttt{Bot} in $G^\mathrm{prog}_{\mathrm{PD}}$ is extended to a family of \texttt{TauBot}$(\alpha)$ in $G^\tau_{\mathrm{PD}}$, where $\alpha \in [0,1]$ is the linear threshold parameter and $\tau$ is a transparency map, common to both players. 

\subsection{Experimental Setup}\label{app:transparency:expr:setup}
The setup has two ingredients: a family of transparency maps indexed by a temperature (\cref{apx:TransExp-Setup-Maps}), and a common scale on which maps from different families can be compared (\cref{apx:TransExp-Setup-Levels}).

\subsubsection{Transparency Maps}\label{apx:TransExp-Setup-Maps}
We build transparency maps from a notion of distance between programs: the belief a program receives about its opponent concentrates on the programs closest to the one actually submitted.

Fix a finite program library $\Sigma = \{\pi_1, \ldots, \pi_N\}$ of admissible programs and assume that $\mathrm{outcome}(\pi_j,\pi_\ell)$ is known for every ordered pair. We consider softmax-based transparency maps induced by a pseudo-distance on $\Sigma$.

\begin{definition}[Softmax-based transparency maps]\label{def:tau-softmax-Family}
Let $d : \Sigma \times \Sigma \to \mathbb{R}_{\ge 0}$ be a pseudo-distance.\footnote{A pseudo-distance is a nonnegative, symmetric distance function that is zero from any point to itself and satisfies the triangle inequality.} For a temperature $T > 0$, the transparency map $\tau^{T}_{d}\colon \Sigma \to \Delta(\Sigma)$ is the softmax of $-d/T$:
\[
  \tau^{T}_{d}(\pi_j)(\pi_\ell)
  := \frac{\exp\bigl(-d(\pi_j,\pi_\ell)/T\bigr)}
          {\sum_{m=1}^{N} \exp\bigl(-d(\pi_j,\pi_m)/T\bigr)},
  \qquad \pi_j, \pi_\ell \in \Sigma.
\]
As is usual, we extend to $T = \infty$ by $\tau^{\infty}_{d}(\pi_j) = \mathrm{unif}_{\Sigma}$ and to $T= 0$ by $\tau^{0}_{d}(\pi_j) = \delta_{\pi_j}$, the full transparency of \cref{sec:osgt}.
\end{definition}

Throughout the experiments, we use two different pseudo-distances that yield the behavioral and syntactic families:

\paragraph{Behavioral Family.}

For each program $\pi \in \Sigma=\{\pi_1, \dots, \pi_N\}$, define its outcome row as
\[
  \mathrm{row}(\pi)
  := \bigl(\mathrm{outcome}(\pi,\pi_m)_1\bigr)_{m=1}^{N} \in \{C, D\}^N,
\]
so that the $m$th entry records the action that $\pi$ takes against $\pi_m$. We define the behavioral pseudo-distance as the Hamming distance between two outcome rows, that is, the number of opponents against which the two programs take different actions:
\[
  d_{\mathrm{bvr}}(\pi_j,\pi_\ell)
  := \bigl|\{m \in \{1,\ldots,N\} :
      \mathrm{row}(\pi_j)_m \neq \mathrm{row}(\pi_\ell)_m\}\bigr|
  \in \{0,\ldots,N\}.
\]
Distinct programs can have identical outcome rows and hence have distance zero. For each $T \in [0,+\infty]$, we write $\tau_{\mathrm{bvr}}^T$ for the transparency map induced by $d_{\mathrm{bvr}}$ according to \cref{def:tau-softmax-Family}.

\paragraph{Syntactic Family.}

For each program $\pi \in \Sigma$, let $\mathrm{ast}(\pi)$ denote the abstract syntax tree of its Lean source code, with nodes labeled by the corresponding \texttt{Prog} or \texttt{Formula} constructors, the action constants $C$ and $D$, and the names of referenced programs; budget arguments are omitted. For a tree $t$, let $|t|$ denote its number of nodes. We define the syntactic pseudo-distance as the Zhang--Shasha tree edit distance \citep{zhang1989simple} between two such trees, normalized by the number of nodes in the larger tree:
\[
  d_{\mathrm{syn}}(\pi_j,\pi_\ell)
  := \frac{
    \mathrm{ted}\bigl(\mathrm{ast}(\pi_j),\mathrm{ast}(\pi_\ell)\bigr)
  }{
    \max\bigl\{|\mathrm{ast}(\pi_j)|,|\mathrm{ast}(\pi_\ell)|\bigr\}
  },
\]
where $\mathrm{ted}$ is the minimum number of node insertions, deletions, and relabelings needed to transform one tree into the other, with each operation assigned unit cost. Programs that differ only in their budget arguments have identical trees and hence have distance zero. For each $T \in [0,+\infty]$, we write $\tau_{\mathrm{syn}}^T$ for the transparency map induced by $d_{\mathrm{syn}}$ according to \cref{def:tau-softmax-Family}.

As shown in \cref{sec:more-less-transparency}, the setup is flexible enough to cover the full spectrum of transparency, from fully transparent OSGT to the base game. In particular, the behavioral family compares the actions that programs take against each opponent in the library, whereas the syntactic family compares the abstract syntax trees of their Lean source code.

\subsubsection{Transparency Levels}
\label{apx:TransExp-Setup-Levels}
The same temperature $T$ can yield different degrees of transparency in the behavioral and syntactic families because the two pseudo-distances measure different features of a program. To compare the families on a common scale, we measure the fraction of uncertainty about the opponent's identity that a signal drawn from the transparency map removes on average. Formally, let $N=|\Sigma|\geq 2$, draw the opponent's program $B$ uniformly from $\Sigma$, and, conditional on $B=\pi$, draw a signal $S$ from $\tau(\pi)$. We define the transparency level of $\tau$ as
\[
  t(\tau)
  := \frac{I(B;S)}{H(B)}
   = \frac{H(B)-H(B\mid S)}{H(B)}
   = 1-\frac{H(B\mid S)}{\log_2 N}
   \in [0,1],
\]
where $H(B)=\log_2 N$ is the initial uncertainty about the opponent's identity,\footnote{The uniform prior reflects a principle of indifference: in the absence of information favoring one program over another, we assign equal probability to every program in $\Sigma$. This assumption may be inappropriate when some programs are known to occur more frequently than others. In that case, the same ratio $I(B;S)/H(B)$ can be evaluated under a prior that reflects those frequencies.} measured in bits, and $H(B\mid S)$ is the uncertainty that remains after observing the signal, averaged over possible signals. Thus, $t(\tau)=0$ when the signal is independent of the opponent's identity, and $t(\tau)=1$ when the signal determines that identity. We compare the two families by choosing a temperature for each that yields the same transparency level. At a matched level, the signals remove the same fraction of uncertainty on average, although they may distinguish different pairs of programs.

\subsection{Equilibrium under Partial Transparency}
\label{app:equilibrium-partial-transparency}

\Cref{tab:components} reports the program-equilibrium components of the eight-program library $\Sigma$ under behavioral transparency, with floor cells evaluated on the negative side of their hinge as in \cref{sec:exper:full-transp}. Every equilibrium in a component yields the same probability of mutual cooperation, $\Pr[(C,C)]$. Reading across a row shows which outcomes become sustainable as transparency increases at fixed caution. Mutual defection remains an equilibrium throughout.

\begin{table}[!htbp]
  \centering
  \caption{Equilibrium outcomes under behavioral transparency. \textsf{DEF} (white) denotes mutual defection alone. \textsf{BOTH} (orange) denotes mutual defection and full cooperation. \textsf{ALL} (yellow) also includes partial cooperation with $\Pr[(C,C)]=11/18$. The body shows every tenth grid point. The last two columns give the first appearances of full cooperation ($t^*$) and of the partially cooperative equilibrium with $\Pr[(C,C)]=11/18$ ($t^{**}$) on the full grid with spacing $0.01$.}
  \label{tab:components}
  \small
  \setlength{\tabcolsep}{2pt}
  \renewcommand{\arraystretch}{1.15}
  \definecolor{fullcoop}{RGB}{247,203,160}
  \definecolor{partialcoop}{RGB}{255,246,176}
  \newcommand{\eqDef}{\textsf{DEF}}
  \newcommand{\eqBoth}{\cellcolor{fullcoop}\textsf{BOTH}}
  \newcommand{\eqAll}{\cellcolor{partialcoop}\textsf{ALL}}
  \begin{tabular}{@{}r*{11}{c}@{\hspace{0.8em}}cc@{}}
    \toprule
    & \multicolumn{11}{c}{Transparency $t$}
    & \multicolumn{2}{c}{Thresholds} \\
    \cmidrule(lr){2-12}\cmidrule(l){13-14}
    $\alpha$ & 0 & 0.1 & 0.2 & 0.3 & 0.4 & 0.5 & 0.6 & 0.7 & 0.8 & 0.9 & 1
      & $t^*$ & $t^{**}$ \\
    \midrule
    0.50 & \eqDef & \eqDef & \eqBoth & \eqAll & \eqAll & \eqAll & \eqAll & \eqAll & \eqAll & \eqAll & \eqAll & 0.14 & 0.26 \\
    0.60 & \eqDef & \eqDef & \eqDef & \eqBoth & \eqAll & \eqAll & \eqAll & \eqAll & \eqAll & \eqAll & \eqAll & 0.29 & 0.35 \\
    0.70 & \eqDef & \eqDef & \eqDef & \eqDef & \eqDef & \eqAll & \eqAll & \eqAll & \eqAll & \eqAll & \eqAll & 0.44 & 0.46 \\
    0.80 & \eqDef & \eqDef & \eqDef & \eqDef & \eqDef & \eqBoth & \eqBoth & \eqBoth & \eqAll & \eqAll & \eqAll & 0.48 & 0.71 \\
    0.90 & \eqDef & \eqDef & \eqDef & \eqDef & \eqDef & \eqDef & \eqDef & \eqBoth & \eqBoth & \eqAll & \eqAll & 0.66 & 0.88 \\
    1.00 & \eqDef & \eqDef & \eqDef & \eqDef & \eqDef & \eqDef & \eqDef & \eqDef & \eqDef & \eqDef & \eqAll & 1.00 & 1.00 \\

    \bottomrule
  \end{tabular}
\end{table}

At $\alpha=0.8$, for example, full cooperation first appears at $t=0.48$. The partially cooperative component appears later, at $t=0.71$, and coexists with both mutual defection and full cooperation thereafter. At $\alpha=1$, the two thresholds coincide. Both cooperative outcomes appear only at full transparency.

\Cref{fig:transparency-selection} shows this selection in the behavioral family. Mutual defection remains an equilibrium at every point shown, yet at and above the threshold all initial populations attain the maximum equilibrium cooperation. Each dot marks the threshold of one caution level, where the curves rise by $1/6$. The small gaps between the two curves at low transparency are the three windows just mentioned.

\begin{figure}[htbp]
  \centering
  \includegraphics[width=0.54\linewidth]{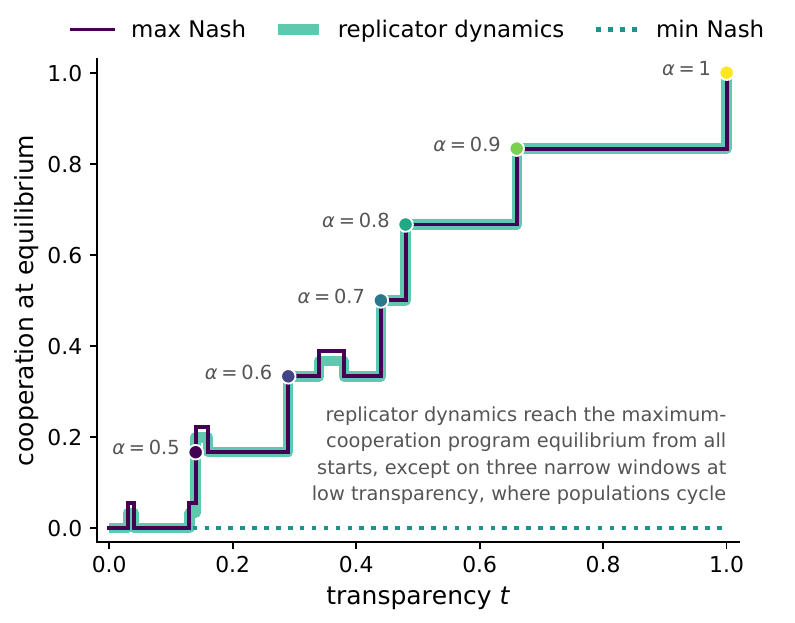}
    \caption{Equilibrium cooperation and selection under behavioral transparency. The thin solid and dotted lines show the maximum and minimum cooperation $\Pr[(C,C)]$ at equilibrium, averaged over the six caution levels $\{0.5, 0.6, 0.7, 0.8, 0.9, 1 \}$ at each $t$. The thick trace shows the outcome of replicator dynamics, the mean of $\Pr[(C,C)]$ over the sampled populations, taken as a time average where they cycle. Dots mark the threshold $t^*$ of each caution level, with the colors of \cref{fig:transparency-thresholds}. The thick trace coincides with the maximum except on three narrow windows at low transparency.}
  \label{fig:transparency-selection}
  \Description{Maximum and selected cooperation rise together in six steps of one sixth from zero to one, at transparency 0.14, 0.29, 0.44, 0.48, 0.66, and 1. Below 0.4 there are three small bumps where the maximum lies slightly above the selected value. Minimum cooperation remains zero.}
\end{figure}

Whenever full cooperation is an equilibrium outcome, all simulated populations converge under replicator dynamics to a fully cooperative equilibrium. This holds even in the \textsf{ALL} cells, where three equilibrium outcomes coexist. Below $t^*$, populations converge to mutual defection, with three exceptions that fall between the grid points of \cref{tab:components}: at $t=0.03$ for $\alpha=0.7$, for $t\in[0.13,0.15]$ at $\alpha=0.8$, and for $t\in[0.34,0.37]$ at $\alpha=0.9$. There, two further equilibria exist, with $\Pr[(C,C)]=1/6$ and $1/3$, and no population settles: the populations cycle among \texttt{CupodBot}, \texttt{DupocBot}, \texttt{EBot}, and \texttt{TitForTatBot}, with a long-run average of $\Pr[(C,C)]$ of about $0.2$. \Cref{fig:transparency-selection} shows that the selected probability of mutual cooperation otherwise coincides with the maximum across program equilibria, and \cref{fig:transparency-thresholds}(a) shows the switch at each caution level.

\subsection{Sensitivity to the Transparency Map and the Hinge}
\label{app:partial-transparency-hinge}

In the syntactic family, the analysis shows the same results as the behavioral family, without the partially cooperative windows where the populations cycle: at each caution level, the same fully cooperative equilibrium first appears at a critical transparency $t^*$ and remains available at every higher level, while every sampled population converges to mutual defection below $t^*$. Only the thresholds differ: $t^*=0.07$, $0.20$, $0.37$, $0.57$, $0.79$, and $1$ for $\alpha=0.5,\dots,1$, against $0.14$, $0.29$, $0.44$, $0.48$, $0.66$, and $1$ in the behavioral family (\cref{fig:transparency-thresholds}(b)). Both sequences increase with caution.

All results on partial transparency are for the negative side of the hinge (\cref{def:floor}). On the positive side, full cooperation is still sustained and selected at high transparency, but not through a single threshold: at the higher caution levels, cooperative equilibria also appear on intervals of low transparency, separated from the upper region by intervals with no fully cooperative equilibrium. We therefore do not claim the threshold structure on that side.

\section{Definitions of the Programs}
\label{app:bot-definitions}

This appendix defines the twelve programs in
\cref{table:outcome-matrix-full}; constant programs, simulation-based programs, and proof-based programs, followed by the two trolls and the two implication-based programs. Throughout, $\pi$ denotes the opponent's program, and $\mathrm{run}(\pi \| \pi')$ denotes the action returned by $\pi$ when given $\pi'$ as its opponent. 

For proof-based programs, $k\in\mathbb{N}$ is the proof-length bound, and \texttt{proof\_search}$(k,\pi,\varphi)$ succeeds exactly when $S$ has a proof of $\varphi$ of length at most $k$. All proof-based programs in a matchup use the same
$k$, as in \cref{table:outcome-matrix-full}.

\subsection{Constant Programs}

\paragraph{CooperateBot.}
\texttt{CooperateBot} always cooperates, regardless of its opponent's program. It also serves as the unconditional cooperation probe used by the simulation-based programs.

\begingroup
\upshape
\begin{lstlisting}[style=pdcode, mathescape=true]
def CooperateBot$(\pi)$:
    return C
\end{lstlisting}
\endgroup

\paragraph{DefectBot.}
\texttt{DefectBot} always defects, regardless of its opponent's program. It also serves as the unconditional defection probe used by the simulation-based programs.

\begingroup
\upshape
\begin{lstlisting}[style=pdcode, mathescape=true]
def DefectBot$(\pi)$:
    return D
\end{lstlisting}
\endgroup

\subsection{Simulation-Based Programs}

\paragraph{TitForTatBot.}
\texttt{TitForTatBot} simulates its opponent against \texttt{CooperateBot} and returns the same action. Thus it responds to the opponent's treatment of an unconditional cooperator, without requiring a history of repeated play.

\begingroup
\upshape
\begin{lstlisting}[style=pdcode, mathescape=true]
def TitForTatBot$(\pi)$:
    if $\mathrm{run}(\pi \| \mathrm{CooperateBot}) = C$:
        return C
    else:
        return D
\end{lstlisting}
\endgroup

\paragraph{DBot.}
\texttt{DBot} simulates its opponent against \texttt{DefectBot} and returns the opposite action. It defects against opponents that cooperate with an unconditional defector, and cooperates with those that defect against it.

\begingroup
\upshape
\begin{lstlisting}[style=pdcode, mathescape=true]
def DBot$(\pi)$:
    if $\mathrm{run}(\pi \| \mathrm{DefectBot}) = C$:
        return D
    else:
        return C
\end{lstlisting}
\endgroup

\paragraph{OBot.}
\texttt{OBot} cooperates only if its opponent cooperates with both \texttt{CooperateBot} and \texttt{DefectBot}. It tests these responses in that order, defecting as soon as either probe returns $D$.

\begingroup
\upshape
\begin{lstlisting}[style=pdcode, mathescape=true]
def OBot$(\pi)$:
    if $\mathrm{run}(\pi \| \mathrm{CooperateBot}) = C$:
        if $\mathrm{run}(\pi \| \mathrm{DefectBot}) = C$:
            return C
        else:
            return D
    else:
        return D
\end{lstlisting}
\endgroup

\paragraph{EBot.}
\texttt{EBot} first defects if its opponent cooperates with \texttt{DefectBot}. Otherwise, it cooperates if the opponent cooperates with \texttt{CooperateBot} or, failing that, with \texttt{MirrorBot}; if neither probe returns $C$, it defects.

Here \texttt{MirrorBot} is an auxiliary probe that simulates its own opponent against \texttt{MirrorBot} and returns the resulting action. It is used within \texttt{EBot}'s definition and is not an additional member of the twelve-program library.

\begingroup
\upshape
\begin{lstlisting}[style=pdcode, mathescape=true]
def MirrorBot$(\pi')$:
    return $\mathrm{run}(\pi' \| \mathrm{MirrorBot})$

def EBot$(\pi)$:
    if $\mathrm{run}(\pi \| \mathrm{DefectBot}) = C$:
        return D
    elif $\mathrm{run}(\pi \| \mathrm{CooperateBot}) = C$:
        return C
    elif $\mathrm{run}(\pi \| \mathrm{MirrorBot}) = C$:
        return C
    else:
        return D
\end{lstlisting}
\endgroup

\subsection{Proof-Based Programs}

\paragraph{DupocBot.}
\texttt{DupocBot} (``Defect Unless Proof of Cooperation'') searches for a proof that its opponent cooperates against it. It cooperates if it finds such a proof within length $k$, and defects otherwise.

\begingroup
\upshape
\begin{lstlisting}[style=pdcode, mathescape=true]
def DupocBot$(k)(\pi)$:
    if proof_search($k, \pi, \mathrm{run}(\pi \| \mathrm{DupocBot}(k)) = C$):
        return C
    else:
        return D
\end{lstlisting}
\endgroup

\paragraph{CupodBot.}
\texttt{CupodBot} (``Cooperate Unless Proof of Defection'') searches for a proof that its opponent defects against it. It defects if it finds such a proof within length $k$, and cooperates otherwise.

\begingroup
\upshape
\begin{lstlisting}[style=pdcode, mathescape=true]
def CupodBot$(k)(\pi)$:
    if proof_search($k, \pi, \mathrm{run}(\pi \| \mathrm{CupodBot}(k)) = D$):
        return D
    else:
        return C
\end{lstlisting}
\endgroup

\subsection{Troll Programs}

The equalities in the following proof searches express syntactic identity of programs, rather than behavioral equivalence.

\paragraph{CupodTrollBot.}
\texttt{CupodTrollBot} searches for a proof that its opponent is exactly \texttt{CupodBot}$(k)$. It defects if it finds such a proof within length $k$, and
cooperates otherwise.

\begingroup
\upshape
\begin{lstlisting}[style=pdcode, mathescape=true]
def CupodTrollBot$(k)(\pi)$:
    if proof_search($k, \pi, \pi = \mathrm{CupodBot}(k)$):
        return D
    else:
        return C
\end{lstlisting}
\endgroup

\paragraph{DupocTrollBot.}
\texttt{DupocTrollBot} searches for a proof that its opponent is exactly \texttt{DupocBot}$(k)$. It defects if it finds such a proof within length $k$, and cooperates otherwise.

\begingroup
\upshape
\begin{lstlisting}[style=pdcode, mathescape=true]
def DupocTrollBot$(k)(\pi)$:
    if proof_search($k, \pi, \pi = \mathrm{DupocBot}(k)$):
        return D
    else:
        return C
\end{lstlisting}
\endgroup

\subsection{Implication-Based Programs}

\paragraph{CIMCIC.}
\texttt{CIMCIC} (``Cooperate If My Cooperation Implies Cooperation'') searches for a proof that its own cooperation implies its opponent's cooperation in the current encounter. It cooperates if it finds such a proof within length $k$, and defects otherwise.

\begingroup
\upshape
\begin{lstlisting}[style=pdcode, mathescape=true]
def CIMCIC$(k)(\pi)$:
    $\varphi$ = $\smash{\big(}\mathrm{run}(\mathrm{CIMCIC}(k) \| \pi) = C \Longrightarrow \mathrm{run}(\pi \| \mathrm{CIMCIC}(k)) = C\smash{\big)}$
    if proof_search($k, \pi, \varphi$):
        return C
    else:
        return D
\end{lstlisting}
\endgroup

\paragraph{DIMDID.}
\texttt{DIMDID} (``Defect If My Defection Implies Defection'') searches for a proof that its own defection implies its opponent's defection in the current encounter. It defects if it finds such a proof within length $k$, and cooperates otherwise.

\begingroup
\upshape
\begin{lstlisting}[style=pdcode, mathescape=true]
def DIMDID$(k)(\pi)$:
    $\varphi$ = $\smash{\big(}\mathrm{run}(\mathrm{DIMDID}(k) \| \pi) = D \Longrightarrow \mathrm{run}(\pi \| \mathrm{DIMDID}(k)) = D\smash{\big)}$
    if proof_search($k, \pi, \varphi$):
        return D
    else:
        return C
\end{lstlisting}
\endgroup

\section{Definitions of the Tau Programs}
\label{app:tau-bot-definitions}

The eight tau programs are obtained by applying a common construction to the definitions in \cref{app:bot-definitions}. Let \texttt{Bot} range over \texttt{CooperateBot}, \texttt{DefectBot}, \texttt{TitForTatBot}, \texttt{DBot}, \texttt{OBot}, \texttt{EBot}, \texttt{DupocBot}, and \texttt{CupodBot}. We denote its lift by \texttt{TauBot}$(k,\alpha)$, where $k\in\mathbb{N}$ is the common proof-length bound and $\alpha\in[0,1]$ is the caution parameter. The individual names are obtained by prefixing \texttt{Tau} to the base name, as in \texttt{TauDupocBot}. For uniformity, we retain $k$ in the notation for every lift, although only the lifts of \texttt{DupocBot} and \texttt{CupodBot} invoke proof search directly.

\paragraph{The single-hypothesis decision procedure.}
For a candidate opponent $\pi'$, let $B^\delta_{k,\alpha}$ be the decision procedure obtained from \texttt{Bot} by the following substitutions:
\begin{enumerate}
    \item The opponent reference denotes $\pi'$, and every reference to the program itself denotes \texttt{TauBot}$(k,\alpha)$.
    \item Every named probe $\texttt{Bot'}$ is replaced by its corresponding lift \texttt{TauBot'}$(k,\alpha)$. Thus a call that runs the opponent
    against $\texttt{Bot'}$ becomes
    \[
        \mathrm{run}\!\left(
            \pi' \| \delta_{\texttt{TauBot'}(k,\alpha)}
        \right).
    \]
    When $\texttt{Bot'}$ is a self-reference, its lift is
    \texttt{TauBot}$(k,\alpha)$.
    \item The same substitutions are made inside formulas passed to
    \texttt{proof\_search}. Constants, branch order, and proof-search
    budgets are preserved: simulations remain simulations, and proof
    searches remain proof searches.
\end{enumerate}
Here $\delta_\rho$ is the point-mass belief assigning probability one to program $\rho$. 

\paragraph{Aggregation.}
On input belief $\mu$, \texttt{TauBot}$(k,\alpha)$ executes the
entire single-hypothesis procedure for each $\pi'$ in the support of
$\mu$. It cooperates exactly when the total probability assigned to
hypotheses for which that procedure returns $C$ is at least $\alpha$.

\begingroup
\upshape
\begin{lstlisting}[style=pdcode, mathescape=true]
def TauBot$(k, \alpha)(\mu)$:
    if $\alpha = 0$:
        return C
    mass = 0
    for $\pi'$ in $\operatorname{supp}(\mu)$:
        if $\mathrm{run}(\smash{B^\delta_{k,\alpha}} \| \pi') = C$:
            mass = mass + $\mu(\pi')$
    if mass $\geq \alpha$:
        return C
    else:
        return D
\end{lstlisting}
\endgroup

\end{document}